%% file: draft_O2_boot.tex
\documentclass[12pt]{article}
\usepackage[utf8]{inputenc}
\pdfoutput=1
\usepackage{amssymb,amsmath,dsfont,verbatim}
\usepackage[normalem]{ulem}
\usepackage{slashed}
\usepackage{epsfig}
\usepackage{epstopdf}
\usepackage{latexsym}
\usepackage{graphicx}
\usepackage{booktabs}
\usepackage{bbm}
\usepackage{xspace}
\usepackage{cancel}
\usepackage{braket}
\usepackage[babel]{csquotes}

\usepackage{enumitem}
\usepackage[numbers,compress]{natbib}
\usepackage[margin=20pt,small]{caption}
\usepackage{subfig}
\usepackage{float}
\usepackage[toc]{appendix}
\usepackage[usenames,dvipsnames]{xcolor}
\usepackage{float}
\usepackage{datetime}
\usepackage[
colorlinks=false,
linkcolor=darkblue,  
urlcolor=blue,    
filecolor=blue,     
citecolor=red,
linktocpage=true,
bookmarksopen=true,
hidelinks
]{hyperref}
\numberwithin{equation}{section}

\ifpdf
\fi
\newcommand*{\boxedcolor}{red}
\makeatletter
\renewcommand{\boxed}[1]{\textcolor{\boxedcolor}{%
		\fbox{\normalcolor\m@th$\displaystyle#1$}}}
\makeatother

\newcommand{\D}{\Delta}
\newcommand{\hD}{\D}
\newcommand{\Disp}{\text{D}}

\newcommand{\id}{{\mathds 1}}

\newcommand{\pb}{\bar{\phi}}

\newcommand{\Oop}{{\mathcal O}}

\newcommand{\veps}{\varepsilon}

\definecolor{cardinal}{rgb}{0.6,0,0}
\definecolor{darkgreen}{rgb}{0,0.5,0}
\definecolor{golden}{rgb}{0.92, 0.7, 0}
\definecolor{midnight}{rgb}{0, 0, 0.5}
\definecolor{darkblue}{rgb}{0.2, 0, 0.8}

\def\sl{\frak{sl}}

\def\Nm{{\mathcal{N}}}
\def\Om{{\mathcal{O}}}

\def\Sm{{\mathcal{S}}}

\def\DDb{{(\Disp \bar{\Disp})}}
\def\DD{{\Disp^2}}
\def\Dispb{{\bar{\Disp}}}
\def\ttb{{(t \bar{t})}}
\def\TT{{t^2}}

\def\tD{{(t \Disp)}}

\def\pspsb{{(\Psi \bar{\Psi})}}

\usepackage{array}
\newcommand{\PreserveBackslash}[1]{\let\temp=\\#1\let\\=\temp}
\newcolumntype{C}[1]{>{\PreserveBackslash\centering}p{#1}}

\usepackage{tikz}
\tikzset{
  vtx/.style={
    circle,
    draw=blue,
    fill=blue,
    inner sep=1pt
  },
  wcirc/.style={
    circle,
    draw=white,
    fill=white,
    inner sep=2pt
  },
  bcirc/.style={
    circle,
    draw=black,
    fill=black,
    inner sep=1pt
  },
  dcirc/.style={
    circle,
    draw=blue,
    fill=blue,
    inner sep=1pt
  },
  rcirc/.style={
    circle,
    draw=red,
    fill=red,
    inner sep=1pt
  },
  phi/.style={
    thick
  },
  sigma/.style={
    thick,
    dashed
  },
  vl1/.style={
    thick,
    blue
  },
  vl2/.style={
    thick,
    dashed,
    blue
  },
  valign/.style={
    baseline={([yshift=-.55ex]current bounding box.center)}
  }
}

\newcommand\phiHphiHdiagA{
\begin{tikzpicture}[baseline,valign]
  \draw[thick] (0, 0) -- (1.4, 0);
  \draw[dashed] (0.2, 0) to[out=90,in=90] (1.2, 0);
\end{tikzpicture} 
}

\newcommand\phiHphiHdiagB{
\begin{tikzpicture}[baseline,valign]
  \draw[thick] (0, 0) -- (1.4, 0);
  \draw[dashed] (0.2, 0) to[out=90,in=90] (1.2, 0);
  \draw[dashed] (0.5, 0) -- (0.7, 0.25);
  \draw[dashed] (0.9, 0) -- (0.7, 0.25);
  \node at (0.7, 0.28) [bcirc] {};
  \node at (0.5, 0.00) [dcirc] {};
  \node at (0.9, 0.00) [dcirc] {};
\end{tikzpicture} 
}

\begin{document}  
	
	\begin{titlepage}
		
		\medskip
		\begin{center} 
			{\Large \bf Bootstrapping line defects with $O(2)$ global symmetry}

			\bigskip
			\bigskip
			\bigskip
			
			{\bf  Aleix Gimenez-Grau$^{1}$,  Edoardo Lauria$^{2}$, Pedro Liendo$^{1}$ and Philine van Vliet$^1$\\ }
			\bigskip
			\bigskip
			${}^{1}$
			DESY Hamburg, Theory Group, \\
			Notkestraße 85, D-22607 Hamburg, Germany
			\vskip 5mm
			${}^{2}$
			Centre de Physique Th\'eorique (CPHT), Ecole Polytechnique\\91128 Palaiseau Cedex, France
			\vskip 5mm

\texttt{edoardo.lauria@polytechnique.edu,~pedro.liendo@desy.de,\\
	aleix.gimenez@desy.de,philine.vanvliet@desy.de;} \\
		\end{center}
		
		\bigskip
		\bigskip
		
		\begin{abstract}
			\noindent We use the numerical bootstrap to study conformal line defects with $O(2)$ global symmetry. Our results are very general and capture in particular conformal line defects originating from bulk CFTs with a continuous global symmetry, which can either be preserved or partially broken by the presence of the defect. We begin with an agnostic approach and perform a systematic bootstrap study of correlation functions between two canonical operators on the defect: the displacement and the tilt. We then focus on two interesting theories: a monodromy line defect and a localized magnetic field line defect. To this end, we combine the numerical bootstrap with the $\varepsilon$-expansion, where we complement existing results in the literature with additional calculations. For the monodromy defect our numerical results are consistent with expectations, with known analytic solutions sitting inside our numerical bounds. For the localized magnetic field line defect our plots show a series of intriguing cusps which we explore.            
		\end{abstract}

		\noindent

	\end{titlepage}
	
	
	\setcounter{tocdepth}{2}	

	\tableofcontents
	\newpage


\input{sections/introduction}

\input{sections/setup}

\input{sections/epsexp}
\input{sections/numerics}

\input{sections/conclusions}

\section*{Acknowledgements }

We would like to thank A. Antunes, N. Bobev, I. Buri\'c, G. Cuomo, A. Kaviraj, D.~Maz\'a\v c, J. Rong, V. Schomerus, E. Trevisani and B. van Rees for discussions. 
Also, we would like to thank the organizers of the conference `Bootstrap 2022' for providing a place for fruitful discussions. 
EL would like to thank A. Antunes and B. van Rees for their collaboration and discussions on related projects.
EL is supported by the Simons Foundation grant $\#$488659 (Simons Collaboration on the non-perturbative bootstrap). 
AG , PL and PvV acknowledge support from the DFG through the Emmy Noether research group `The Conformal Bootstrap Program' project number 400570283, and through the German-Israeli Project Cooperation (DIP) grant `Holography and the Swampland'.

\appendix

\input{sections/app_monodromy}
\input{sections/app_crosseqs}

\providecommand{\href}[2]{#2}\begingroup\raggedright\endgroup

\end{document}

%% file: sections/introduction.tex
\section{Introduction}\label{sec:intro}

Defects offer an ideal bridge between low-energy and high-energy physics, experiments and theory. 
On the one hand defects such as boundaries and impurities are common in low-energy systems, like ferromagnetic materials and spin systems, where they induce a variety of curious phenomena, including the famous Kondo effect~\cite{Affleck:1995ge,Billo:2013jda,sachdev2000quantum,2000,Cuomo:2021kfm,Cuomo:2022xgw}. On the other hand, line and surface operators, boundary conditions and interfaces offer helpful theoretical tools to investigate the structure of Quantum Field Theories (QFTs). Prototypical examples are Wilson and 't Hooft operators in 4d gauge theories, which provide the order parameter for confinement~\cite{Wilson:1974sk,tHooft:1977nqb,Kapustin:2005py}, and vortices in superfluids and superconductors, which can confine or proliferate, signaling distinct phases of the bulk~\cite{Dias:2013bwa}. Beside specific examples, the language of defects can be used to characterize generalized global symmetries~\cite{Gaiotto:2014kfa}, to investigate dualities between 4d gauge theories~\cite{Aharony:2013hda}, and to study information-theoretic aspects of QFTs~\cite{Bianchi:2015liz,Bianchi:2016xvf}. Furthermore, defects offer alternative strategies to engineer lower-dimensional strongly-coupled QFTs, for example Gaiotto-Witten $\mathcal{T}[G]$ theories~\cite{Gaiotto:2008sa,Gaiotto:2008ak} as conformal boundary conditions for $\mathcal{N}=4$ SYM, 3d abelian gauge theories as conformal boundary conditions for the 4d Maxwell field~\cite{DiPietro:2019hqe}, and $O(N)$ vector models as conformal boundary conditions for the 4d free massless scalar field~\cite{DiPietro:2020fya}. Finally, defects with higher co-dimensions can be used to obtain higher-dimensional descriptions of theories with long-range interactions, for example the long-range Ising model~\cite{Paulos:2015jfa}, as well as the long-range $O(N)$ models~\cite{Giombi:2019enr}.

In this paper we explore the space of co-dimension two conformal line defects with $O(N)$ global symmetry. We base our approach on the modern conformal bootstrap of \cite{Rattazzi:2008pe} (see e.g. \cite{Poland:2018epd,Chester:2019wfx,Bissi:2022mrs} for three recent reviews on the subject), and we focus exclusively on the 1d theory living on the line defect. This work is then a natural extension of~\cite{Gaiotto:2013nva}, where line defects with a $\mathbb{Z}_2$ global symmetry were studied using similar techniques. Other setups where this strategy has been successful include line defects in supersymmetric models in~\cite{Liendo:2018ukf,Gimenez-Grau:2019hez} and in the study of the long-range Ising model in various dimensions in~\cite{Behan:2018hfx}. More recently it has also been employed in order to carve out the space of conformal boundary conditions for a free massless scalar in~\cite{Behan:2020nsf,Behan:2021tcn}.

By definition, a co-dimension two conformal line defect with $O(N)$ global symmetry preserves a `little' conformal group $SL(2,\mathbb{R})$ along the line times a residual `transverse rotations' group $SO(2)_T$ about the line.\footnote{Such `transverse rotations' may be broken e.g. by spinning conformal defects, see e.g.~\cite{Kobayashi:2018okw}.} We can think of $O(N)$ to be the remnant bulk global symmetry after the introduction of the defect in the homogeneous 3d CFT with a global symmetry $G$ so that
\begin{align}\label{symmgroup}
SO(4,1)\times G\longrightarrow	SL(2,\mathbb{R})\times SO(2)_T\times O(N)~, \quad O(N) \subseteq G~.
\end{align}
The residual symmetry of the theory allows us to define local defect operators that are primaries with respect to the `little' conformal group. Such operators will then be conveniently labeled by their scaling dimensions $\D_{}$, which are non-negative in unitary theories, their transverse $SO(2)_T$ spins $s$ and their $SO(N)$ charges $r_i$. Since all connected rotations along the line are trivial, we just need to distinguish between parity-odd and parity-even scalar defect operators, whenever parity on the line is preserved.\footnote{In the context of the 3d Ising model, this parity has been called $\mathcal{S}$-parity~\cite{Gaiotto:2013nva,Billo:2013jda}.} 

Correlation functions between defect operators are akin to those of a one-dimensional CFT and must be crossing symmetric in the usual sense. Hence, while on the one hand these correlation functions enjoy `positivity' in unitary theories and so they can be bootstrapped using semi-definite programming,\footnote{See~\cite{Gliozzi:2015qsa,Gliozzi:2016cmg} for an alternative numerical bootstrap method that does not require positivity.} on the other hand they know little about the bulk. While it remains a very interesting open problem to understand how, for generic bulk CFTs, the bulk information can be encoded into defect correlation functions,\footnote{Some progress in this direction can be found in~\cite{Lauria:2020emq,Behan:2020nsf,Behan:2021tcn} in the context of conformal boundaries and defects for the free massless scalar field and in~\cite{Herzog:2022jqv} in the context of surface defects for the 4d Maxwell field.} at the same time we find that a systematic exploration of co-dimension two line defects with $O(N)$ global symmetry purely based on the numerical conformal bootstrap technique is still missing.\footnote{See~\cite{Ghosh:2021ruh} and references therein for works that combine analytic functionals in 1d with the numerical conformal bootstrap.}
In the present paper we aim at filling this gap, and in doing so we will be starting from co-dimension two defects with $O(2)_F$ global symmetry.\footnote{Here and below (whenever necessary) we will denote the global symmetry group as $O(2)_F$, to be distinguished from the group of transverse rotations about the defect that is denoted as $SO(2)_T$. }

In section~\ref{sec:numerics} we will present a systematic bootstrap study of certain universal defect observables such as mixed correlation functions involving the displacement operator and the tilt operator. As we will review in section~\ref{sec:setup}, these observables capture the breaking of certain bulk local symmetries by the presence of the defect. Along with the agnostic bootstrap and whenever possible, we will try to isolate known defect theories by making gap assumptions inspired by $\varepsilon$-expansion predictions in specific models. There are indeed quite a few instances of interesting conformal defects of this sort which should be allowed by our bootstrap bounds. The preeminent example is the so-called localized magnetic field line defect or magnetic line defect (see~\cite{Parisen_Toldin_2017,Cuomo:2021kfm} and references therein), which for our purposes can be defined in the $O(3)$ CFT and breaks the bulk global symmetry down to $O(2)_F$, i.e.
\begin{align}
S_{\text{LML}}= S_{O(3)}	+ h \int_{-\infty}^\infty d\tau \, \phi_1(\tau)~,
\end{align}
where $\phi_1$ is one component of the fundamental $O(3)$ vector.
This symmetry breaking implies the existence of a tilt in the spectrum, namely a defect primary operator with protected scaling dimension $\Delta_t = 1$ and transforming as a vector of $O(2)_F$. A review of known results on the magnetic line defect, along with original computations of correlators relevant to our study in the $\varepsilon$-expansion will be presented in section~\ref{sec:epsexpTot}. Other interesting known examples of co-dimension two line defects with continuous global symmetry are of the monodromy type, i.e. they can be thought of as boundaries of co-dimension one topological operators that implement a bulk global symmetry transformation $g\in G$, see e.g.~\cite{Billo:2013jda,Gaiotto:2013nva}. For our purposes we can consider a bulk CFT with global symmetry $G$ and a complex scalar field $\Phi$ charged under $U(1)_F\in G$. We can then define an $U(1)_F$-preserving monodromy defect for any element $g=e^{2 \pi i v}\in U(1)_F$ by requiring $\Phi$ to be single valued after a $SO(2)_T$ rotation only up to $g$, i.e.~\cite{Soderberg:2017oaa,Giombi:2021uae,Bianchi:2021snj}
\begin{equation}
 \Phi (r, \theta + 2\pi, \vec{x}) = e^{2 \pi i v} \Phi(r,\theta,\vec{x})\:, \quad v \sim v + 1~,\quad v \in [0,1)~,
\end{equation}
where $(r,\theta)$ are the polar coordinates in the transverse plane with respect to the defect.
For both $v = 0$ (i.e. no monodromy) and $v = 1/2$, the internal $U(1)_F$ symmetry of $\Phi$ will get enhanced to $O(2)_F$, which includes complex conjugation of $\Phi$. 
Now, the specific choice for the monodromy has broken the original symmetry $G$ down to $U(1)_F$, and therefore the resulting defect may feature a tilt operator as well. In our study however we will focus on the $O(2)$ bulk global symmetry, for which the defect spectrum does not contain a tilt operator. We will present a small review on $\varepsilon$-expansion results relevant to our study in section~\ref{sec:epsexpTot}.

%% file: sections/setup.tex
\section{Line defects with global symmetry}\label{sec:setup}
In this section we discuss some universal properties of co-dimension two line defects with global symmetry. We start our analysis with a discussion of the discrete symmetries that characterize a line defect. We then introduce the two main characters of our bootstrap analysis: the displacement operator and the tilt operator. 
We conclude by presenting the crossing equations for the correlators that we will study in section~\ref{sec:numerics}.

\subsection{Discrete symmetries and parity}

In addition to the continuous part of the symmetry, e.g. the group in eq.~\eqref{symmgroup}, we are also interested in the case where the symmetry group involves improper reflections along the parallel or transverse directions with respect to the line, i.e.
\begin{align}
O^+(2,1)\times O(2)_T \times O(2)_F\:, \qquad O^+(2,1)\times SO(2)_T \times O(2)_F\:.
\end{align}
We can think about the $O^+(2,1)$ parity as the improper rotations in 1d, such that spin-odd primaries in higher dimensions become parity-odd in 1d. This is the $\mathcal{S}$-parity of~\cite{Billo:2013jda,Gaiotto:2013nva},
\begin{align}
\mathcal{S}:\qquad \tau\rightarrow -\tau~,\quad	\mathcal{S}(\psi(\tau))=(-1)^{S_\psi}\psi(-\tau)~,\quad S_\psi=0,1~,
\end{align}
so that invariance under $\mathcal{S}$-parity of defect correlation functions implies that $(\tau_i<\tau_{i+1})$
\begin{align}
	\langle \psi_1 (\tau_1)\psi_2 (\tau_2) \psi_3 (\tau_3) \rangle &= (-1)^{S_1+S_2+S_3} \langle \psi_3 (-\tau_3) \psi_2 (-\tau_2) \psi_1 (-\tau_1) \rangle\:.
\end{align}
For the conformal three-point correlation functions, invariance under $\mathcal{S}$-parity means that~\cite{Gaiotto:2013nva}
\begin{equation}\label{eq:Sparity}
	\lambda_{123} =(-1)^{S_1 + S_2 + S_3} \lambda_{213}~.
\end{equation}
Hence, only $\mathcal{S}$-parity even operators are allowed to appear in the fusion of two identical local defect operators.\footnote{Note that the connected component of the conformal group does not change the cyclic order of operators insertions, so  $\lambda_{123} = \lambda_{231}= \lambda_{312}$.}  Throughout this paper we will assume $\mathcal{S}$-parity invariant defects and so $\mathcal{S}$ will play an important role in our numerical bootstrap study of section~\ref{sec:numerics}. 

The second parity assignment is for $O(2)_T$.
We will adopt the convention of~\cite{Gaiotto:2013nva} and denote the action of the $O(2)_T$-parity with $\mathcal{B}$.
The action $\mathcal{B}$ is a reflection in a plane perpendicular to the defect~\cite{Billo:2013jda} which flips the sign of one of the transverse coordinates\footnote{Flipping the sign of both transverse coordinates would be a transformation with $\text{det}(\mathcal{B}^{'}) = 1$ and is part of the connected part of the group $O(2)_T$ instead of the disconnected part.} and reverses the $O(2)_T$ charge (as it follows from the anti-symmetric properties of the generator of the rotations around the defect, $\mathcal{M}_{xy}$)
\begin{align}\label{eq:O2Tparity}
\mathcal{B}:\qquad (x,y)\rightarrow (-x,y)~,\quad	\mathcal{B}(\psi_s(\tau))= b_{\psi_s} \psi_{-s}(\tau)~.
\end{align}
The coefficient $b_{\psi_s}$ determines the parity of the operator. Without loss of generality we can choose a basis of operators such that operators with $s \neq 0$ are even under $O(2)_T$-parity, while operators with $s = 0$ can be both even or odd~\cite{Gaiotto:2013nva}. Throughout this paper we will not require $O(2)_T$ parity to be a symmetry of the defect CFT.

\subsection{Universal defect operators}

\subsubsection{The displacement operator}\label{sec:dispSpec}
When considering a local $d$-dimensional bulk CFT, i.e. with a stress-energy tensor $T^{\mu\nu}$, conservation of $T^{\mu\nu}$ is generically violated by terms localized on the defect~\cite{Gaiotto:2013nva,Billo:2016cpy}:
\begin{align}\label{eq:disp}
	\partial_\mu T^{\mu i} =- \delta^{(q)}(\mathcal{D}) \Disp^i~.
\end{align}
Here $\delta^{(q)}(\mathcal{D})$ is a Dirac delta function with support on the co-dimension $q=d-p$ defect, and $i=1,\ldots,q$ is an index in the directions orthogonal to the defect.
The operator $\Disp$ on the right-hand side of the eq.~\eqref{eq:disp} is the displacement operator, i.e. a defect primary operator of scaling dimension $\hD=p+1$ and a vector under $SO(2)_T$. 
For the co-dimension two case we use the notation
\begin{align}
	\Disp\equiv \Disp^1 + i \Disp^2~,\quad \bar{\Disp}\equiv \Disp^1 - i \Disp^2\:,
\end{align}
to denote the positive and negative spin components.

When taking the correlator of four displacement operators, there are two OPEs that we need to consider. In a theory where $\mathcal{S}$-parity is preserved, the $\Disp \times \bar{\Disp}$ OPE contains the identity operator $\id$, as well as $SO(2)_T$ singlet operators, which can be either even or odd under $\mathcal{S}$-parity as dictated by eq.~\eqref{eq:Sparity}. We denote these operators as $\DDb^{\pm}$. The $\Disp \times \Disp $ OPE exchanges operators with transverse spin $s=2$ and positive $\mathcal{S}$-parity, denoted $\DD$.
All in all, we get
\begin{equation}\label{eq:dispOPE}
	\Disp \times \bar{\Disp} \sim \id +  \DDb^{+} + \DDb^{-} + \cdots \:, \quad \Disp \times \Disp \sim \DD+ \cdots\:.
\end{equation}

\subsubsection{The tilt operator}\label{sec:tiltSpec}
In analogy to the case of the displacement operator, a conformal defect that breaks a local continuous global symmetry of the bulk must feature a tilt operator.\footnote{Here the term `local' refers to the existence of a conserved current in the bulk that can be used to define conserved charges.} Consider the case where the bulk global symmetry $G$ is broken down to a subgroup $H$. If $J^\mu_A$ is the conserved current of the symmetry $G$, then for each symmetry generator broken by the defect we have~\cite{Bray:1977,Cuomo:2021cnb,Padayasi:2021sik}
\begin{align}
	\partial_\mu J^\mu_A=& \delta^{(q)}(\mathcal{D})t_{A}~,\quad A \in G/H~.
\end{align}
The tilt $t_A$ has protected scaling dimension $ \Delta_{t} = p$ and is a scalar under rotations. Furthermore, the tilt consists of $\text{dim}(G)-\text{dim}(H)$ components, which are organized into irreducible representations of the preserved subgroup $H$.
The example we consider in this work is $G=O(N)$ and $H=O(N-1)$, when the tilt transforms in the vector representation of $H$.
For the particular case $N=3$, such that the preserved sugroup is $O(2)_F$, the tilt consists of two components $t^1$ and $t^2$, which can be expressed in complex notation as
\begin{equation}
 t \equiv t^1 + i t^2\:, \qquad \bar{t} \equiv t^1 - i t^2\:.
\end{equation}
The symmetries allow us to define the following OPEs:
\begin{align}\label{ttOPE}
 &t \times \bar{t} \sim \id + (t\bar{t})^++(t\bar{t})^-+\dots ,\quad t \times t \sim t^2 +\dots\:.
\end{align}
In the expressions above, the operators $(t\bar{t})^{\pm}$ are $O(2)_F \times SO(2)_T$ singlets and even (odd) under $\mathcal{S}$-parity.
The operators $t^2$ are charged under $O(2)_F$, and they are even under $\mathcal{S}$-parity.

\subsection{Crossing equations}
\label{sec:crossing}

We are interested in the study of one-dimensional conformal defects in three-dimensional bulk CFTs, but because we restrict our attention to four-point functions on the defect, the crossing equations are identical to the ones for regular one-dimensional CFTs.
In particular, a general four-point function takes the form
\begin{align}
 \langle \phi_i(\tau_1) \phi_j(\tau_2) \phi_k(\tau_3) \phi_l(\tau_4) \rangle
 = \frac{G_{ijkl}(\xi)}
        {|\tau_{12}|^{\Delta_i + \Delta_j} |\tau_{34}|^{\Delta_k + \Delta_l}}
  \left( \frac{|\tau_{24}|}{|\tau_{14}|} \right)^{\Delta_i-\Delta_j}
  \left( \frac{|\tau_{14}|}{|\tau_{13}|} \right)^{\Delta_k-\Delta_l} \: ,
 \label{eq:four-pt-def}
\end{align}
where $\tau$ is the coordinate along the defect, and we introduced the cross-ratio
\begin{align}
 \label{eq:cross-ratio}
 \xi = \frac{\tau_{12} \tau_{34}}{\tau_{13} \tau_{24}} \: .
\end{align}
The external operators are ordered along the line $\tau_1 < \tau_2 < \tau_3 < \tau_4$, such that the cross-ratio takes the values $0 < \xi < 1$.
In the limit when $\tau_1$ approaches $\tau_2$, or equivalently when $\xi \to 0$, the correlation function $G_{ijkl}(\xi)$ admits an expansion in s-channel conformal blocks~\cite{Dolan:2011dv}
\begin{align}\label{eq:OPEexp}
 G_{ijkl}(\xi)
 = \sum_{\Om} \lambda_{ij\Om} \lambda_{kl\Om} 
   g_{\Delta}^{\Delta_{ij},\Delta_{kl}}(\xi) \: , \quad
g_{\Delta}^{\Delta_{ij},\Delta_{kl}}(\xi)
 = \xi^\Delta {}_2F_1(\Delta - \Delta_{ij}, \Delta + \Delta_{kl}; 2\Delta; \xi) \: ,
\end{align}
where $\lambda_{ij \Om}$ are three-point OPE coefficients.
It is natural to think of the one-dimensional CFT as living on a circle, which is the conformal compactification of the real line.
On the circle, it is clear that the correlator should respect cyclicity $\langle \phi_i \phi_j \phi_k \phi_l \rangle = \langle \phi_l \phi_i \phi_j \phi_k \rangle$, which, including the prefactor in eq.~\eqref{eq:four-pt-def}, leads to
\begin{align}
 (1-\xi)^{\Delta_j + \Delta_k} G_{ijkl}(\xi)
 = \xi^{\Delta_i + \Delta_j} G_{lijk}(1-\xi) \:.
\end{align}
The crossing equation is obtained by requiring consistency between cyclicity and the conformal block decomposition.
Using standard manipulations, see for example~\cite{Kos:2014bka}, we can write the crossing equations as 
\begin{align}
 \sum_{\Om} \left[
      \lambda_{ij\Om} \lambda_{kl\Om} F_{\mp,\Delta}^{ij,kl}(\xi)
  \pm \lambda_{kj\Om} \lambda_{il\Om} F_{\mp,\Delta}^{kj,il}(\xi)
 \right] = 0 \: ,
 \label{eq:build-block-cross}
\end{align}
where $F_{\pm, \Delta}^{ij,kl}$ are defined similarly as for higher-dimensional CFTs:
\begin{align}
 F_{\pm,\Delta}^{ij,kl}(\xi) 
 \equiv (1-\xi)^{\Delta_k + \Delta_j} g_{\Delta}^{\Delta_{ij},\Delta_{kl}}(\xi)
 \pm \xi^{\Delta_k + \Delta_j} g_{\Delta}^{\Delta_{ij},\Delta_{kl}}(1-\xi) \:.
\end{align}
With the help of these results, the process of writing all crossing equations becomes straightforward.
Given a set of external operators, one simply lists all non-vanishing four-point functions.
Then, for each inequivalent ordering, eq.~\eqref{eq:build-block-cross} gives the relevant crossing equations.

\subsubsection{Comment on complex notation}

In this work we consider scalar operators charged under $O(2)$ symmetry, so in order to write the crossing equations we should take global-symmetry tensor structures into account. Alternatively, we can exploit complex notation, which means that for an $O(2)$ vector instead of working with a real field $\phi^i$ with a two-valued index $i = 1,2$, we work with a single complex field $\phi = \phi^1 + i \phi^2$ and its complex conjugate $\bar \phi \equiv \phi^*$. In a completely analogous way, we can construct operators with arbitrary charge.
The advantage of complex notation is that it eliminates the need to keep track of global-symmetry indices and tensor structures.
Using again the example of a vector under $O(2)$, the OPE $\phi^i \times \phi^j$ contains the singlet representation ($S$), the antisymmetric representation ($A$), and the symmetric-traceless representation ($T$), each with an associated tensor structure.
Instead, in complex notation we have the OPE $\phi \times \bar \phi$ with $\Sm$-parity even operators corresponding to ($S$), the OPE $\phi \times \bar \phi$ with $\Sm$-parity odd operators corresponding to ($A$), and the OPE $\phi \times \phi$ with $\Sm$-parity even operators corresponding to ($T$).

\subsubsection{One complex scalar}

The simplest case of crossing that we can consider is for one complex scalar $\phi$.
From the defect CFT perspective, this setup has three applications depending on the interpretation given to $\phi$.
In the first application, which we consider in section~\ref{sec:SCdisp}, we take $\Delta_\phi = 2$ and interpret $\phi = \Disp^1 + i \Disp^2 \equiv \Disp$ as the displacement operator of a co-dimension two defect.
In the second application, which we consider in section~\ref{sec:SCtilt}, we take $\Delta_\phi = 1$ and interpret $\phi = t^1 + i t^2 \equiv t$ as the tilt operator.
In the third application, which we consider in section~\ref{sec:monodromy}, we can think of $\phi = \psi_{n+v}\:$ with $n \in \mathbb{Z}$ as a defect mode in a monodromy defect.

Regardless of the interpretation of $\phi$, we can use eq.~\eqref{eq:build-block-cross} on the orderings $\langle \phi \bar \phi \phi \bar \phi \rangle$ and $\langle \phi \phi \bar \phi \bar \phi \rangle$. We find a system of three crossing equations which in vector notation read~\cite{Poland:2011ey}
\begin{align}
 \sum_{\Om^{\pm}} (\lambda_{\phi\bar\phi \Om})^2 \vec V^{\phi\bar\phi}_{\Delta,S}
 + \sum_{\Om^+} |\lambda_{\phi\phi \Om}|^2 \vec V^{\phi\phi}_{\Delta}
 = 0 \, ,
 \label{eq:cross-one-cplx}
\end{align}
where the crossing vectors are given explicitly in eq.~\eqref{eq:single-scalar-vecs}.
The leftmost sum runs over defect primaries of both $\mathcal{S}$-parities, while the rightmost one include only $\mathcal{S}$-parity even defect primaries.
The contribution of the identity operator has not been separated explicitly, but is given by $\vec V_{0,0}^{\phi\bar\phi}$.

\paragraph{One complex generalized free field.}

A notable solution to the crossing equations in eq.~\eqref{eq:cross-one-cplx} is complex generalized free field theory, based on a complex scalar field $\phi$ with scaling dimension $\Delta_\phi$, and its complex conjugate $\bar \phi$.
Using Wick's theorem, the four-point function in the notation of eq.~\eqref{eq:four-pt-def} reads
\begin{align}
	G_{\phi \pb \phi \pb}(\xi)
	& = 1 + \left(\frac{\xi}{1-\xi}\right)^{2 \D_{\phi}} 
	= 1 + \sum_{p=0}^\infty c_p(\Delta_\phi, \Delta_\phi) g_{2\Delta_\phi + p}^{0,0}(\xi)
	\:, \label{eq:gff-PPbPPb} \\
	G_{\phi \phi \pb \pb}(\xi)
	&= \alpha \xi^{2 \D_{\phi}} + \left(\frac{\xi}{1-\xi}\right)^{2 \D_{\phi}} 
	= \sum_{p=0}^\infty \big( 1 + \alpha (-1)^p \big) 
	c_p(\Delta_\phi, \Delta_\phi) g_{2\Delta_\phi + p}^{0,0}(\xi) ~, 
	\label{eq:gff-PPPbPb}
\end{align}
with~\cite{Fitzpatrick:2011dm,Gaiotto:2013nva}
\begin{align}
	c_p(\Delta_1, \Delta_2)
	= \frac{(2 \Delta_1)_p (2 \Delta_2)_p}{p! (2 \Delta_1 + 2 \Delta_2 + p - 1)_p}\: .
\end{align}
The second correlator above contains a parameter $\alpha$, such that $-1 \leq \alpha \leq 1$ in unitary theories. For $\alpha = 1$ the solution corresponds to a generalized free boson (GFB), while for $\alpha = -1$  it corresponds to a generalized free fermion (GFF).

It was noted in~\cite{Gaiotto:2013nva} that the solution to the crossing equation of the generalized real free fermion $\langle \psi(\tau_1) \psi(\tau_2) \psi(\tau_3) \psi(\tau_4) \rangle$ can be extended to a solution of the crossing equations in eq.~\eqref{eq:cross-one-cplx}, if the gap on the scaling dimension of the first operator in the traceless-symmetric representation does not exceed $2\D_{\psi}$. 
The spectra for the $\Sm$-parity even and odd operators are equal in this case. 
For the displacement operator, this happens for 
$\D_{\DD} < 2 \times 2  = 4$, while for the tilt operator the solution appears for $\D_{t^2} < 2 \times 1 = 2$. 

\subsubsection{Tilt and displacement}
Next we consider a mixed system between the tilt operator $t$ and the displacement operator $\Disp$, where we again use complex notation. In order to write down the crossing equations we note that the $\mathcal{S}$-parity even singlets can appear both in $t \times \bar t$ and $\Disp \times \bar \Disp$, while the $\mathcal{S}$-parity odd channels are different in the two OPEs, because the tilt and displacement transform under different $O(2)$ groups.
All in all, crossing reads
\begin{align}
 \sum_{\Om^+}
 (\lambda_{t\bar t\Om} & \; \lambda_{\Disp\bar \Disp\Om})
 \vec V^+_{\Delta}
 \begin{pmatrix} 
    \lambda_{t\bar t\Om} \\ \lambda_{\Disp\bar \Disp\Om} 
 \end{pmatrix}
 + \sum_{\Om^-}
 |\lambda_{t\bar t\Om}|^2 \vec V^{t\bar t,-}_{\Delta}
 + \sum_{\Om^-}
 |\lambda_{\Disp\bar \Disp\Om}|^2 \vec V^{\Disp\bar \Disp,-}_{\Delta} \notag \\
&+ \sum_{\Om^+}
 |\lambda_{tt\Om}|^2 \vec V^{tt}_{\Delta}
 + \sum_{\Om^+}
 |\lambda_{\Disp\Disp\Om}|^2 \vec V^{\Disp\Disp}_{\Delta}
 + \sum_{\Om^\pm}
 |\lambda_{t \Disp\Om}|^2 \vec V^{t\Disp}_{\Delta,S}
 = 0 \: .
 \label{eq:cross-eq-tilt-disp}
\end{align}
Once again the crossing vectors are presented in appendix~\ref{sec:cv-tilt-disp}.

\subsubsection{One real scalar and the tilt}

The third setup we study is crossing for one real scalar $\phi_1$ and a tilt operator $t$, for which we use complex notation.
This has applications to the magnetic line defect of section~\ref{sec:epsexp}, where $\phi_1$ corresponds to the scalar that breaks the bulk symmetry, and the tilt operator must be present due to the symmetry breaking.
The numerical results for this setup will be discussed in section~\ref{sec:MCpinning}.
The crossing equations can be obtain in a similar way as before, but we now obtain seven independent equations that in vector notation read
\begin{align}
 \sum_{\Om^+}
 & (\lambda_{\phi_1\phi_1\Om} \; \lambda_{t\bar t\Om})
 \vec V^+_{\Delta}
 \begin{pmatrix} 
    \lambda_{\phi_1\phi_1\Om} \\ \lambda_{t\bar t\Om} 
 \end{pmatrix}
 + \sum_{\Om^-} (\lambda_{t\bar t\Om})^2
 \vec V^{-}_{\Delta} 
 + \sum_{\Om^+}   |\lambda_{tt\Om}|^2 \vec V^{tt}_{\Delta}
 + \sum_{\Om^\pm} |\lambda_{\phi_1t\Om}|^2 \vec V^{\phi_1 t}_{\Delta,S}
 = 0 \: .
 \label{eq:cross-eq-real-cplx}
\end{align}
The crossing vectors are found in appendix~\ref{sec:cv-real-cplx}.
\emph{Mutatis mutandis}, the same system of crossing equations can be used to study the mixed correlators of $\phi_1$ and $\Disp$.
We will leave this idea for future exploration.

\subsubsection{Two complex scalars}

Finally we study the crossing equations of two unequal complex scalars $\phi_1$ and $\phi_2$.
We apply this to monodromy defects, see section~\ref{sec:MCmono}, where $\phi_1 = \psi_{n_1+v}$ and $\phi_2 = \psi_{n_2+v}$ are two different defect modes of a bulk scalar field.
In total there are twelve crossing equations and five different OPE channels~\cite{Lemos:2015awa}
\begin{align}
 \sum_{\Om^\pm}
 (\lambda_{\phi_1\bar\phi_1\Om} & \; \lambda_{\phi_2\bar\phi_2\Om})
 \vec V_{\Delta,S}
 \begin{pmatrix} 
    \lambda_{\phi_1\bar\phi_1\Om} \\ \lambda_{\phi_2\bar\phi_2\Om} 
 \end{pmatrix}
 + \sum_{\Om^+}
 |\lambda_{\phi_1\phi_1\Om}|^2 \vec V^{11}_{\Delta}
 + \sum_{\Om^+}
 |\lambda_{\phi_2\phi_2\Om}|^2 \vec V^{22}_{\Delta} \notag \\
&+ \sum_{\Om^\pm}
 |\lambda_{\phi_1\phi_2\Om}|^2 \vec V^{12}_{\Delta,S}
 + \sum_{\Om^\pm}
 |\lambda_{\phi_1\bar\phi_2\Om}|^2 \vec V^{1\bar2}_{\Delta,S}
 = 0 \: .
 \label{eq:cross-eq-cplx-cplx}
\end{align}
The crossing vectors can be found in appendix~\ref{sec:cv-cplx-cplx}.

\paragraph{Two complex generalized free fields.}

The OPE coefficients that contain only $\phi_1$ or only $\phi_2$ follow from our discussion above.
The new information is contained in the $\phi_1 \times \phi_2$ OPE, which can be analyzed from the following four-point function
\begin{align}
	\begin{split}
		G_{\phi_1 \phi_2 \pb_2 \pb_1}(\xi) 
		&= \frac{\xi^{\D_1 + \D_2}}{(1-\xi)^{2 \Delta_2}}
		= \sum_{p=0}^\infty c_p(\Delta_1, \Delta_2) 
		g_{\Delta_1 + \Delta_2 + p}^{\Delta_{12},\Delta_{21}}(\xi) \: . \label{eq:gff-MC}
	\end{split}
\end{align}
From here we immediately read off $|\lambda_{\phi_1\phi_2\Om}|^2$, and by sending $\phi_2 \to \bar \phi_2$ we find that the same formula applies to $|\lambda_{\phi_1\bar\phi_2\Om}|^2$.

%% file: sections/epsexp.tex
\section{Defect theories in the \texorpdfstring{$\veps$}{eps}-expansion}\label{sec:epsexpTot}

In this section we use perturbation theory to study two important examples of conformal line defects: the $SO(2)_F$-preserving monodromy defect and the $O(3)$-breaking magnetic line defect. 
We start by reviewing known results on the $SO(2)_F$ monodromy defect in the $\varepsilon$-expansion, which has been studied in great detail in~\cite{Soderberg:2017oaa,Giombi:2021uae,Bianchi:2021snj}. In view of the comparison to the numerical bootstrap results, we add a few OPE coefficients to the CFT data already available in the literature, which can be straightforwardly extracted from the results of~\cite{Giombi:2021uae}.
Then, we study the magnetic line defect in the $\varepsilon$-expansion, and present new results which complement the study performed in~\cite{Cuomo:2021kfm}.
We compare these perturbative results with the predictions from the numerical conformal bootstrap in section~\ref{sec:numerics}.

\subsection{Monodromy defects with $SO(2)_F$ symmetry}
\label{sec:monodromy}

\paragraph{Monodromy defects in the free $O(2)$ model.}
The first example we discuss is a $U(1)_F$-preserving monodromy defect in free theory. This defect, first considered in~\cite{Soderberg:2017oaa,Giombi:2021uae,Bianchi:2021snj}, generalizes the $\mathbb{Z}_2$ twist defect defined in~\cite{Billo:2013jda,Gaiotto:2013nva}.
Following~\cite{Soderberg:2017oaa}, we start from a set of $N=2$ free real scalars $\phi^i$ in the bulk that satisfy
\begin{equation}
 \phi^{i} (r, \theta + 2\pi , \vec{x}) = g^{ij} \phi^{j} (r, \theta, \vec{x})\:, \quad g^{ij} \in O(2)_F\:.
\end{equation}
The scalars $\phi^i$ either get mixed into each other, obtain a minus sign, or remain unchanged when going around the monodromy defect. In terms of the complex combination
\begin{equation}
 \Phi = \phi^{1} + i \phi^{2}\:,
\end{equation}
the most general $U(1)_F$ monodromy becomes\footnote{As explained in~\cite{Bianchi:2021snj} this monodromy can be thought of as a large and constant $U(1)$ background gauge transformation for $\Phi$.}
\begin{equation}\label{eq:monTrans}
 \Phi (r, \theta + 2\pi, \vec{x}) = e^{2 \pi i v} \Phi(r,\theta,\vec{x})\:, \quad v \sim v + 1~,\quad v \in [0,1)~.
\end{equation}
The complex scalar $\Phi$ has an internal $U(1)_F$ symmetry that gets enhanced to $O(2)_F$ for $v = 0$ (the trivial defect) and for $v = \frac{1}{2}$ (the $\mathbb{Z}_2$ monodromy defect).
For these values of $v$, the transformation $\Phi \to \bar{\Phi}$, which belongs to $O(2)_F$ but not $U(1)\simeq SO(2)_F$, is a symmetry. 
As a consequence of the monodromy, the local primary operators allowed to appear in the bulk-defect expansion of $\Phi$ will generically have non-integer spin $s \in \mathbb{Z} + v$, i.e.
\begin{align}
	\Phi(r,\theta,\vec{x})\underset{r\rightarrow 0}{\sim}\sum_{\Psi_s}\sum_{s \in \mathbb{Z} + v} \frac{e^{-i \theta s }}{r^{\Delta_{\Phi}-\hD_{\Psi_s}}}\Psi_s (\vec{x})+\text{c.c.}
\end{align}
The scaling dimensions of the defect modes of $\Phi$ are completely fixed as a consequence of the bulk free equation of motion, and read (see e.g.~\cite{Billo:2013jda,Gaiotto:2013nva,Billo:2016cpy})
\begin{align}
\hD_{\Psi_s} = \Delta_\Phi + |s| = \frac{d-2}{2} + |s| \,.
\end{align}
In terms of real bulk scalar $\phi^i\sim \sum_s e^{i s \theta} \psi_s^i +\text{c.c.}$, the reality condition is that $\bar{\psi}_s^i=\psi_{-s}^i$ and so $\Psi_s = \psi^{1}_{s} + i \psi^{2}_{s}$  satisfies $\bar{\Psi}_{s} \equiv \bar{\psi}^{1}_{s} - i \bar{\psi}^{2}_{s} = \psi^{1}_{-s} - i \psi^{2}_{-s}$.

\paragraph{Monodromy defects in the interacting $O(2)$ model.}
The simplest example of an interacting $U(1)_F$ monodromy defect is obtained by imposing the condition of eq.~\eqref{eq:monTrans} on the fundamental vector of the critical 3d $O(2)$ vector model. In perturbation theory this example is tractable in the standard framework of $\varepsilon$-expansion (with fixed co-dimension $q = 2$).
The bulk is tuned to the Wilson-Fisher fixed point with coupling\footnote{There is a slight difference in notation with respect to~\cite{Giombi:2021uae}, where the symmetry group in the bulk and on the defect is $O(2N)$, while we will denote it by $O(N)$ with $N = $ even, and in particular take $N = 2$, to avoid differences in notation throughout the paper.}
\begin{equation}\label{eq:WFlambda}
 \lambda^{*} = \frac{8 \pi^2}{10} + O(\veps^2)\:.
\end{equation}
The scaling dimensions of the defect modes $\Psi_s$ are found to be~\cite{Soderberg:2017oaa,Giombi:2021uae}
\begin{equation}\label{eq:dimPsiMon}
	\D_{\Psi_s} = 1 + |s| - \frac{\varepsilon}{2} + \frac{1}{5}\frac{v (v-1)}{|s|}\varepsilon + O(\veps^2)\:.
\end{equation}
For the monodromy defect, the displacement operator $\Disp$ appears in the OPE of two defect modes with spins $|s|=v$ and $|s|=v-1$:
\begin{align}
\Psi_{v} \times \bar{\Psi}_{v - 1} \sim \Disp~,\quad\bar{\Psi}_{v} \times {\Psi}_{v - 1} \sim \Dispb~.
\end{align}
In appendix~\ref{app:moremonodromy} we show that it appears in the $\Psi_{v} \times \bar{\Psi}_{v - 1} $ OPE with (squared) OPE coefficient
\begin{equation}
	|\lambda_{\Psi_v \bar{\Psi}_{v-1}\Disp}|^2 = 1 + \frac{\veps}{10} \left(2 H^{1-v}+2 H^{v}-3\right)\:,
\end{equation}
where $H^v$ is the analytic continuation of the harmonic number.
We will also need the scaling dimensions and correponding OPE coefficients of the leading singlets in 
\begin{align}
\Psi_v     \times \bar{\Psi}_v     \sim \id + \mathcal{O}_0+\dots,\quad 
\Psi_{v-1} \times \bar{\Psi}_{v-1} \sim \id + \mathcal{O}_0+\dots.
\end{align}
In appendix~\ref{app:moremonodromy} we show that
\begin{align}\label{epsO0}
	&\Delta_{\mathcal{O}_0} = 2 + 2v - \veps + \veps \left(\frac{4}{5(1 + 2v)} + \frac{2 (v-1)}{5}\right)~.
\end{align}
In section~\ref{sec:MCmono} we will compare the $\varepsilon$-expansion predictions to the numerical bootstrap results. 

It would be interesting to compute the correlator $\langle \Disp \Disp \Dispb \Dispb \rangle$ at the first non-trivial order in  $\varepsilon$-expansion to compare this to our single-correlator numerical bounds for the displacement operator. We will leave this for future work.

\subsection{Localized magnetic field line defect}\label{sec:epsexp}
Let us continue with the determination of three-point OPE coefficients of defect operators for the magnetic line defect.
The calculation is based on and extends the work~\cite{Cuomo:2021kfm}, which focused on the scaling dimensions of low-lying defect operators.
We obtain these results from a Feynman diagram expansion, keeping terms up to order $O(\veps)$ in the $\veps$-expansion.
In order to obtain properly normalized OPE coefficients, we need to determine both two- and three-point functions.
At the end, we also calculate several four-point functions, that by means of the conformal block decomposition allow us to check our results and obtain further coefficients.

\subsubsection{Overview}
\label{sec:localized-overview}

To study the magnetic line defect we use a Lagrangian description that couples the $O(N)$ model to a magnetic field localized along an infinite line.
Using rotation invariance we take the line to be oriented as $x^\mu(\tau) = (\tau, \vec 0)$, while using $O(N)$ invariance, we choose the magnetic field to be in the $\phi_1$ direction.
All in all, the action is
\begin{align}
 S = \int d^dx \left( \frac12 (\partial_\mu \phi_a)^2 
   + \frac{\lambda_0}{4!} (\phi_a^2)^2 \right)
   + h_0 \int_{-\infty}^\infty d\tau \, \phi_1(x(\tau)) \, , \qquad
 a = 1, \ldots, N \, .
 \label{eq:action-pinning-def}
\end{align}

Let us start reviewing the results of~\cite{Cuomo:2021kfm}, which motivate our bootstrap setup in section~\ref{sec:MCpinning}.
Because the defect in eq.~\eqref{eq:action-pinning-def} breaks the global symmetry $O(N) \to O(N-1)$, there exists a tilt operator besides the displacement operator.
In perturbation theory, the tilt and displacement operators are identified as
\begin{align}
 t_{\hat a} \propto \phi_{\hat a} \, ,  \qquad
 \Disp \propto \nabla \phi_1 \, , \qquad
 \hat a = 2, \ldots, N \, ,
\end{align}
and as usual they have protected dimension $\Delta_t = 1$ and $\Delta_\Disp = 2$.
After the tilt, the operator with the second-lowest dimension is the localized magnetic field $\phi_1$.
The scaling dimension of $\phi_1$ to two-loop order in the $\veps$-expansion reads
\begin{align}
 \Delta_{\phi_1}
 = 1 
 + \veps 
 - \frac{3 N^2+49 N+194}{2 (N+8)^2} \veps^2
 + O(\veps^3) 
 \;\; \xrightarrow{\text{Pad\'e}} \; 1.55 \, . \label{eq:dimPhi}
\end{align}
Furthermore, inputting information from $d=2$ in a Pad\'e approximant, the authors of~\cite{Cuomo:2021kfm} estimated the value $\Delta_{\phi_1} \approx 1.55$, which is also consistent with their $1/N$ results and Monte-Carlo simulations~\cite{2017PhRvB95a4401P,2014arXiv14123449A}.

Since the two lowest-dimensional operators on the defect are $t_{\hat a}$ and $\phi_1$, a natural candidate for a bootstrap study is the mixed correlator involving them, which is the one-dimensional analog of~\cite{Kos:2015mba,Kos:2016}.
In order to motivate gap assumptions in the numerical study, let us look at the lowest-lying operators in the different OPE channels.
In the singlet channel $(S)$, which appears for $\phi_1 \times \phi_1$ or $(t_{\hat a} \times t_{\hat b})_{S}$, the lowest dimension operators are $\phi_1 + s_- +  s_+ + \ldots$.
Here $s_\pm$ are linear combinations of $\phi_1^2$ and $\phi_a^2$ with scaling dimension
\begin{align}
 \Delta_{s_\pm}
 & = 2 
 + \veps \, \frac{3N + 20 \pm \sqrt{N^2 + 40N + 320}}{2(N+8)}
 + O(\veps^2) \, .
 \label{eq:delta-spm}
\end{align}
Similarly, vector operators appear in the OPE $\phi_1 \times t_{\hat a} = t_{\hat a} + V_{\hat a} + \ldots$.
The leading vector is $V_{\hat a} \propto \phi_1 \phi_{\hat a}$, and its dimension reads
\begin{align}
 \Delta_V 
 & = 2 + \veps \, \frac{N+10}{N+8} + O(\veps^2) \: . \label{eq:dimV}
\end{align}
Finally, in the OPE $t_{\hat a} \times t_{\hat b}$ there is an antisymmetric channel $(A)$, where the lowest-dimensional operator is $A_{\hat a\hat b} \propto \phi_{[\hat a} \phi_{\hat b]}$, and a symmetric-traceless channel $(T)$ with the lowest-lying operator $T_{\hat a \hat b} \propto  \phi_{\hat a} \phi_{\hat b}$.
Their dimensions are given by
\begin{align}
 \Delta_A
 = 3 + O(\veps^2) \, , \qquad 
 \Delta_T
 = 2 + \frac{2 \veps}{N+8} + O(\veps^2) \: . \label{eq:dimAT}
\end{align}

Besides scaling dimensions, the numerical conformal bootstrap can also probe the three-point OPE coefficients of these operators.
The goal of the rest of the section will be to compute these OPE coefficients to leading order in the $\veps$-expansion, see table~\ref{tab:OPE-coeffs} for a summary of the main results. Besides two- and three-point functions, we also compute several four-point functions, which thanks to the conformal block decomposition, contain information of many other OPE coefficients.

\begin{table}
 \centering
 {\renewcommand{\arraystretch}{1.3}
 \begin{tabular}{ C{4em} | C{4em} | C{4em} | C{4em} | C{4em} | C{4em} | C{4em} }
   $\lambda_{\phi_1\phi_1\phi_1}$ &
   $\lambda_{tt\phi_1}$ &
   $\lambda_{\phi_1\phi_1s_\pm}$ &
   $\lambda_{tts_\pm}$ &
   $\lambda_{\phi_1tV}$ &
   $\lambda_{ttA}$ &
   $\lambda_{ttT}$ \\ \hline
  ~\eqref{eq:lambda-PPP} &
  ~\eqref{eq:lambda-PPP} &
  ~\eqref{eq:lambda-PPS} &
  ~\eqref{eq:lambda-TTS} &
  ~\eqref{eq:lambda-VandT} &
  ~\eqref{eq:block-Anti} &
  ~\eqref{eq:lambda-VandT} 
 \end{tabular}}
 \caption{Summary of the most important OPE coefficients computed in this section. Further coefficients appear in~\eqref{eq:lambda-grad} or can be extracted from the four-point functions in section~\ref{sec:four-point}.}
 \label{tab:OPE-coeffs}
\end{table}

\subsubsection{Conventions}
\label{sec:epsexp-conv}

Throughout this section we follow the conventions of~\cite{Cuomo:2021kfm}, so in particular the action is given by eq.~\eqref{eq:action-pinning-def} and the scalar propagator in free theory is
\begin{align}
 \begin{tikzpicture}[baseline,valign]
  \draw[dashed] (0, 0) -- (0.8, 0);
 \end{tikzpicture} 
 \;\, \equiv \;
 \langle \phi_a(x_1) \phi_b(x_2) \rangle_{\lambda_0=h_0=0}
 \, = \;
 \frac{\kappa \delta_{ab}}{(x_{12}^2)^{1 - \frac{\veps}{2}}} \, ,
 \qquad
 \kappa
 = \frac{\Gamma \! \left(\frac{d}{2}\right)}{2 \pi ^{d/2} (d-2)} \, .
 \label{eq:free-prop}
\end{align}
In perturbative expansions there is a bulk four-point vertex and a vertex that couples a bulk operator to the defect
\begin{align}
 \begin{tikzpicture}[baseline,valign]
  \draw[dashed] ( 0.2, 0.2) -- (-0.2, -0.2);
  \draw[dashed] (-0.2, 0.2) -- ( 0.2, -0.2);
  \node at (0,0) [bcirc] {};
 \end{tikzpicture} 
 \;\; \equiv \; 
 - \lambda_0 \int d^d x \ldots \, , \qquad
  \begin{tikzpicture}[baseline,valign]
  \draw[thick] (-0.1, 0) -- (0.7, 0);
  \draw[dashed](0.3, 0) -- (0.3, 0.6);
  \node at (0.3, 0) [dcirc] {};
 \end{tikzpicture}
 \;\; \equiv \;
 - h_0 \int_{-\infty}^\infty d \tau \ldots \, .
\end{align}
Notice that only $\phi_1$ couples to the line, and not all $\phi_{\hat a}$ for $\hat a = 2, \ldots, N$.
We work in dimensional regularization with minimal subtraction, so the bare couplings $\lambda_0$ and $h_0$ are related to the renormalized ones as
\begin{align}
 \lambda_0 
 = \lambda M^\veps \left( 1
 + \frac{\lambda}{(4\pi)^2} \frac{N+8}{3 \veps}
 + O(\lambda^2)
 \right) \, , \quad
 h_0
 = h M^{\veps/2} \left( 1
 + \frac{\lambda}{(4\pi)^2} \frac{h^2}{12 \veps}
 + O(\lambda^2)
 \right) \, .
\end{align}
The renormalized couplings depend on the renormalization scale $M$ as
\begin{align}
 \beta_\lambda 
&= M \frac{d\lambda}{dM}
 = -\lambda \veps
 + \frac{\lambda^2}{(4 \pi )^2} \frac{N+8}{3}
 + O(\lambda^3) \, , \nonumber\\
 \beta_h 
&= M \frac{d h}{dM}
 = -\frac{h \veps}{2}
 + \frac{\lambda }{(4 \pi)^2} \frac{h^3}{6} 
 + O(\lambda^2) \, ,
\end{align}
so there exists a non-trivial fixed point where we evaluate most of our results: 
\begin{align}
 \frac{\lambda_*}{(4\pi)^2} 
 = \frac{3\veps}{N+8} + O(\veps^2) \, , \qquad
 h_*^2 
 = N+8 + O(\veps) \, .
\end{align}
Since we are interested in obtaining OPE coefficients at order $O(\veps)$, we shall consider diagrams with at most one bulk vertex insertion $\lambda_* \sim O(\veps)$.
On the other hand, we have to allow an arbitrary number of defect insertions because $h_* \sim O(1)$.
However, in practice only a finite number of diagrams will contribute at any given order in $\veps$.

We often split the coordinates into a direction $\tau$ parallel to the defect and $\vec x \in \mathbb{R}^{d-1}$ directions orthogonal to the defect.
We shall only consider correlation functions of operators that live on the defect, for which we use the shorthand notation $\Om(\tau) = \Om(\tau, \vec x = 0)$.

\subsubsection{Two-point functions}
\label{sec:two-pt}

We obtain all two-point functions of operators of the schematic form $\phi, \, \phi^2, \, \nabla \phi$, where $\phi$ is the fundamental scalar and $\nabla$ are derivatives orthogonal to the defect. Besides rederiving their scaling dimensions, which appeared previously in~\cite{Cuomo:2021kfm}, we obtain the overall normalization, which is needed in the calculation of higher-point functions.

\subsubsection*{The correlator \texorpdfstring{$\langle \, \phi \, \phi \, \rangle$}{< phi phi >}}
\label{sec:2pt-PP}

Let us start with the simplest example: the two-point function of the fundamental field. 
To order $O(\veps)$ only two diagrams contribute:
\begin{align}
 & \langle \, \phi_a(\tau) \, \phi_b(0) \, \rangle
 \; = \; \phiHphiHdiagA 
 \; + \; \phiHphiHdiagB 
 \; + \; \ldots \, .
 \label{eq:two-pt-fund}
\end{align}
The first diagram is obtained setting $\vec x_1 = \vec x_2 = 0$ in the free propagator~\eqref{eq:free-prop}.
For the integral that enters the second diagram, we first integrate the two defect insertions using
\begin{align}
 & \int_{-\infty}^\infty \frac{d\tau}{(|\vec x|^2 + \tau^2)^\Delta}
 = \frac{\sqrt{\pi} \, \Gamma \! \left(\Delta -\frac{1}{2}\right)}{\Gamma(\Delta )}
   \frac{1}{|\vec x^2|^{\Delta-\frac{1}{2}}} \, ,
\end{align}
and then the bulk vertex using
\begin{align}
 & \int d^d x_3 \, \frac{|\vec x_3|^{\veps}}{(x_{13}^2 x_{23}^2)^{1-\frac{\veps}{2}}}
 = \frac{\sqrt{\pi} \, \Gamma \! \left(1-\frac{3 \veps}{2}\right) 
         \Gamma(\veps)^2 \, \Gamma \! \left(\frac{\veps+1}{2}\right)}
        { 2 \Gamma \! \left(1-\frac{\veps}{2}\right)^2 \Gamma (2 \veps)}
 \frac{1}{|\tau_{12}^2|^{1-\frac{3 \veps}{2}}} \, ,
\end{align}
which follows from Schwinger parametrization.
Using these results, it is a simple bookkeeping exercise to obtain the contribution of each diagram:
\begin{align}
 & \phiHphiHdiagA = \frac{\kappa \delta_{ab}}{\tau^{2 - \veps}} \, , \nonumber\\
 & \phiHphiHdiagB 
 = - \frac{\lambda_0 h_0^2}{(4\pi)^4} \,
 \frac{2 \delta_{1a} \delta_{1b} + \delta_{ab}}{\tau^{2 - 3 \veps}}
 \left( \frac{2}{3 \veps} - \frac{1}{3} + \aleph + O(\veps) \right) \, .
\end{align}
To simplify the notation, we have introduced the constant 
\begin{align}
 \aleph \equiv 1 + \gamma_E + \log \pi \, ,
\end{align}
which appears repeatedly in the calculations below but drops out of physical quantities such as scaling dimensions and OPE coefficients.
We now introduce renormalized fields, which transform irreducibly under the unbroken symmetry group $O(N-1)$:
\begin{align}
 \phi_1 \equiv Z_{\phi_1} [\phi_1] \, , \qquad
 \phi_{\hat a} \equiv Z_{t} \, t_{\hat a} \, .
\end{align}
The operator $t_{\hat a}$ is the tilt operator from section~\ref{sec:tiltSpec}.
The two renormalization factors are obtained demanding that poles in $\veps$ cancel in~\eqref{eq:two-pt-fund}, giving
\begin{align}
 Z_{\phi_1}
 = 1 - \frac{\lambda}{(4\pi)^2} \frac{h^2}{4 \veps} + O(\lambda^2) \, , \qquad
 Z_{t}
 = 1 - \frac{\lambda}{(4\pi)^2} \frac{h^2}{12 \veps} + O(\lambda^2) \, .
\end{align}
From the renormalization factor one can immediately obtain the anomalous dimension at the critical point using $\gamma_\Om = M \frac{d}{dM} \log Z_\Om$.
The results have been summarized in section~\ref{sec:localized-overview}, and are in perfect agreement with~\cite{Cuomo:2021kfm}.
Finally, note that the two-point functions at the critical point read
\begin{align}
 \langle \, [\phi_1](\tau) \, [\phi_1](0) \, \rangle
 & = \frac{\Nm_{\phi_1}^2}{\tau^{2\Delta_{\phi_1}}} \, , \quad \quad \;
 \Nm_{\phi_1}^2
 = \kappa \left(1-\frac{3 \aleph }{2} \, \veps + O(\veps^2) \right) , \nonumber\\
 \langle \, t_{\hat a}(\tau) \, t_{\hat b}(0) \, \rangle
 & = \delta_{\hat a \hat b} \,
     \frac{\Nm_t^2}{\tau^2} \, , \quad \quad \;
 \Nm_{t}^2
 = \kappa \left(1-\frac{\aleph}{2} \, \veps + O(\veps^2) \right) .
\end{align}

\subsubsection*{The correlator \texorpdfstring{$\langle \, \nabla \phi \, \nabla \phi \, \rangle$}{< dphi dphi >}}

Similarly, we can consider the two-point function  $\langle \nabla_i \phi_a \nabla_j  \phi_b \rangle$, which again consists of two diagrams at this order:
\begin{align}
 & \phiHphiHdiagA = \kappa (d-2) \frac{\delta_{ij}\delta_{ab}}{\tau^{4 - \veps}} \, , \nonumber\\
 & \phiHphiHdiagB 
 = - \frac{\lambda_0 h_0^2}{(4\pi)^4} \,
 \frac{\delta_{ij}(2 \delta_{1a} \delta_{1b} + \delta_{ab})}{\tau^{4 - 3 \veps}}
 \left( \frac{4}{9 \veps} - \frac{8}{27} + \frac 23 \aleph + O(\veps) \right) \, .
\end{align}
The integrals are computed as before, but one needs to be careful to first take transverse derivatives with respect to $\vec x_1$ and $\vec x_2$ and then sending $\vec x_1, \vec x_2 \to 0$.
Once again we introduce renormalized fields
\begin{align}
 \nabla \phi_{1} \equiv Z_{\Disp} \, \Disp \, , \qquad
 \nabla \phi_{\hat a}
 \equiv Z_{\nabla \phi} \, [\nabla \phi_{\hat a}] \, , 
\end{align}
where $\Disp$ is the displacement operator which transforms as a vector under $SO(d-1)$ transverse rotations.
Cancelling poles in $\veps$ we find
\begin{align}
 Z_\Disp
 = 1 
 - \frac{\lambda}{(4\pi)^2} \frac{h^2}{12 \veps} 
 + O(\lambda^2) \, , \qquad
 Z_{\nabla\phi}
 = 1 
 - \frac{\lambda}{(4\pi)^2} \frac{h^2}{36 \veps} 
 + O(\lambda^2) \, .
\end{align}
It follows that the scaling dimension of the displacement is protected, and we obtain the scaling dimension $\Delta_{\nabla \phi} = 2 - \frac{1}{3} \veps + O(\veps^2)$, in agreement with~\cite{Cuomo:2021kfm}.
The defect-defect two-point functions at the critical point read
\begin{align}
 \langle \, \Disp_i(\tau) \, \Disp_j(0) \, \rangle
 & = \delta_{ij} \,
     \frac{\Nm_\Disp^2}{\tau^4} \, , \quad \quad \quad \, \;
 \Nm_\Disp^2
 = 2 \kappa
 \left(1-\frac{4 + 3 \aleph}{6} \, \veps + O(\veps^2) \right) , \nonumber\\
 \langle \, [\nabla_i \phi_{\hat a}](\tau) \, 
            [\nabla_j \phi_{\hat b}](0) \, \rangle
 & = \delta_{\hat a \hat b} \delta_{ij} \,
     \frac{\Nm_{\nabla\phi}^2}
     {\tau^{2\Delta_{\nabla\phi}}} \, , \quad
 \Nm_{\nabla\phi}^2
 = 2 \kappa \left(1-\frac{10 + 3\aleph}{18} \, \veps + O(\veps^2) \right) .
\end{align}

\subsubsection*{The correlator \texorpdfstring{$\langle \, \phi^2 \, \phi^2 \, \rangle$}{< phi2 phi2 >}}

Finally, the last type of operators we are interested in are composites of two fundamental fields:
\begin{align}
 \langle \,
   \phi_a \phi_b(\tau) \, 
   \phi_c \phi_d(\tau) \, \rangle
 \;\; = \;\;
 \begin{tikzpicture}[baseline,valign]
  \draw[thick] (0, 0) -- (1.4, 0);
  \draw[dashed] (0.2, 0) to[out=90,in=90] (1.2, 0);
  \draw[dashed] (0.2, 0) to[out=90,in=-180] (0.7, 0.5) to[out=0,in=90] (1.2, 0);
\end{tikzpicture}
\;\; + \;\;
 \begin{tikzpicture}[baseline,valign]
  \draw[thick] (0, 0) -- (1.4, 0);
  \draw[dashed] (0.2, 0) to[out=90,in=90] (1.2, 0);
  \draw[dashed] (0.2, 0) to[out=30,in=-120](0.7, 0.28) to[out=-60,in=150] (1.2, 0);
  \node at (0.7, 0.28) [bcirc] {};
\end{tikzpicture} 
\;\; + \;\;
 \begin{tikzpicture}[baseline,valign]
  \draw[thick] (0, 0) -- (1.4, 0);
  \draw[dashed] (0.2, 0) to[out=90,in=90] (1.2, 0);
  \draw[dashed] (0.2, 0) to[out=90,in=-180] (0.7, 0.5) to[out=0,in=90] (1.2, 0);
  \draw[dashed] (0.5, 0) -- (0.7, 0.25);
  \draw[dashed] (0.9, 0) -- (0.7, 0.25);
  \node at (0.7, 0.28) [bcirc] {};
  \node at (0.5, 0.00) [dcirc] {};
  \node at (0.9, 0.00) [dcirc] {};
\end{tikzpicture}
 \;\; + \;\; \ldots
\end{align}
The first and third diagrams are computed as in section~\ref{sec:2pt-PP}, whereas the second is a chain diagram, for which the integral is well known (see e.g.~\cite{kleinert2001critical}):
\begin{align}
   \int \frac{d^d x_3}{(x_{13}^2)^{\Delta_1} (x_{23}^2)^{\Delta_2}}
 = \frac{\pi^{\frac{d}{2}}}{(x_{12}^2)^{\Delta_1 + \Delta_2 - \frac{d}{2}}}
   \frac{\Gamma \! \left( \frac{d}{2} - \Delta_1 \right)}{\Gamma(\Delta_1)}
   \frac{\Gamma \! \left( \frac{d}{2} - \Delta_2 \right)}{\Gamma(\Delta_2)}
   \frac{\Gamma \! \left( \Delta_1 + \Delta_2 - \frac{d}{2} \right)}
        {\Gamma(d-\Delta_1-\Delta_2)} \, .
\end{align}
Once again, we are interested in reducing $\phi_a \phi_b$ into irreducible components.
On one hand, we can form a vector and a symmetric-traceless operator as follows:
\begin{align}
 Z_V V_{\hat a} = \phi_1 \phi_{\hat a} \, , \qquad
 Z_T T_{\hat a \hat b} 
 = \phi_{\hat a} \phi_{\hat b}  
 - \frac{\delta_{\hat a \hat b}}{N-1} \phi_{\hat c}^2 \, ,
\end{align}
with the following renormalization factors:
\begin{align}
 Z_V 
 = 1
 - \frac{\lambda }{(4 \pi)^2} \frac{h^2 + 2}{3 \veps}
 + O(\lambda^2) \, , \qquad
 Z_T 
 = 1
 - \frac{\lambda }{(4 \pi)^2} \frac{h^2 + 4}{6 \veps}
 + O(\lambda^2) \, .
\end{align}
Again, it is straightforward to extract the anomalous dimensions at the critical point and the normalization of the two-point functions:
\begin{align}
 \langle \, V_{\hat a}(\tau) \, V_{\hat b}(0) \, \rangle
&= \delta_{\hat a \hat b} \, \frac{\Nm_V^2}{\tau^{2\Delta_V}} \, , \quad\quad
 \Nm_V^2 
 = \kappa^2 \left(1-\frac{2N+18}{N+8} \, \aleph \, \veps + O(\veps^2)\right) \, , \nonumber \\
 \langle \, T_{\hat a \hat b}(\tau) \, T_{\hat c \hat d}(0) \, \rangle
&= \mathbf T_{\hat a \hat b, \hat c \hat d} \, 
   \frac{\Nm_T^2}{\tau^{2\Delta_T}} \, , \quad
 \Nm_T^2
 = 2 \kappa^2 \left(1-\frac{N+10}{N+8}\, \aleph \, \veps + O(\veps^2)\right) \, .
\end{align}
Here we have introduced a symmetric-traceless tensor which will be useful later:
\begin{align}
 \mathbf T_{\hat a \hat b, \hat c \hat d}
 = \frac{1}{2} \delta_{\hat a \hat c} \delta_{\hat b \hat d}
 + \frac{1}{2} \delta_{\hat a \hat d} \delta_{\hat b \hat c} 
 - \frac{\delta_{\hat a \hat b} \delta_{\hat c \hat d}}{N-1} \, .
 \label{eq:tens-struct}
\end{align}
On the other hand, we can form two independent scalars $\phi_1^2$ and $\phi_a^2$, which mix at $O(\veps)$ due to quantum corrections.
The renormalized fields $s_\pm$ are defined as
\begin{align}
 \begin{pmatrix} \phi_1^2 \\ \phi_a^2 \end{pmatrix}
 = Z_s
 \begin{pmatrix} s_- \\ s_+ \end{pmatrix} \, ,
 \label{eq:mixing-problem}
\end{align}
where the renormalization factor $Z_s$ is a two-by-two matrix. 
To determine $Z_s$ one requires that three-point functions of renormalized fields
\begin{align}
 \langle \, [\phi_1] \, [\phi_1] \, s_\pm \,  \rangle \, , \quad
 \langle \, [\phi_1] \, t_{\hat a} \, s_\pm \, \rangle \, ,  \quad
 \langle \, t_{\hat a} \, t_{\hat b} \, s_\pm \, \rangle \, ,  \quad
\end{align}
have no poles in $\veps$. 
We explain how to compute these three-point functions in section~\ref{sec:three-pt-PPP2}.
Furthermore, one should require that $Z_s$ is such that $s_{\pm}$ have anomalous dimensions that do not mix.
Demanding this we find the anomalous-dimension matrix
\begin{align}
 \gamma
 = Z^{-1}_s \frac{\partial Z_s}{\partial \log M} \Big|_{\text{fixed point}}
 = \left(
\begin{array}{cc}
 \frac{5 N + 36 - \sqrt{N^2+40 N+320}}{2N + 16} & 0 \\
 0 & \frac{5 N + 36 + \sqrt{N^2+40 N+320}}{2N + 16} \\
\end{array}
\right) \veps + O(\veps^2) \, .
\end{align}
Our choice of $Z_s$ guarantees that the two-point functions are orthogonal, as one usually requires in CFT:
\begin{align}
 \langle \, s_{\pm}(\tau) \, s_{\pm}(0) \, \rangle
 = \frac{\Nm_{s_\pm}^2}{\tau^{2\Delta_{s_\pm}}} \, , \qquad
 \langle \, s_{\pm}(\tau) \, s_\mp(0) \, \rangle
 = 0 \, .
\end{align}
The formulas for $Z_s$ and $\Nm_{s_\pm}^2$ are somewhat complicated and not particularly illuminating, so instead of writing them explicitly we attach them in a notebook.

\subsubsection{Three-point functions}
\label{sec:three-pt}

Having determined the scaling dimension and normalization of the defect operators of interest, we are ready to compute some of their three-point OPE coefficients $\lambda_{\Om_1\Om_2\Om_3}$.
Here we shall focus on parity-even operators, that have three-point functions of the form
\begin{align}
 \langle \Om_1(\tau_1) \Om_2(\tau_2) \Om_3(\tau_3) \rangle
 = \frac{\Nm_{\Om_1} \Nm_{\Om_2} \Nm_{\Om_3} \, \lambda_{\Om_1\Om_2\Om_3}}
        {|\tau_{12}|^{\Delta_1+\Delta_2-\Delta_3} 
         |\tau_{13}|^{\Delta_1+\Delta_3-\Delta_2}
         |\tau_{23}|^{\Delta_2+\Delta_3-\Delta_1}} \, .
 \label{eq:three-pt}
\end{align}
By choosing different external operators there are many three-point functions that can be computed.
Here we focus on a subset of them, which can be compared to our numerical study or in future works.

\subsubsection*{The correlator \texorpdfstring{$\langle \, \phi \, \phi \, \phi \, \rangle$}{< phi phi phi >}}

The simplest three-point functions are the ones that involve only the fundamental scalar.
These are zero at tree level, but receive a contribution at order $O(\veps)$ from the following diagram:
\begin{align}
 \langle \, \phi_a (\tau_1) \, \phi_b(\tau_2) \, \phi_c(\tau_3) \, \rangle
 \;\; = \;\;
 \begin{tikzpicture}[baseline,valign]
  \draw[thick] (0, 0) -- (1.4, 0);
  \draw[dashed] (0.2, 0) to[out=90,in=90] (1.2, 0);
  \draw[dashed] (0.5, 0) -- (0.7, 0.25);
  \draw[dashed] (0.9, 0) -- (0.7, 0.25);
  \node at (0.7, 0.28) [bcirc] {};
  \node at (0.9, 0.00) [dcirc] {};
\end{tikzpicture}
 \;\; + \;\; \ldots
 \label{eq:diag-3pt-PPP}
\end{align}
Because this diagram is proportional to a factor $\lambda_* \propto \veps$, it suffices to evaluate the integral in exactly $d=4$:
\begin{align}
 \int \frac{d\tau_4 \, d^4 x_5}{x^2_{15} x^2_{25} x^2_{35} x^2_{45} }
 = \frac{2 \pi ^4}{\sqrt{\tau^2_{12} \tau^2_{13} \tau^2_{23}}} \, .
 \label{eq:int-3pt-1}
\end{align}
To compute this integral, we exploit that it is invariant under the one-dimensional conformal group, so we use a frame where $\tau_1 = 0$, $\tau_2 = 1$ and $\tau_3 = \infty$.
We start integrating over $\tau_4$, then the orthogonal directions $\vec x_5^2$, and finally over $\tau_5$, all the steps being elementary.
Splitting the diagram~\eqref{eq:diag-3pt-PPP} into irreducible components under $O(N-1)$, and keeping track of all normalization factors, we find the two OPE coefficients
\begin{align}
 \lambda_{\phi_1 \phi_1 \phi_1} 
 = \frac{3 \pi \veps}{\sqrt{N+8}} 
 + O(\veps^2) \, , \qquad
 \lambda_{t t \phi_1} 
 = \frac{\pi  \veps}{\sqrt{N+8}} 
 + O(\veps^2) \, .
 \label{eq:lambda-PPP}
\end{align}

\subsubsection*{The correlator \texorpdfstring{$\langle \, \nabla \phi \, \nabla \phi \, \phi \, \rangle$}{< dphi dphi phi >}}

We can also consider a three-point function where two of the external operators are orthogonal derivatives $\langle \nabla \phi_a (\tau_1) \nabla \phi_b(\tau_2) \phi_c(\tau_3) \rangle$.
The same diagram as in~\eqref{eq:diag-3pt-PPP} contributes, but now it leads to the integral
\begin{align}
 \lim_{\vec x_1, \vec x_2 \to 0} 
 \frac{\partial}{\partial \vec x_{1,i}} \frac{\partial}{\partial \vec x_{2,j}}
 \int \frac{d\tau_4 \, d^4 x_5}{x^2_{15} x^2_{25} x^2_{35} x^2_{45} }
 = \frac{4 \pi^4 / 3}{\sqrt{\tau^6_{12} \tau^2_{13} \tau^2_{23}}} \, ,
 \label{eq:int-3pt-2}
\end{align}
which has been computed as before in eq.~\eqref{eq:int-3pt-1}.
Splitting the result into irreducible components under $O(N-1)$ and keeping track of all normalization factors, we find the three OPE coefficients
\begin{align}
 \lambda_{DD\phi_1}
 =\frac{\pi  \veps}{\sqrt{N+8}} + O(\veps^2) \, , \quad
 \lambda_{\Disp\nabla\phi t}
 =\frac{\pi  \veps}{3 \sqrt{N+8}} + O(\veps^2)  \, , \quad 
 \lambda_{\nabla\phi \nabla\phi \phi_1}
 =\frac{\pi  \veps}{3 \sqrt{N+8}} + O(\veps^2) \, .
 \label{eq:lambda-grad}
\end{align}

\subsubsection*{The correlator \texorpdfstring{$\langle \, \phi \, \phi \, \phi^2 \, \rangle$}{< phi phi phi2 >}}
\label{sec:three-pt-PPP2}

The last type of OPE coefficient we consider involves two fundamental fields and a composite one:
\begin{align}
 \langle \, \phi_a(\tau_1) \, \phi_b(\tau_2) \, 
            \phi_c \phi_d(\tau_3) \, \rangle
 \;\; = \;\;
 \begin{tikzpicture}[baseline,valign]
  \draw[thick] (-0.4, 0) -- (1.4, 0);
  \draw[dashed] (-0.2, 0) to[out=90,in=90] (1.2, 0);
  \draw[dashed] (0.2, 0) to[out=90,in=90] (1.2, 0);
\end{tikzpicture}
\;\; + \;\;
 \begin{tikzpicture}[baseline,valign]
  \draw[thick] (-0.4, 0) -- (1.4, 0);
  \draw[dashed] (-0.2, 0) to[out=90,in=90] (1.2, 0);
  \draw[dashed] ( 0.2, 0) -- (0.5, 0.4) -- (1.2, 0);
  \node at (0.5, 0.4) [bcirc] {};
\end{tikzpicture} 
\;\; + \;\;
 \begin{tikzpicture}[baseline,valign]
  \draw[thick] (-0.4, 0) -- (1.4, 0);
  \draw[dashed] ( 0.2, 0) to[out=90,in=90] (1.2, 0);
  \draw[dashed] (-0.2, 0) to[out=90,in=90] (1.2, 0);
  \draw[dashed] (0.5, 0) -- (0.7, 0.25);
  \draw[dashed] (0.9, 0) -- (0.7, 0.25);
  \node at (0.7, 0.28) [bcirc] {};
  \node at (0.5, 0.00) [dcirc] {};
  \node at (0.9, 0.00) [dcirc] {};
\end{tikzpicture}
 \;\; + \;\; \ldots
 \label{eq:three-pt-PPP2}
\end{align}
The first and third diagrams are elementary, and have been computed in section~\ref{sec:2pt-PP}.
The second diagram contains a new integral that reads
\begin{align}
 \int 
 \frac{d^d x_4 }{(x^2_{14} x^2_{24})^{1-\frac{\veps}{2}} (x^2_{34})^{2-\veps}}
 = \frac{\pi^2}{\tau^2_{13} \tau^2_{23}}
   \left(\frac{2}{\veps} + 3 - \aleph
 + \log \left(\frac{\tau^4_{13} \tau^4_{23}}{\tau^2_{12}}\right)
 + O(\veps) \right) \, .
\end{align}
To compute it, we first rewrite the integral in parametric form using Schwinger's representation, and then we partial integrate as described in~\cite{Panzer:2014gra,Panzer:2015ida}.\footnote{Although we have not used \texttt{HyperInt}~\cite{Panzer:2014caa}, the package automatizes this in the function \texttt{dimregPartial}.}
The result is expanded to order $O(\veps^0)$, and each of the terms in the expansion is a convergent integral that can be solved with elementary methods.

Once again, the correlator in eq.~\eqref{eq:three-pt-PPP2} contains several OPE coefficients. 
If we let the third operator be a scalar under $O(N-1)$, then we have the following OPE coefficients:
\begin{align}
 \lambda_{\phi_1 \phi_1 s_{\pm}}
 & = \frac{\pm 2 \sqrt{N-1}}{
 \sqrt{N^2+40N +320 \mp (N + 18) \sqrt{N^2 + 40N + 320}}} \Bigg(
    1 
    \pm \frac{(95 N-640)\veps}{64\sqrt{N^2+40 N+320}}
    \notag \\ & \quad
    \pm \frac{17 \veps \sqrt{N^2+40 N+320}}{64(N+8)}
    - \frac{7 N^3+182 N^2+1160 N+2280}{4 (N+8) \left(N^2+40 N+320\right)} \veps 
    + O(\veps^2)
 \Bigg) \, , \label{eq:lambda-PPS} \\
 \lambda_{tts_{\pm}}
 & = \frac{\left(\mp (N+18) + \sqrt{N^2+40 N+320}\right)(N-1)^{-1/2}}
          {\sqrt{N^2+40 N+320 \mp (N+18) \sqrt{N^2+40 N+320}}}
 \Bigg(
   1 
   \mp \frac{(95 N-640) \veps}{64 \sqrt{N^2+40 N+320}}
   \notag \\ & \quad
   \mp \frac{49 \veps \sqrt{N^2+40 N+320}}{64 (N+8)}
   - \frac{\left(11 N^3+374 N^2+3720 N+12520\right) \veps}{4 (N+8) \left(N^2+40 N+320\right)}
   + O(\veps^2)
 \Bigg) \, . \label{eq:lambda-TTS}
\end{align}
If we instead let the third operator be a vector or a symmetric-traceless tensor, we find:
\begin{align}
 \lambda_{\phi_1 t V} 
 = 1-\frac{\veps}{N+8} + O(\veps^2) \, , \qquad
 \lambda_{ttT} 
 = \sqrt{2} \left(1-\frac{\veps}{N+8} + O(\veps^2)\right) \, .
 \label{eq:lambda-VandT}
\end{align}
The normalization of $\lambda_{ttT}$ is chosen such that the tensor structure in eq.~\eqref{eq:tens-struct} multiplies the three-point function in eq.~\eqref{eq:three-pt}.

\subsubsection{Four-point functions}
\label{sec:four-point}

In this final section, we turn our attention to four-point functions of operators formed by the fundamental field and possibly one transverse derivative.
Because of the OPE, four-point functions contain information about three-point OPE coefficients, so we will be able to check some calculations of section~\ref{sec:three-pt} and obtain new results.

\subsubsection*{The correlator \texorpdfstring{$\langle \, \phi \, \phi \, \phi \, \phi \, \rangle$}{< phi phi phi phi >}}

The simplest four-point function is that of the fundamental field:
\begin{align}
 \langle \, \phi_a(\tau_1) \, \phi_b(\tau_2) 
         \, \phi_c(\tau_3) \, \phi_d(\tau_4) \, \rangle
 \;\; = \;\;
 \begin{tikzpicture}[baseline,valign]
  \draw[thick] (-0.4, 0) -- (1.4, 0);
  \draw[dashed] (-0.2, 0) to[out=90,in=90] (0.4, 0);
  \draw[dashed] (0.6, 0) to[out=90,in=90] (1.2, 0);
\end{tikzpicture}
\;\; + \;\;
 \begin{tikzpicture}[baseline,valign]
  \draw[thick] (-0.4, 0) -- (1.4, 0);
  \draw[dashed] (-0.2, 0) -- (0.5, 0.7) -- (1.2, 0);
  \draw[dashed] ( 0.3, 0) -- (0.5, 0.7) -- (0.7, 0);
  \node at (0.5, 0.7) [bcirc] {};
\end{tikzpicture} 
\;\; + \;\;
 \begin{tikzpicture}[baseline,valign]
  \draw[thick] (-0.4, 0) -- (1.6, 0);
  \draw[dashed] (-0.2, 0) to[out=90,in=90] (0.4, 0);
  \draw[dashed] ( 0.6, 0) -- (1, 0.7) -- (1.4, 0);
  \draw[dashed] ( 0.85, 0) -- (1, 0.7) -- (1.15, 0);
  \node at (1.0, 0.70) [bcirc] {};
  \node at (0.85, 0.00) [dcirc] {};
  \node at (1.15, 0.00) [dcirc] {};
\end{tikzpicture}
 \;\; + \;\; \ldots
 \label{eq:4pt-PPPP}
\end{align}
The first and third contributions lead to the disconnected part of the correlator, while the non-trivial part is given by the second diagram.
Since the second diagram is multiplied by a coupling $\lambda_* = O(\veps)$, it suffices to evaluate the integral in $d=4$:
\begin{align}
 \lim_{\vec x_n \to 0}
 \int \frac{d^4 x_5}{x^2_{15} x^2_{25} x^2_{35} x^2_{45}}
 = - 2 \pi ^2 \frac{I_1(\xi)}{\tau^2_{12} \tau^2_{34}} \, .
\end{align}
Since this integral preserves the one-dimensional conformal group, we can evaluate it in the frame $(\tau_1, \tau_2, \tau_3, \tau_4) = (0, \xi, 1, \infty)$, where the cross-ratio $\xi$ is defined in eq.~\eqref{eq:cross-ratio}.
We integrate first over orthogonal directions $\vec x_5^2$ and then over $\tau_5$, considering separately the intervals $\tau_5 \in (-\infty, 0)$, $\tau_5 \in (0,\xi)$, $\tau_5 \in (\xi,1)$ and $\tau_5 \in (1,\infty)$.
The final result reads
\begin{align}
 I_1(\xi)
 = \xi \log ( 1-\xi ) 
 + \frac{\xi^2}{1-\xi} \log \xi  \, .
 \label{eq:4pt-int-1}
\end{align}
As usual, we decompose the correlator in eq.~\eqref{eq:4pt-PPPP} into irreducible components, and using the appropriate renormalization factors we obtain three inequivalent correlators:
\begin{align}
 G_{\phi_1 \phi_1 \phi_1 \phi_1}(\xi)
&= 1
 + \xi ^{2 \Delta_{\phi_1}}
 + \left(\frac{\xi }{1-\xi }\right)^{2 \Delta_{\phi_1}} 
 + \frac{6 \veps I_1(\xi)}{N+8}
 + O(\veps^2) 
 \, , \nonumber\\
 G_{\phi_1 \phi_1 t_{\hat a} t_{\hat b}}(\xi)
&= \delta_{\hat a \hat b} 
 + \veps \, \frac{2 \delta_{\hat a \hat b}}{N+8} I_1(\xi) 
 + O(\veps^2)  \, , \nonumber\\
 G_{t_{\hat a} t_{\hat b} t_{\hat c} t_{\hat d}}(\xi)
&= \delta_{\hat a \hat b} \delta_{\hat c \hat d}
 + \delta_{\hat a \hat c} \delta_{\hat b \hat d} \xi^2
 + \delta_{\hat a \hat d} \delta_{\hat b \hat c} 
   \left(\frac{\xi }{1-\xi }\right)^2 
 + \frac{2 \veps I_1(\xi)}{N+8}
   \Big( \delta_{\hat a \hat b} \delta_{\hat c \hat d} + \text{perms} \Big)
 + O(\veps^2) \notag \nonumber\\
&=  \delta_{\hat a \hat b} \delta_{\hat c \hat d}  \, G_{tttt}^{S}(\xi)
 + (\delta_{\hat a \hat d} \delta_{\hat b \hat c} 
 -  \delta_{\hat a \hat c} \delta_{\hat b \hat d}) \, G_{tttt}^{A}(\xi)
 + \mathbf T_{\hat a \hat b, \hat c \hat d} \, G_{tttt}^{T}(\xi) \, .
\end{align}
The correlator of four tilt operators decomposes into singlet $(S)$, antisymmetric ($A$) and symmetric-traceless ($T$) channels with respect to the $O(N-1)$ symmetry, where the last tensor structure is shown in eq.~\eqref{eq:tens-struct}.

The virtue of having four-point functions is that they can be expanded in conformal blocks to obtain anomalous dimensions and three-point coefficients.
For the vector, symmetric-traceless and antisymmetric channels we find:
\begin{align}
 G_{t\phi_1\phi_1t}(\xi)
 & = \left(1-\frac{2 \veps}{N+8}\right) g^{\Delta_{t\phi_1},\Delta_{\phi_1 t}}_2(\xi)
   + \veps \, \frac{N+10}{N+8} \, 
     \partial_\Delta g^{\Delta_{t\phi_1},\Delta_{\phi_1 t}}_2(\xi)
   + \ldots \, , \nonumber\\
 G^{S}_{tttt}(\xi)
 & = \left(2-\frac{4 \veps}{N+8}\right) g_{2}(\xi )
   + \frac{4 \veps}{N+8} \, \partial_\Delta g_{2}(\xi)
   + \ldots \, , \nonumber\\
 G^{A}_{tttt}(\xi)
 & = g_{3}(\xi)
   + \ldots \, . \label{eq:block-Anti}
\end{align}
The ellipses stand for higher dimensional operators, as well as higher order corrections in $\veps$.
The first two lines confirm our computations of $\Delta_V$, $\Delta_T$, $\lambda_{\phi_1 t V}$ and $\lambda_{ttT}$.
From the third line, we conclude that the antisymmetric operator $A_{a b} \sim \phi_{[a} \partial_\tau \phi_{b]}$ has scaling dimension 
$\Delta_A = 3 + O(\veps^2)$ and three-point coefficient $\lambda_{ttA} = 1 + O(\veps^2)$.

Of course, we can also expand correlators in the singlet channel, for example
\begin{align}
 G_{\phi_1 \phi_1 \phi_1 \phi_1}(\xi)
& = 1
  + \left(2-\frac{6 \veps}{N+8}\right) g_{2}(\xi)
  + \veps \, \frac{4N+38}{N+8} \, \partial_\Delta g_{2}(\xi)
  + \ldots
\end{align}
However, the above contributions are due to two operators $s_\pm$ with nearly-degenerate dimension, so the expansion does not fix the scaling dimension or OPE coefficient, but instead it relates them in a non-trivial way:
\begin{align}
 &(\Delta_{s_+}-2) \lambda_{\phi_1\phi_1s_+}^2
 + (\Delta_{s_-}-2) \lambda_{\phi_1\phi_1s_-}^2 
 = \veps \, \frac{4N+38}{N+8} + O(\veps^2)~,\nonumber\\
&\lambda_{\phi_1\phi_1s_+}^2 + \lambda_{\phi_1\phi_1s_-}^2
 = 2-\frac{6 \veps}{N+8} + O(\veps^2) \,.
\end{align}
It is reassuring that eq.~\eqref{eq:delta-spm} and eq.~\eqref{eq:lambda-PPS} indeed satisfy these relations.
In a similar way, we can expand $G_{tttt}^{\text{sing}}$ and $G_{\phi_1\phi_1tt}$, finding again perfect agreement with our results.
Note that the four-point functions do not capture the OPE coefficients in eq.~\eqref{eq:lambda-PPP} at this order in $\veps$.

\subsubsection*{The correlators \texorpdfstring{$\langle \, \nabla \phi \, \nabla \phi \, \phi \, \phi \, \rangle$}{< dphi dphi phi phi >} and \texorpdfstring{$\langle \, \nabla \phi \, \nabla \phi \, \nabla \phi \, \nabla \phi \, \rangle$}{< dphi dphi dphi dphi >}}

In a similar way, one can consider four-point functions that include derivative operators $\nabla \phi_a$, and the only non-trivial contribution is a contact diagram.
For the case of two derivative operators, the relevant integral is
\begin{align}
& \lim_{\vec x_n \to 0}
  \frac{\partial^2}{\partial \vec x_{3,i} \partial \vec x_{4,j}} 
 \int \frac{d^4 x_5}{x^2_{15} x^2_{25} x^2_{35} x^2_{45}}
 = \frac{2 \pi ^2}{3} \frac{I_2(\xi)}{\tau_{12}^2 \tau_{34}^4} \delta^{ij} \, , \nonumber\\
& I_2(\xi)
  = \frac{\xi ^2 }{1-\xi }
  - (\xi +2) \xi  \log ( 1-\xi )
  + \frac{\xi ^4 }{(1-\xi)^2} \log \xi \, ,
\end{align}
which has been computed using the same technique as in eq.~\eqref{eq:4pt-int-1}.
From this, we can read off several correlators, for example
\begin{align}
 G_{\phi_1\phi_1 \Disp_i \Disp_j}(\xi)
&= \delta_{ij} \left(
   1
 - \frac{\veps I_2(\xi)}{N+8}
 + O(\veps^2)
 \right) \, , \nonumber\\
 G_{t_{\hat a} t_{\hat b} \Disp_i \Disp_j}(\xi)
&= \delta_{\hat a \hat b} \delta_{ij} \left(
   1
 - \frac{\veps I_2(\xi)}{3 (N+8)}
 + O(\veps^2)
 \right) \, .
\end{align}
We can also obtain correlators with $\nabla \phi_{\hat a}$, but we do not write them for compactness.
Finally, if we consider four derivative operators the relevant integral is
\begin{align}
& \lim_{\vec x_n \to 0}
  \frac{\partial^4}{\partial \vec x_{1,i} \partial \vec x_{2,j} \partial \vec x_{3,k} \partial \vec x_{4,l}} 
  \int \frac{d^4 x_5}{x^2_{15} x^2_{25} x^2_{35} x^2_{45}}
  = -\frac{8\pi^2}{15} \frac{I_3(\xi)}{\tau_{12}^4 \tau_{34}^4}
    \Big( \delta_{ij}\delta_{kl} + \delta_{ik}\delta_{jl} + \delta_{il}\delta_{jk} \Big) 
    \, , \\[0.3em]
& I_3(\xi)
  = \frac{\left(\xi ^2-\xi +1\right) \xi ^2}{(1-\xi)^2}
  + \frac{1}{2} \left(2 \xi ^2+\xi +2\right) \xi  \log ( 1-\xi ) 
  + \frac{\left(2 \xi ^2-5 \xi +5\right) \xi ^4}{2 (1-\xi)^3} \log \xi \, .
  \notag
\end{align}
From here we extract the four-point function of the displacement operator 
\begin{align}
 G_{\Disp^i\Disp^j\Disp^k\Disp^l}(\xi)
&= \delta^{ij} \delta^{kl}
 + \delta^{ik} \delta^{jl} \xi^4
 + \delta^{il} \delta^{jk} 
   \Big(\frac{\xi }{1-\xi }\Big)^4
 + \frac{2 \veps I_3(\xi)}{5 (N+8)}
   \Big(\delta^{ij} \delta^{kl} + \text{perms} \Big)
   + O(\veps^2) \, .
\end{align}
In order to extract CFT data from this correlator, we decompose it in terms of singlet $(S)$, antisymmetric $(A)$ and symmetric-traceless $(T)$ channels under $SO(d-1)_T$:
\begin{align}\label{resultsmagneticline}
 G_{\Disp\Disp\Disp\Disp}^{S}(\xi)
 & = 1
   +\left(\frac{2}{3} + \frac{(4 N+37) \veps}{18 (N+8)}\right) f_{4}(\xi)
   + \frac{5 \veps}{3 (N+8)} \partial_\Delta f_{4}(\xi)
   + \ldots \, , \nonumber\\
 G_{\Disp\Disp\Disp\Disp}^{A}(\xi)
 & = 2 f_{5}(\xi) + \ldots \, , \nonumber\\
 G_{\Disp\Disp\Disp\Disp}^{T}(\xi)
 & = \left(2 + \frac{\veps}{3 (N+8)}\right) f_{4}(\xi)
   + \frac{2 \veps}{N+8} \partial_\Delta f_{4}(\xi)
   + \ldots \, .
\end{align}
Notice that these operators are not necessarily the lowest-dimensional operators in each channel.
For instance, in the $(\Disp \times \Disp)_{S}$ channel we have $\phi_1$ with $\Delta_{\phi_1} = 1+\veps+O(\veps^2)$, but it does not appear because $\lambda^2_{DD\phi_1} = O(\veps^2)$, see eq.~\eqref{eq:lambda-grad}.
Similarly, in the $(\Disp \times \Disp)_{T}$ channel the lowest-dimension operator is $\partial_i \partial_j \phi_1$ with $\Delta = 3 + O(\veps)$, but it is invisible in the four-point function at this order.
Only in the channel $(\Disp \times \Disp)_{A}$ we expect the lowest-lying operators to be $\partial_{[i} \phi_1 \partial_\tau \partial_{j]} \phi_1$ and $\partial_{[i} \phi_{a} \partial_\tau \partial_{j]} \phi_{a}$, with nearly-degenerate dimension $\Delta = 5 + \ldots$. 
However, in order to disentangle these operators one should do an analysis similar to the one we did for $s_{\pm}$ below eq.~\eqref{eq:mixing-problem}.

%% file: sections/numerics.tex
\section{Numerical results}\label{sec:numerics}

In this section we use numerical conformal bootstrap and the semidefinite program solver SDPB~\cite{SimmonsDuffin:2015qma} to carve out the space of line defects with $O(2)_F$ global symmetry. 
We consider:
\begin{enumerate}
	\item Line defects without a tilt operator $(t)$ that preserve a $U(1)_F$ subgroup of the $O(2)_F$ symmetry in the bulk;
	\item Line defects that break local $O(3)_F$ to $U(1)_F\simeq SO(2)_F$ and therefore feature a tilt operator transforming in the vector representation of $U(1)_F$.
\end{enumerate}
In section~\ref{sec:singleMCcorrgeneral} we study correlation functions involving $\Disp$ and $t$. Due to the universal nature of these operators, the numerical bounds presented in section~\ref{sec:singleMCcorrgeneral} are valid for a large class of conformal defects.
In sections~\ref{sec:MCmono} and~\ref{sec:MCpinning} we focus on the specific models already announced in section~\ref{sec:epsexpTot}: the magnetic line defect and the monodromy line defect. Here we use the $\varepsilon$-expansion results of section~\ref{sec:epsexpTot} as guidance, in order to zoom in on specific regions of the parameter space, where we expect these models to live.

\subsection{Bounds on universal correlators}\label{sec:singleMCcorrgeneral}

\subsubsection{Single-correlator with the displacement operator}\label{sec:SCdisp}

We start bootstrapping correlation functions of the displacement operator $\Disp$. Recall from section~\ref{sec:dispSpec} that $\Disp$ is a defect primary of scaling dimension $\D_\Disp=2$, transforming as a vector under $SO(2)_T$ and neutral under $O(2)_F$. Hence, while we cannot impose the global symmetry $O(2)_F$, the charge of the displacement under $SO(2)_T$ implies we restrict to defects of co-dimension $q\geq 2$.
In the complex notation introduced in section~\ref{sec:dispSpec}, there are two non-equivalent orderings of the correlation functions involving the displacement
\begin{equation}
 \langle \Disp (\tau_1) \Disp (\tau_2) \bar{\Disp} (\tau_3) \Dispb (\tau_4) \rangle~,\quad  \langle \Disp (\tau_1) \Dispb (\tau_2) \Disp (\tau_3) \Dispb (\tau_4) \rangle~,
\end{equation}
and the relevant crossing equations can be obtained from eq.~\eqref{eq:cross-one-cplx} upon setting $\D_\Disp=2$.
The leading non-identity defect primaries in the $\Disp \times \Dispb$ OPE are denoted $\DDb^{\pm}$ in the conventions of section~\ref{sec:dispSpec}, while the leading primary in the $\Disp \times \Disp$ OPE is denoted $\DD$. 

\paragraph{Gap bounds.}
\begin{figure}
	\centering
	\subfloat[]{\includegraphics[width=1\textwidth]{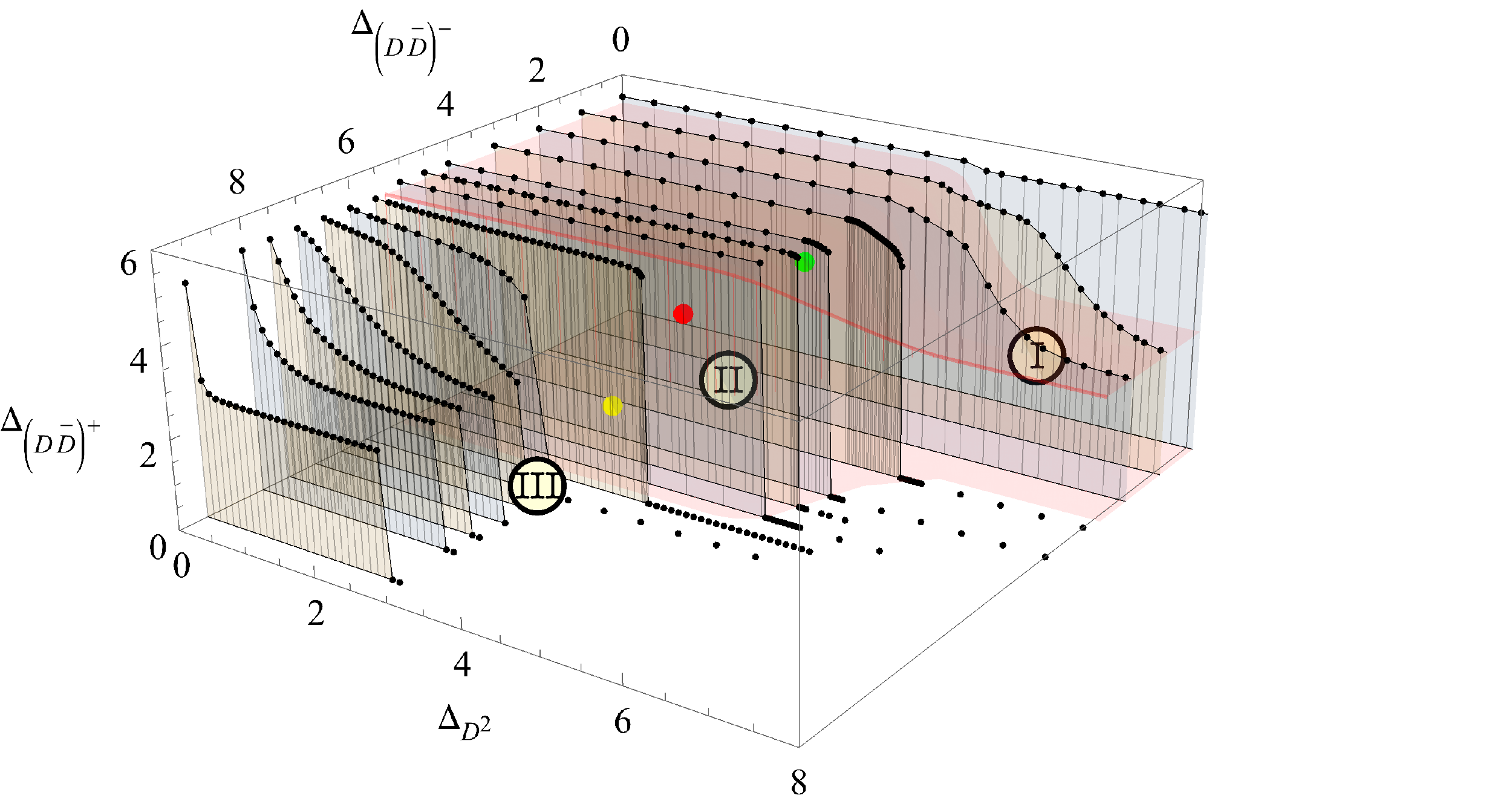}}\\
	\subfloat[]{\includegraphics[width=0.9\textwidth]{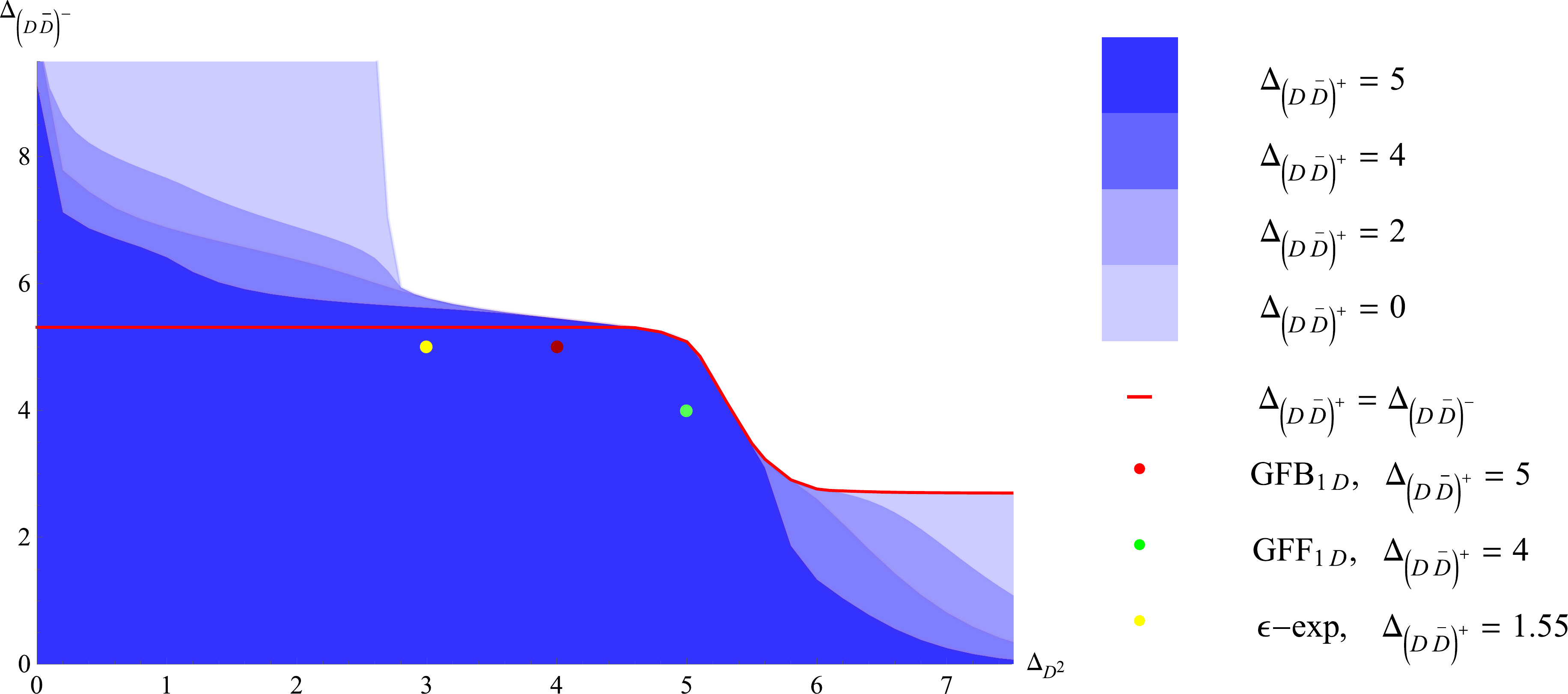}}\\
	\caption{(a) Upper bounds on the dimension of the $\mathcal{S}$-parity even scalar $\DDb^{+}$ as a function of the $\mathcal{S}$-parity odd operator $\DDb^{-}$ and the leading charged operator $\Disp^2$. (b) Projection of the three-dimensional allowed region in the $(\D_{\DDb^{-}}, \D_{\DD})$ plane for different values of $\D_{\DDb^{+}}$. All points are computed with $\Lambda = 33, P = 53$.  The green and red dots correspond to the GFF and GFB solutions respectively. The yellow dot is the extrapolation to $\varepsilon=1$ of the $\veps$-expansion predictions for the magnetic line defect. The solid red line in (b) is the `agnostic' bound for $\D_{\DDb^{+}}=\D_{\DDb^{-}}$.}
	\label{fig:SCBosDDDD_L33P53}
\end{figure}

We start computing the upper bound on the scaling dimension of the leading $\mathcal{S}$-parity even scalar $\D_{\DDb^{+}}$ as we vary $\D_{\Disp^2}$ and $\D_{{\DDb}^{-}}$. The result is shown in the 3d plot of figure~\ref{fig:SCBosDDDD_L33P53}(a), where the light-red shaded region represents the `agnostic' bound obtained imposing the same gap on the dimension of the $\mathcal{S}$-parity odd and $\mathcal{S}$-parity even scalars. The term `agnostic' here refers to the fact that by holding $\D_{\DDb^{+}}=\D_{\DDb^{-}}$ we are putting a bound on the lowest-lying singlet, whether it is parity-even or parity-odd. In order to help visualizing the constant-$\D_{\DDb^{+}}$ slices, we included figure~\ref{fig:SCBosDDDD_L33P53}(b). 
There are three notable regions in figure~\ref{fig:SCBosDDDD_L33P53}(a):
\begin{enumerate}
	 \item[I.] This is the region with the weakest $\D_{\DDb^{-}}$ gap assumptions, i.e. $\D_{\DDb^{-}}\simeq 0\div 3 $. In this region the upper bound is smooth: it is saturated by the `agnostic' bound for both small and large values of $\D_{\Disp^2}$, with a cross-over behavior at around $\D_{\Disp^2}\simeq 6$.
	 For low values of $\D_{\Disp^2}$, the agnostic bound approaches the GFF value for a real fermion $\D_{\DDb} = 2 \D_{\Disp} + 1=5$. This is because a single real GFF satisfies the crossing equations for a single complex scalar as long as the gap in the charged sector is $\D_{\Disp^2}<2 \D_{\Disp}= 4$, see e.g.~\cite{Paulos:2015jfa}.	
	 
     \item[II.] This is the region with intermediate values of $\D_{\DDb^{-}}\simeq 3\div 5$. The upper bound remains constant for $\D_{\Disp^2} \lesssim 6$, while it drops to zero for $\D_{\Disp^2} \gtrsim 6$. These vertical drops can be interpreted as an upper bound on the dimension of $\D_{\Disp^2}$ as a function on the gap on $\D_{\DDb^{-}}$. As we increase the gap on $\D_{\DDb^{-}}$, the upper bound on $\D_{\Disp^2}$ becomes stronger. This region includes the GFF and GFB solutions -- see eq.~\eqref{eq:gff-PPPbPb}-- which are indicated in the figure by green and red dots respectively, as well as the (extrapolated) $\veps$-expansion prediction for the magnetic line defect -- see eqs.~\eqref{eq:dimPhi} and~\eqref{resultsmagneticline} -- which is $(\D_\DD,\D_{\DDb^{-}},\D_{\DDb^{+}})\simeq (3,5,1.55)$ and is marked in yellow.
 	  Interestingly, the yellow dot is close to saturating the upper bound on $\D_{\DDb^{-}}$, as can be seen more clearly from figure~\ref{fig:SCBosDDDD_L33P53}(b).
	  The $U(1)_F$ monodromy defect should also live in this region, and it would be interesting to verify this by computing e.g. the correlator $\langle \Disp \Disp \Dispb \Dispb \rangle$ at the first non-trivial order in the $\veps$-expansion.	 
	  
	 \item[III.] This is the region with the strongest $\D_{\DDb^{-}}$ gap assumptions, i.e. $\D_{\DDb^{-}}\simeq 5\div 8$. As we enter this region from region II, we observe that for values $6 \lesssim\D_{\DDb^{-}} \lesssim 7 $, the convexity of the bound changes and a plateau starts forming as we move towards small values of $\D_{\Disp^2}$. If we increase $\D_{\DDb^{-}} $ further, the plateau terminates around $(\D_{\Disp^2}, \D_{\DDb^{+}})\simeq (2.7,2.7)$. The plateau happens to have the same height $\D_{\DDb^{+}} \simeq 2.7$ as the bound for large $\D_{\Disp^2}$ in region I. This indicates that there is a universal upper bound $\Delta_{\DDb^+} \simeq 2.7$ when increasing the gaps on the dimensions of all other operators.
\end{enumerate}

\paragraph{Including the OPE coefficient of $\DD$.}
\begin{figure}[h]
	\centering
	\hspace{-2cm}\includegraphics[width=0.8\textwidth]{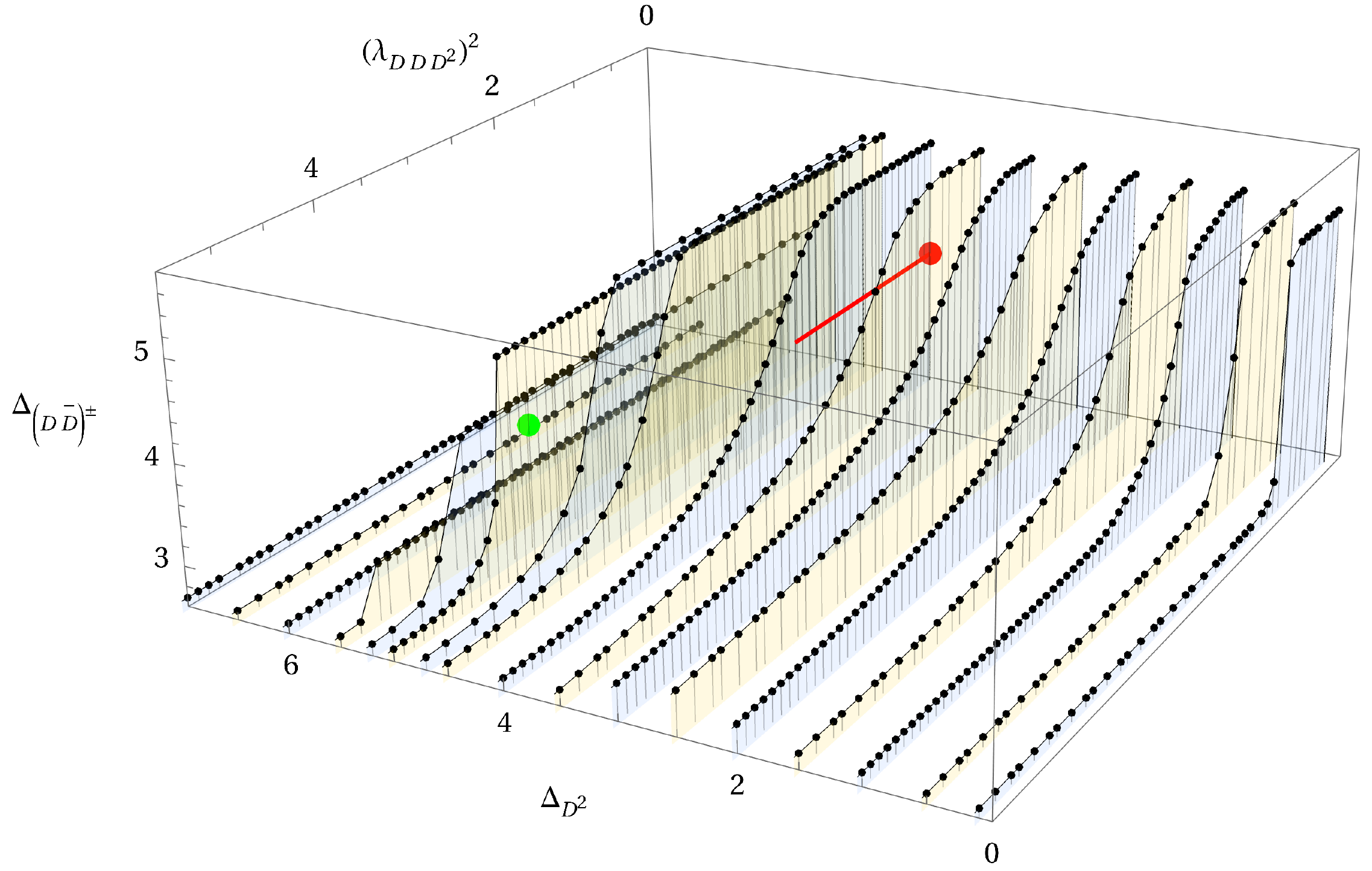}
	\caption{Bounds on the dimension of the first singlet in the $\Disp \times \Dispb$ OPE as a function of the gap on the dimension $\D_{\Disp^2}$ and the OPE coefficient $(\lambda_{\Disp \Disp \Disp^2})^2$ of the first charged operator in the $\Disp \times \Disp$ OPE. The parity-even and -odd singlets are set to have the same gap: $\D_{\DDb^{+}}=\D_{\DDb^{-}}$. The green and red dots correspond to the solutions for GFF and GFB, respectively. $\Lambda = 49, P = 69$.}
	\label{fig:3dplotDDDD}
\end{figure}
The light-red curve in figure~\ref{fig:SCBosDDDD_L33P53} shows a family of solutions to crossing that maximize the gap on the lowest-lying operator in the singlet channel. This curve looks rather smooth in figure~\ref{fig:SCBosDDDD_L33P53}. To further investigate this family of solutions, we repeat the gap maximization procedure but this time we keep $\D_{\DDb^{+}}=\D_{\DDb^{-}}$ and varying both $\D_{\Disp^2}$ and the (squared) OPE coefficient $(\lambda_{\Disp \Disp \Disp^2})^2$. This OPE coefficient is the most straightforward to implement, and a similar choice was made in~\cite{Gaiotto:2013nva,Ghosh:2021ruh}. 
The results are shown in figure~\ref{fig:3dplotDDDD}, whose features we now describe.
The free theory solutions are marked in the figure by green (GFF) and red (GFB) dots. We recall that these free theory solutions are given in eqs.~\eqref{eq:gff-PPbPPb} and~\eqref{eq:gff-PPPbPb}, where it is shown that the OPE coefficients depend on the parameter $\alpha$, which is $\alpha = -1$ for GFF and $\alpha = 1$ for GFB. The red line in the figure represents the solutions for intermediate $\alpha \in (-1,1)$; it is well inside the allowed region and appears to be parallel to the upper bound. The GFF solution lies outside the red line, which can be understood from the fact that for an anti-commuting fermion $\psi_a$, the leading symmetric traceless representation in the $\psi_a\times \phi_b$ OPE is $\psi_a \partial_\tau \psi_b$ and has scaling dimension $2 \D_{\psi} + 1 = 5$, while for a boson $\phi_a$ it is $\phi_a \phi_b$ and has scaling dimension $2 \D_{\phi} = 4$. 
A family of rising kinks appears for values around $\D_{\Disp^2} \simeq 5$, i.e. close to the GFF solution. 
Although we have not studied the evolution of these kinks when increasing the number of derivatives, it is plausible that they can be explained by the vicinity of the GFF solution.
For $\D_{\Disp^2} \gtrsim 5$, the bound on $\D_{\DDb^{\pm}}$ quickly drops to the value $\D_{\DDb^{\pm}} \simeq 2.7$.
For $\D_{\DD} \lesssim 5$, the kinks move towards smaller values of $(\lambda_{\Disp \Disp \Disp^2})^2$ until around $\D_{\Disp^2} \simeq 4$ and below, the upper bound becomes completely smooth. The magnetic line defect lies well inside the allowed region, since for this defect the first parity-even singlet has dimension $\D_{\DDb^{+}} \simeq 1.55$.

\subsubsection{Single-correlator with the tilt operator}\label{sec:SCtilt}

Next, we consider the four-point functions of the tilt operator $t$. We recall that $t$ is a defect primary of scaling dimension $\D_t=1$, transforming as a vector of $O(2)_F$ and neutral under $SO(2)_T$ (see section~\ref{sec:tiltSpec}). This means that unlike the displacement bootstrap, here we are imposing the existence of a global symmetry $O(2)_F$, although we cannot distinguish between one-dimensional conformal defects with different co-dimension. In the complex notation of section~\ref{sec:tiltSpec}, the two non-equivalent orderings along the line are
\begin{equation}
 \langle t (\tau_1) t (\tau_2) \bar{t} (\tau_3) \bar{t} (\tau_4)\rangle\:, \quad \langle t (\tau_1) \bar{t} (\tau_2) t (\tau_3) \bar{t} (\tau_4)\rangle\:.
\end{equation}
The leading non-identity defect primaries in the $t \times \bar{t}$ OPE are denoted $\ttb^{\pm}$ in the conventions of section~\ref{sec:dispSpec}, while the leading primary in the $t \times t$ OPE is denoted $t^2$. 
The bootstrap equations can be obtained from eq.~\eqref{eq:cross-one-cplx}, setting  $\D_{t} = 1$ as external dimensions.

\paragraph{Gap bounds.}
\begin{figure}
	\centering
\subfloat[]{\hspace{-1cm}\includegraphics[width=1\textwidth]{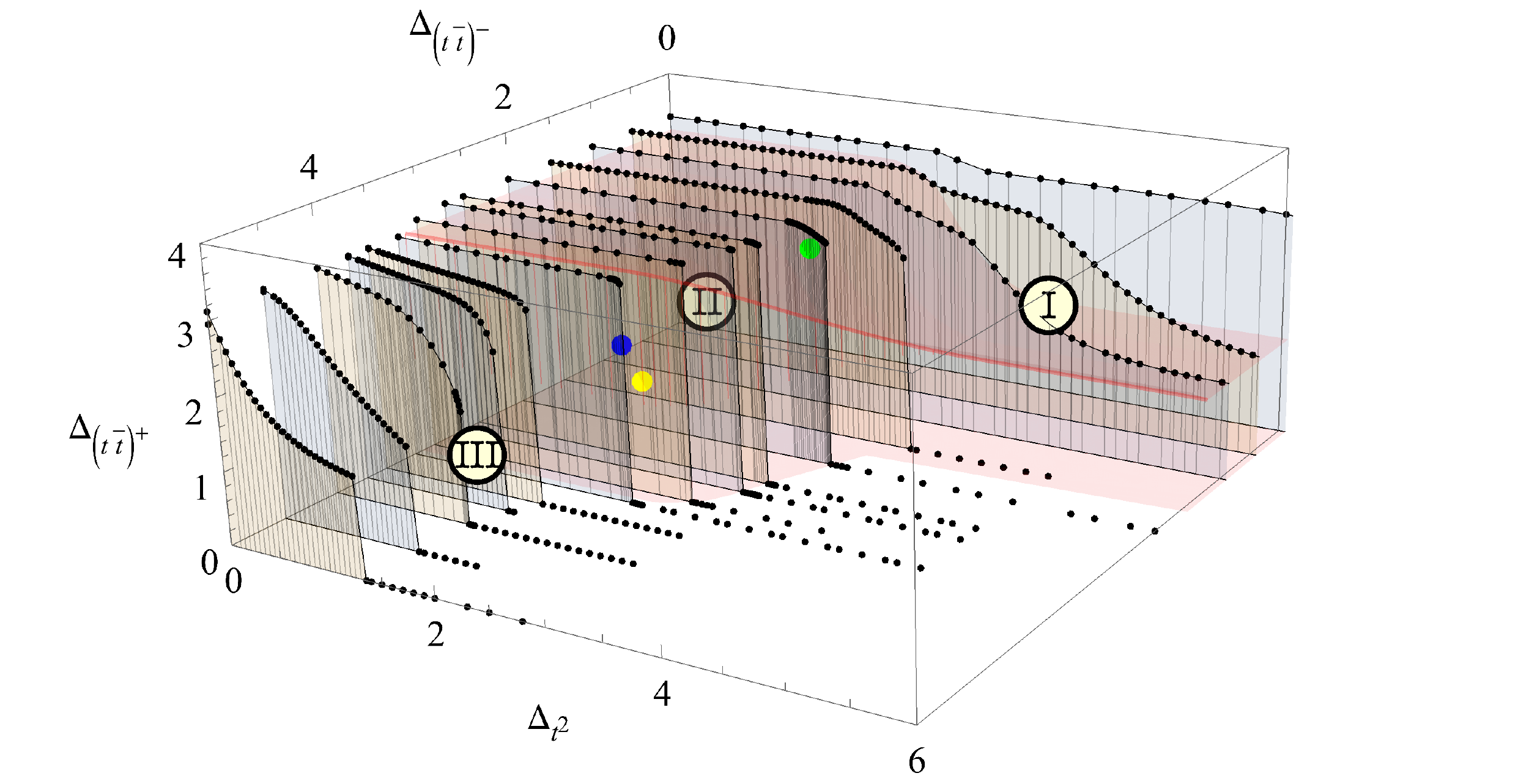}}\\
	\subfloat[]{
			\includegraphics[width=0.9\textwidth]{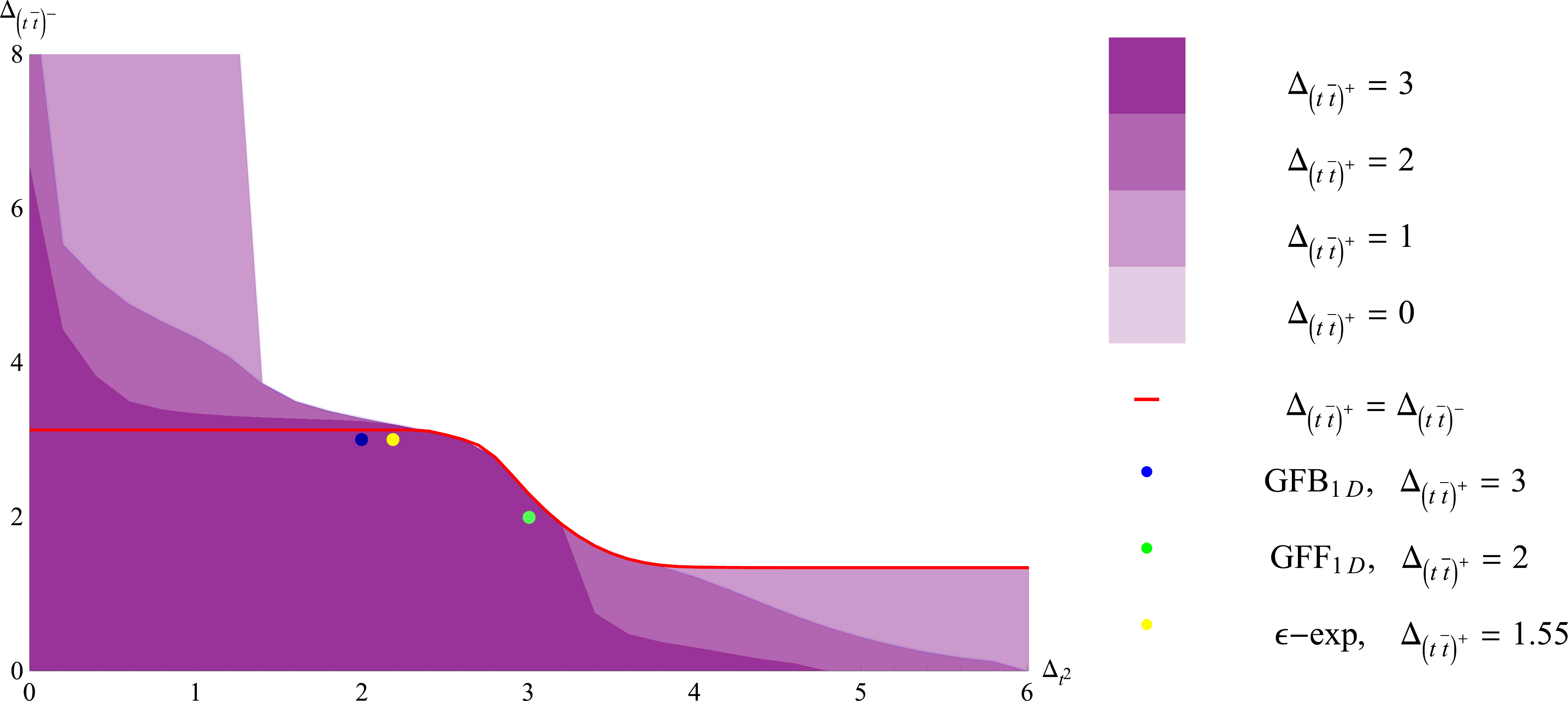}}
	\caption{Bounds on the maximal gap on the dimension of the $\mathcal{S}$-parity even scalar $\ttb^{+}$ (a) or $\mathcal{S}$-parity odd scalar $\ttb^{-}$ (b) vs. the $\mathcal{S}$-parity odd (a) or even (b) scalar gap vs. the gap on the leading charged operator $t^2$. $\Lambda = 33, P = 53$. The green and blue dots correspond to the solutions for GFF and GFB, respectively. The yellow dot is the 1-loop $\varepsilon$-expansion prediction for the magnetic line defect discussed in section~\ref{sec:epsexp}. The solid red line in (b) is the `agnostic' bound for $\D_{\ttb^{+}}=\D_{\ttb^{-}}$. 
	}
	\label{fig:SCBosTTTT_L33P53}
\end{figure}

Here we ask the same question as for the single-correlator displacement bootstrap, namely what is the upper bound on the scaling dimension of the leading $\mathcal{S}$-parity even scalar $\D_{\ttb^{+}}$ as we vary $\D_{t^2}$ and $\D_{{\ttb}^{-}}$. 
The results are shown in the 3d plot of figure~\ref{fig:SCBosTTTT_L33P53}(a), where the light-red shaded region represents the `agnostic' bound obtained by imposing the same gap on the $\mathcal{S}$-parity odd and $\mathcal{S}$-parity even scalars, and in figure~\ref{fig:SCBosTTTT_L33P53}(b).
Not surprisingly, these plots show many similarities with those in figure~\ref{fig:SCBosDDDD_L33P53} because we used the same crossing equations, although for different external scaling dimensions. What changes is the interpretation of the results. There are three notable regions in~\ref{fig:SCBosTTTT_L33P53}(a):

\begin{enumerate}
	\item[I.] This is the region with the weakest $\D_{\ttb^{-}}$ gap assumptions, i.e. $\D_{\ttb^{-}}\simeq 0\div 1.5 $. In this region the upper bound is saturated by the `agnostic' bound for both small and large values of $\D_{t^2}$, which approaches the GFF bound for a real fermion $\D_{\ttb} = 2 \D_{t} + 1 = 3$ for $\D_{t^2}<2$ (cf. previous discussion in the displacement bootstrap). At around $\D_{t^2} \simeq 3$, the bound drops but remains smooth.
	
	\item[II.] This is the region with intermediate values of $\D_{\ttb^{-}}\simeq 1.5\div 3.5$. For  $\D_{t^2} \lesssim 3$, the upper bound still approaches the GFF bound at $\D_{\ttb} = 3$. For $\D_{t^2} \gtrsim 3$ the bound drops to zero, with increasingly sharper drops until $\D_{\ttb^{-}} = 3$, after which the drops remain but become smoother. As we noted previously, these vertical drops are due to the existing upper bound on $\D_{t^2}$, as a function of $\D_{\ttb^{-}}$.
	This region also contains the GFF (green) and GFB (blue) solutions -- see eqs.~\eqref{eq:gff-PPbPPb} and~\eqref{eq:gff-PPPbPb} --, as well as the $\varepsilon$-expansion results $(\D_{t^2},\D_{\ttb^-},\D_{\ttb^+}) \simeq (2 + 2/11,3,1.55)$ for the magnetic line defect (see section~\ref{sec:epsexp}) which are shown as a yellow dot.\footnote{Here, the leading $\mathcal{S}$-parity even singlet is $\ttb^{+}=\phi_1$, with scaling dimension $\simeq 1.55$, see eq.~\eqref{eq:gff-PPbPPb}.} The $\veps$-expansion results are close to saturating the bootstrap bound at $\D_{\ttb^{-}} \simeq 3.2$, which can be most clearly seen in figure~\ref{fig:SCBosTTTT_L33P53}(b). It would be very interesting to know the sign of the $O(\veps^2)$ correction to the scaling dimension of $\ttb^{-}$ in the magnetic line defect.
	
	\item[III.] This is the region with the largest $\D_{\ttb^{-}}$ gap, i.e. $\D_{\ttb^{-}}\simeq 3.5\div 5$. The convexity of the bound changes for $4 \lesssim\D_{\ttb^{-}} \lesssim 5$. The plateau that was clearly visible in the same region in figure~\ref{fig:SCBosDDDD_L33P53} also appears here once we move towards small values of $\D_{t^2}$, but it is less pronounced. For higher $\D_{\ttb^{-}} $, the plateau again terminates around $(\D_{t^2}, \D_{\ttb^{+}})\simeq (1.35,1.35)$. This is the same height $\D_{\ttb^{+}} \simeq 1.35$ as the bound for large $\D_{t^2}$ in region I, which can again be thought of as a universal upper bound on $\D_{\ttb^{+}}$. 
	These kinks are reflected in figure~\ref{fig:SCBosTTTT_L33P53}(b) around $\D_{t^2} \sim 1.35$. The bound on $\D_{\ttb^{-}}$ becomes infinite for $\D_{\ttb^{+}} \lesssim 2$ and $ \D_{t^2} \lesssim 1.35$, while it is saturated by the agnostic $\D_{\ttb^{+}} = \D_{\ttb^{-}}$ bound for $\D_{t^2} >1.35$. 
\end{enumerate}

\paragraph{Including the OPE coefficient of $\TT$.}
\begin{figure}[h]
	\centering
	\includegraphics[width = 0.8\textwidth]{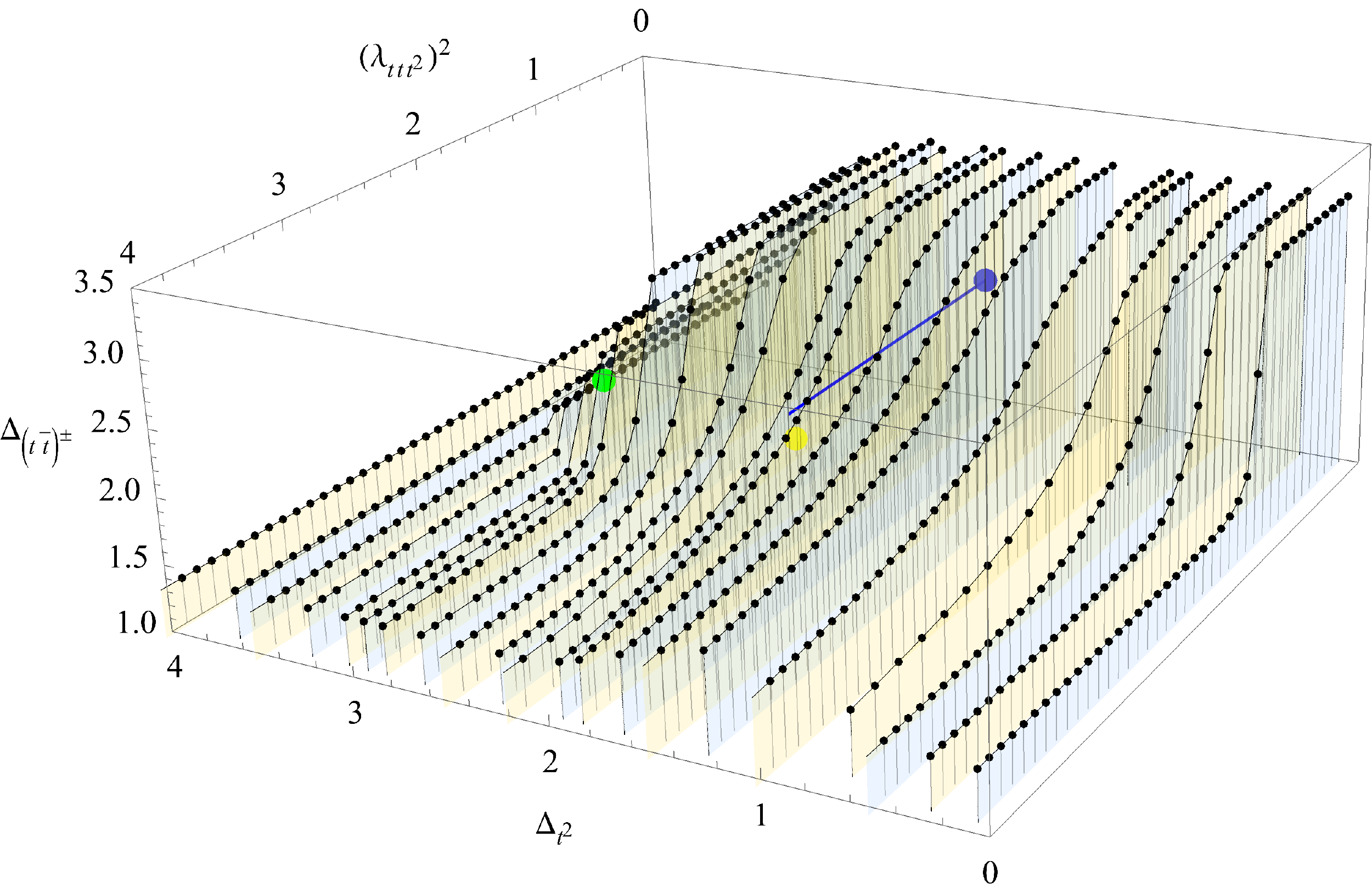}
	\caption{Bounds on the dimension of the first singlet in the $t \times \bar{t}$ OPE as a function of the gap on the dimension $\D_{t^2}$ and the OPE coefficient $(\lambda_{t t t^2})^2$ of the first operator charged under $O(2)_F$ in the $t \times t$ OPE. The gaps on the dimension of the parity-even and -odd operators are set to be equal. 
		The free theory solutions GFF (green), GFB (blue), and their interpolation (blue), presented in eqs.~\eqref{eq:gff-PPbPPb} and~\eqref{eq:gff-PPPbPb} are also shown, as is the $\veps$-expansion result for the magnetic line defect discussed in section~\ref{sec:epsexp} (yellow). $\Lambda = 49, P = 69$.}
	\label{fig:SCBosTTTT_L33P53_OvsSOPE_3d}
\end{figure}

The $\veps$-expansion results for the magnetic line defect from section~\ref{sec:epsexp} are close to saturating the bound in figure~\ref{fig:SCBosTTTT_L33P53}. 
However, since we do not know the sign of the $O(\veps^2)$ correction on $\D_{\ttb^{-}}$, we cannot predict if the point will move closer or further away from the bound.
Let us focus on the agnostic bound and impose the same gap on $\D_{\ttb^{+}}$ and $\D_{\ttb^{-}}$. 
Similarly to what we did for the displacement bootstrap, we look for bounds on $\D_{\ttb^{\pm}}$ as a function of $(\lambda_{t t t^2})^2$ -- the (squared) OPE coefficient of the first charged operator $t^2$ -- and $\D_{t^2}$. 
The results are shown in figure~\ref{fig:SCBosTTTT_L33P53_OvsSOPE_3d}.
The free theory solutions shown in green (GFF), and blue (GFB) -- see eqs.~\eqref{eq:gff-PPbPPb} and~\eqref{eq:gff-PPPbPb} -- are close to saturating the bound. Again, the GFF solution ($\alpha = -1$) is disconnected from the GFB solution ($\alpha = 1$) and the solution for $\alpha \in (-1,1)$ (cf. previous discussion).
The $\veps$-expansion result up to $O(\veps)$ for the magnetic line defect given by $((\lambda_{t t t^2})^2,\D_{t^2}, \D_{\ttb}) \simeq (2 - \frac{4}{11}, 2 + \frac{2}{11}, 1.55)$ -- see eqs.~\eqref{eq:dimPhi}, \eqref{eq:dimAT},\eqref{eq:lambda-VandT} -- and marked with the yellow dot in the figure is below the upper bound.
 There are additional kinks for $2 \lesssim \D_{t^2} \lesssim 4$ around $(\lambda_{tt t^2})^2 \simeq 2$, one of which gets saturated by the GFF solution. For $\D_{t^2} \gtrsim 4$ the bound becomes horizontal and settles at $\D_{\ttb^{\pm}} \simeq 1.35$ .

\subsubsection{Mixed-correlator with tilt and displacement}

After having analyzed the single correlators of either the tilt or the displacement operators, we consider mixed correlators that involve both at the same time:
\begin{equation}
	\langle \Disp (\tau_1) \Dispb (\tau_2)  \Disp (\tau_3) \Dispb (\tau_4) \rangle\:, \quad \langle t (\tau_1) \bar{t} (\tau_2)  t (\tau_3) \bar{t} (\tau_4) \rangle\:, \quad \langle \Disp (\tau_1) \Dispb (\tau_2)  t (\tau_3) \bar{t} (\tau_4) \rangle\:.
\end{equation}
plus all other non-equivalent orderings. 
This is the natural system of correlators to study the magnetic line defect for the $O(3)$ vector model, which features a tilt operator in the vector representation of $O(2)_F$, as well as a displacement in the vector of $SO(2)_T$.
The bootstrap equations can be found in eq.~\eqref{eq:cross-eq-tilt-disp}.

\paragraph{Gap bounds.}
\begin{figure}[h]
\centering
 \includegraphics[width=0.9\textwidth]{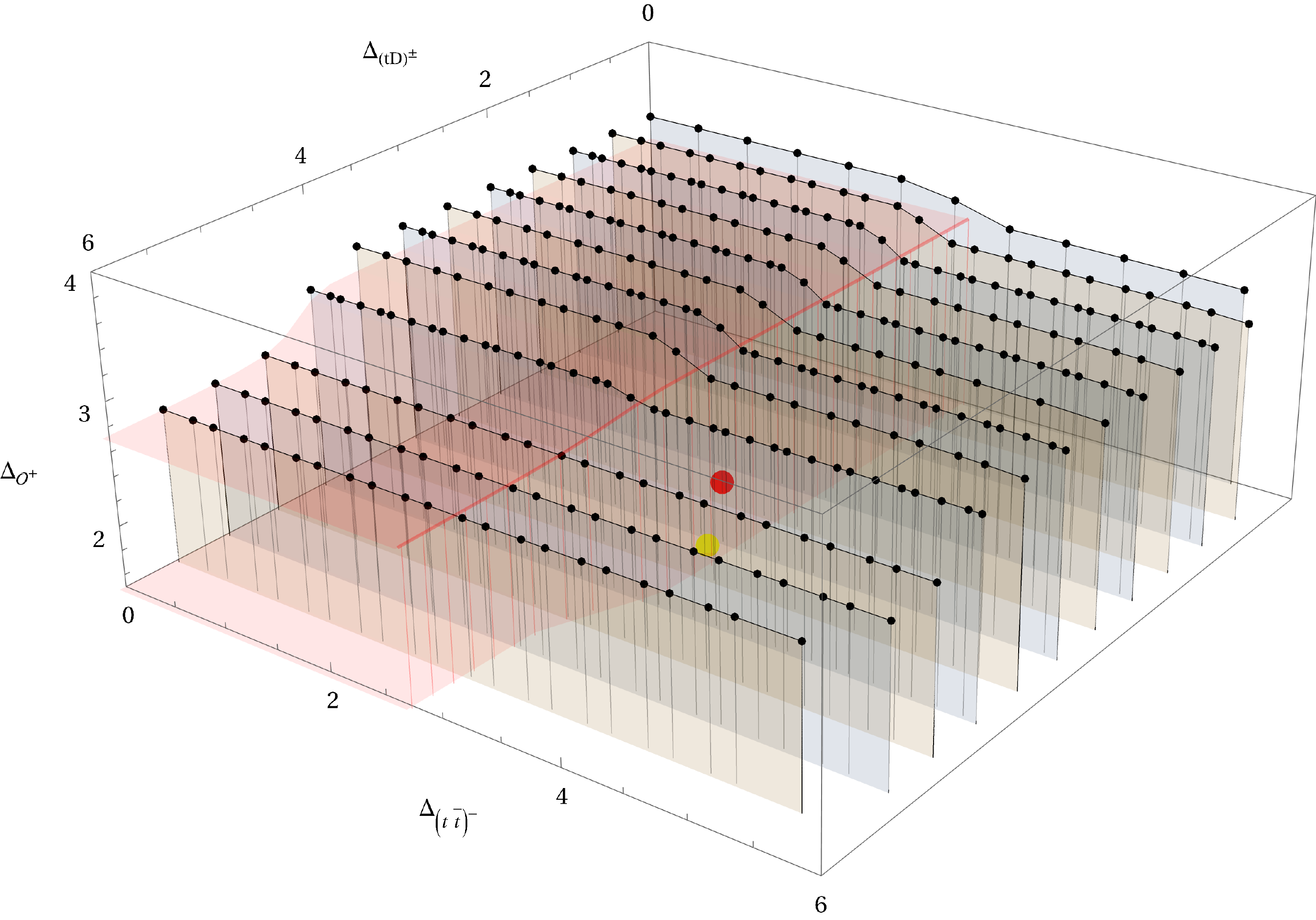}
	\caption{Bounds on the dimension of the first $\mathcal{S}$-parity even singlet $\mathcal{O}^{+}$ in the $t \times \bar{t}$ and the $\Disp \times \Dispb$ OPEs as a function of the scaling dimension of $\tD^{\pm}$ and $\ttb^{-}$.  The GFB solution, given in eq.~\eqref{eq:gff-MC}, is shown in red. The $\veps$-expansion results for the magnetic line defect are given in yellow.  The light-shaded region represents the results for the agnostic bound $\D_{\ttb^{+}} = \D_{\ttb^{-}}$. $ \Lambda = 33, P = 63$. 
	}
	\label{fig:MCBos_ttDD_L21P41_3d}
\end{figure}
There are several operators one can study with these crossing equations.
On the one hand, we have  the leading non-identity $\mathcal{S}$-parity even primary $\mathcal{O}^{+}$ which appears both in the $t \times \bar{t}$ and the $\Disp \times \Dispb$ OPEs.
On the other hand, there are the leading $\mathcal{S}$-parity odd primaries $\ttb^{-}$ and $\DDb^{-}$, which are in general different to each other. Finally, there is the lowest-lying $O(2)_F \times SO(2)_T$ vector in the $t \times \Disp$ channel
\begin{align}
	t \times \Disp \sim \tD^{\pm}+\dots~,
\end{align}
where the superscript $\pm$ denotes the $\mathcal{S}$ parity of the operator.
In figure~\ref{fig:MCBos_ttDD_L21P41_3d} we compute the upper bound on $\D_{\mathcal{O}^+}$, while assuming gaps on the scaling dimensions of $\ttb^{-}$, $\tD^{\pm}$ keeping $\D_{\tD^{+}} = \D_{\tD^{-}}$.
We take all other gaps to be very close to the unitarity bound, concretely we set them to $\D > 0.001$.
If we are interested in the most `agnostic' bound with $\D_{\mathcal{O}^{+}} = \D_{\ttb^{-}}$, then the allowed region shrinks to the light-red region of figure~\ref{fig:MCBos_ttDD_L21P41_3d}, which contains both the GFB (red dot) and the $\veps$-expansion result for the magnetic line defect.
The latter is far from saturating the upper bound and it seems hard to make progress without further assumptions. We will come back to this issue in section~\ref{sec:MCpinning}. 
The bounds are very uniform and show two drops, one in the $\D_{\ttb^{-}}$ direction around $\D_{\ttb^{-}} \simeq 3$ and one in the $\D_{\tD^{\pm}}$ direction around $\D_{\tD^{\pm}} \simeq 5$. For $\D_{\tD^{\pm}} \gtrsim 5$ the bound on $\D_{\mathcal{O}^{+}}$ becomes flat and approaches the value $\D_{\mathcal{O}^{+}}\simeq  2.7$, a result we already found in the single-correlator bootstrap of the displacement operator of section~\ref{sec:SCdisp}. In the upper `cubic-shaped' region we have that $\D_{\mathcal{O}^{+}}\lesssim 3.4$, which approaches the bound for a real GFF $\D_{\mathcal{O}^{+}} = 3$. 

\begin{figure}
\centering
 \includegraphics[width=0.9 \textwidth]{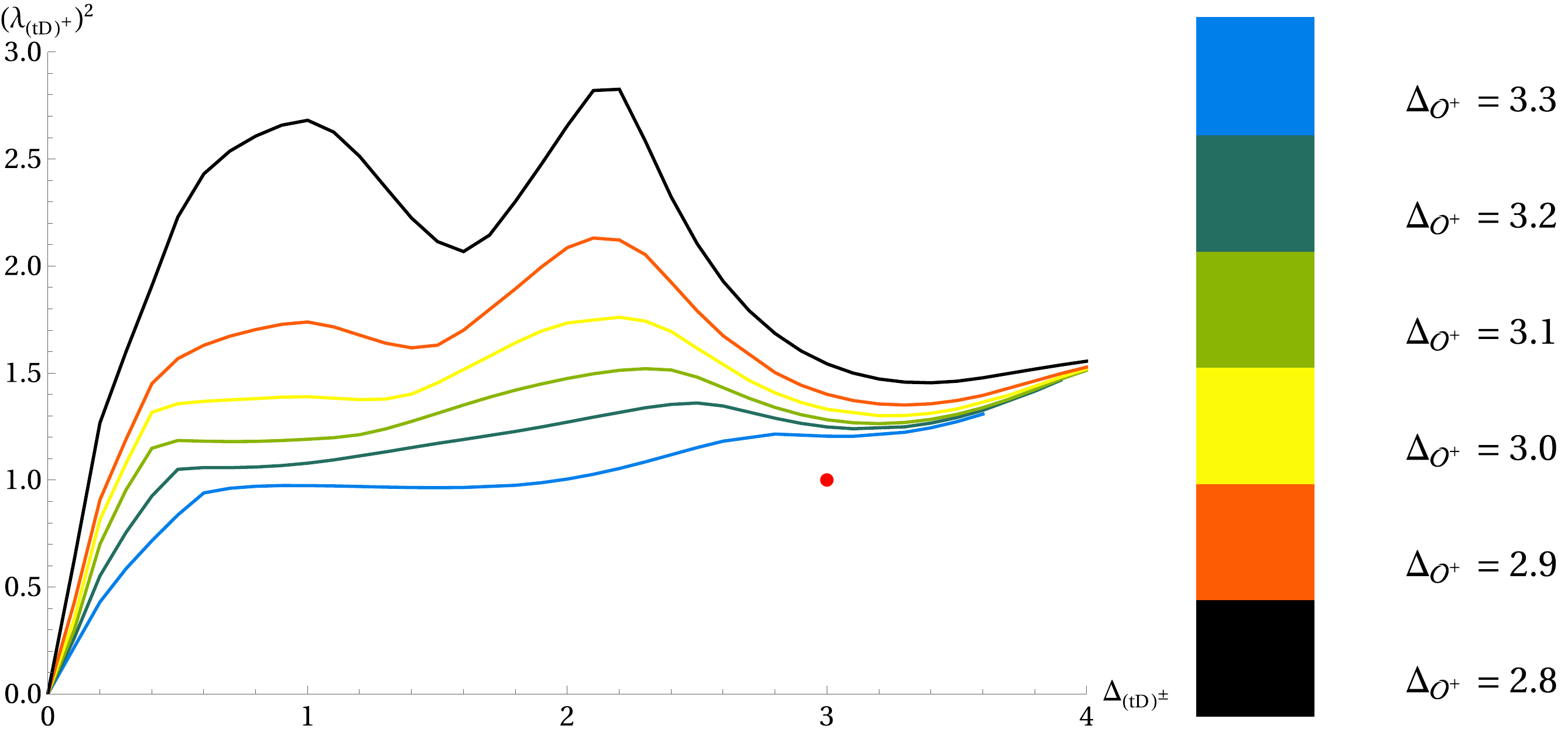}
	\caption{Bounds on $(\lambda_{t \Disp \tD^{+}})^2$ as a function of the scaling dimension of $\Delta_{\tD^{+}}$ and of the scaling dimension of the first parity-even singlet $\D_{\mathcal{O}^{+}}$. The gap on the dimension $\D_{\tD^{-}}$ is set to $\D_{\tD^{+}}$, and all other gaps are set to 0.001. The GFB solution given in eq.~\eqref{eq:gff-MC} is marked by the red dot. $ \Lambda = 33, P = 63$. }
	\label{fig:MCBos_ttDD_L21P41_Phi1b2OPE}
\end{figure}

\paragraph{Including one OPE coefficient.}
In sections~\ref{sec:SCdisp} and~\ref{sec:SCtilt} we have seen that including one OPE coefficient leads to interesting bounds. We repeat this strategy here and bound $(\lambda_{t \Disp \tD^{+}})^2$ while varying the gap $\D_{\tD^{\pm}}$ in the agnostic region $\D_{\ttb^{+}} = \D_{\ttb^{-}}$. It turns out that $(\lambda_{t \Disp \tD^{+}})^2$ is unbounded for $\D_{\mathcal{O}^{+}} \lesssim 2.7$. Above this value and for $\D_{\mathcal{O}^{+}} \lesssim 3.4$ there exists an upper bound which is shown in figure~\ref{fig:MCBos_ttDD_L21P41_Phi1b2OPE}, while for $\D_{\mathcal{O}^{+}} \gtrsim 3.4$ the upper bound $(\lambda_{t \Disp \tD^{+}})^2$ becomes negative, consistently with the results shown in figure~\ref{fig:MCBos_ttDD_L21P41_3d}. 
Since we are assuming a gap on the lowest operator $\D_{\tD^{\pm}}$, but do not make any assumptions on the scaling dimension of the next operator in the $t \times \Disp$ OPE, the lower bound on $(\lambda_{t \Disp \tD^{+}})^2$ is at zero.
The $\veps = 1$ solution of the $\veps$-expansion for the magnetic line defect is not shown, since it lies far within the bounds in the region where the OPE coefficient $(\lambda_{t \Disp \tD^{+}})^2$ is unbounded.
 
\subsection{Bootstrapping the monodromy line defect in the \texorpdfstring{$O(2)$}{O(2)} model}
\label{sec:MCmono}

The approach adopted so far was agnostic, in that we bounded CFT data without committing to any particular model. In this section we pursue a complementary approach, and combine the numerical bootstrap with the $\varepsilon$-expansion with the goal of bootstrapping the monodromy defect in the $O(2)$ model. 
Recall from section~\ref{sec:monodromy} two universal features of this monodromy defect: (i) for generic values of the parameter $v\in[0,1)$ the flavor symmetry of the model is $SO(2)_F$, which gets enhanced to $O(2)_F$ for $v = 0,\frac{1}{2}$; (ii) the defect spectrum contains a family of $\mathcal{S}$-parity even defect primaries $\Psi_s$ with $SO(2)_F$ charge $v$ and transverse spin $s \in \mathbb{Z}+v$. 
These two features can be combined in our numerical bootstrap problem as follows. First, we consider a system of correlation functions that involve the lowest-lying operator with $SO(2)_F$ charge $r=v$ and its partner with charge $r=1-v$. By charge conservation, the non-zero correlation functions are
\begin{align}
	\begin{split}
		\langle \Psi_{v} (\tau_1) \bar{\Psi}_{v} (\tau_2) \Psi_{v} (\tau_3) \bar{\Psi}_{v} (\tau_4) \rangle\:, &\quad  \langle \Psi_{1-v} (\tau_1) \bar{\Psi}_{1-v} (\tau_2) \Psi_{1-v} (\tau_3) \bar{\Psi}_{1-v} (\tau_4) \rangle\:, \\
		\langle \Psi_{v} (\tau_1) \bar{\Psi}_{1-v} &(\tau_2) \Psi_{1-v} (\tau_3) \bar{\Psi}_{v} (\tau_4) \rangle \:, \label{sys:monodr}
	\end{split}
\end{align}
which lead to the crossing equations presented in eq.~\eqref{eq:cross-eq-cplx-cplx}. 
Although for general $v$ the monodromy defect is not invariant under $O(2)_F$ symmetry, but only under $SO(2)_F$ symmetry, we can use the same crossing equations in both cases.
The justification for this appeared in~\cite{Kos:2015mba}, but we repeat it here for convenience.
The tensor $\epsilon_{ij}$ is invariant under $SO(2)$ but not under $O(2)$, so the antisymmetric representation is isomorphic to the singlet representation for $SO(2)$ but not for $O(2)$.
However, this does not lead to additional relations in the crossing equations, because even when the singlet and antisymmetric representations are isomorphic, we can distinguish them since they contain $\Sm$-parity even and odd operators respectively.
\begin{figure}
	\centering
	\subfloat[]{\hspace{-2cm}
\includegraphics[width=0.8\textwidth]{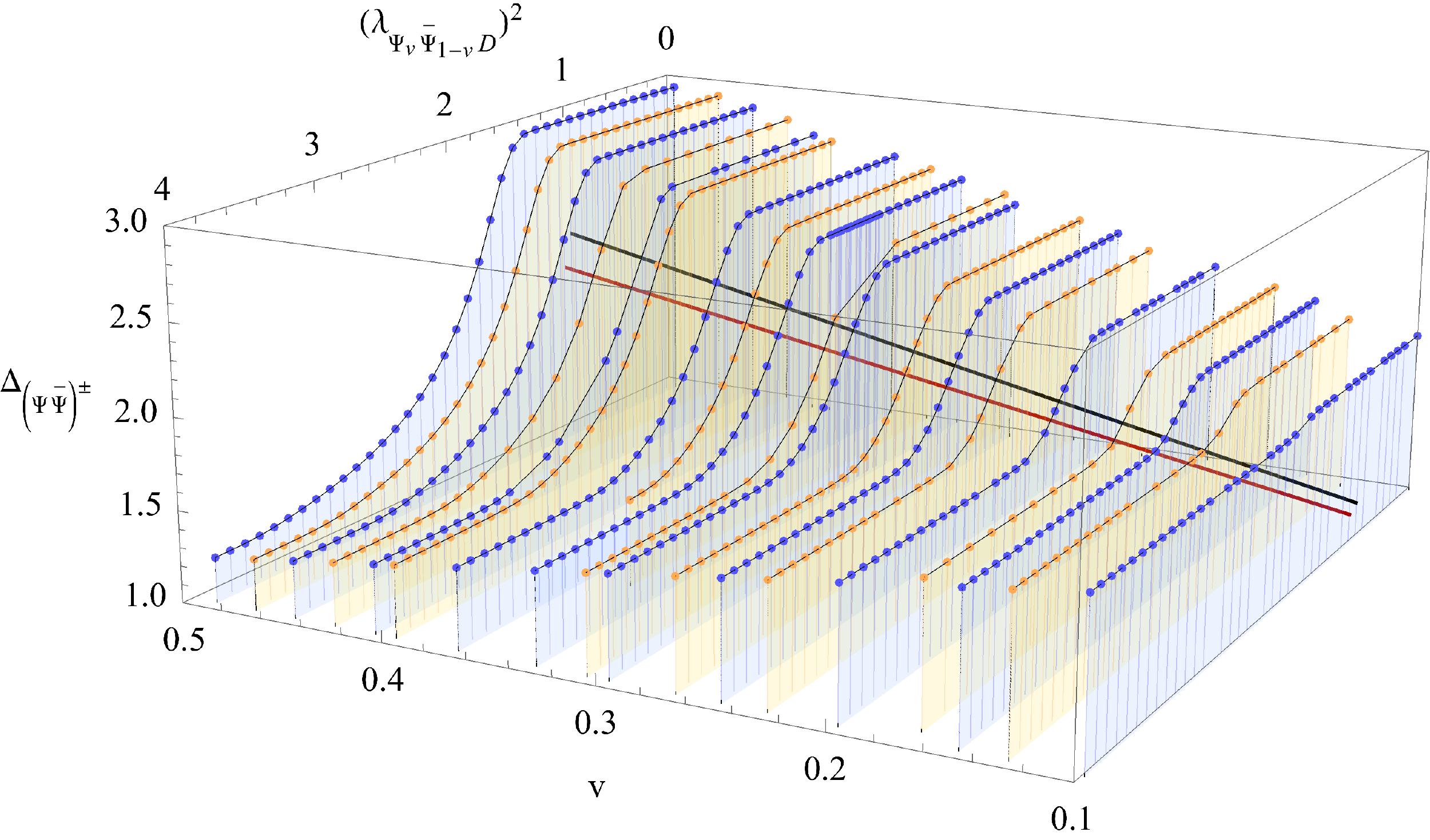}}\\
	\subfloat[]{
			\includegraphics[width=0.8\textwidth]{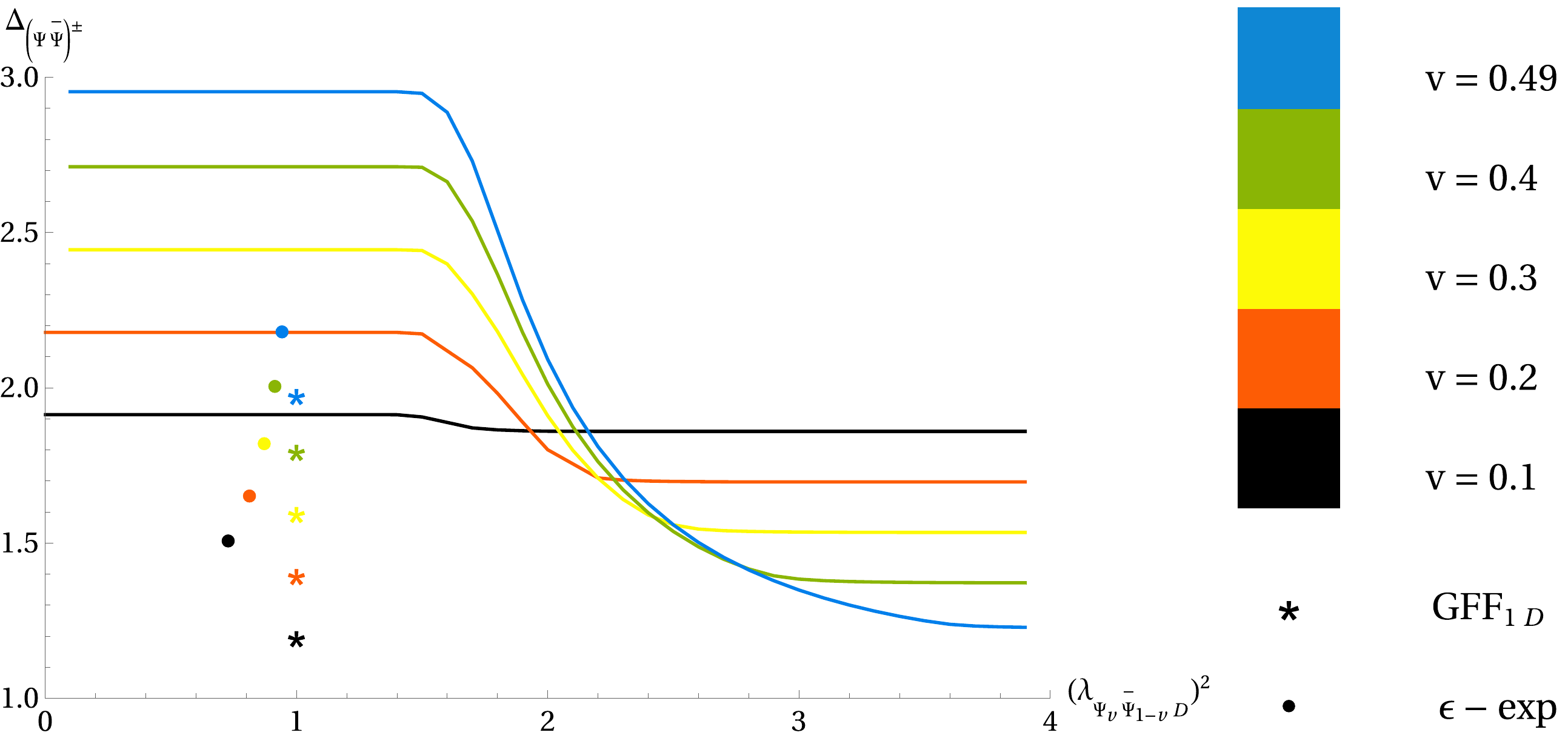}}
	\caption{Bounds on the dimension of the first singlet in the $\Psi_s \times \bar{\Psi}_s$ OPE $\D_{\pspsb^{\pm}}$ versus the OPE coefficient of the displacement operator and the monodromy $v$. The gaps on the dimension of the lowest-lying operators in all other channels is set to the unitarity bound. Parity-even and odd operators are set to have the same gaps. The gray line corresponds to the $\varepsilon$-expansion results. $\Lambda = 21, P = 41$.}
	\label{fig:MCBos_PhiibvsDispOPEvsv_L21P41}
\end{figure} 

Second, we input the scaling dimensions of $\Psi_s$ using the predictions from the $\varepsilon$-expansion in eq.~\eqref{eq:dimPsiMon}, extrapolated to $\veps = 1$.
With this input, we put an upper bound to the dimension of the lowest-lying operator in $\Psi_s \times \bar{\Psi}_s$, i.e.
\begin{align}
	\Psi_s \times \bar{\Psi}_s = \id+\pspsb^{\pm} +\dots,\quad s=v,\text{~or~} v-1~.
\end{align}
Note that $\pspsb^{\pm}$ is a singlet under $SO(2)_T\times SO(2)_F$, and it could be either $\mathcal{S}$-parity even or $\mathcal{S}$-parity odd.
For the $\Psi_{v} \times \bar{\Psi}_{v-1}$ OPE we assume the lowest-lying operator is the displacement operator, which is $\mathcal{S}$-parity even. We will further assume that there is no operator in the $\mathcal{S}$-parity odd channel with dimension smaller than the displacement, namely:
\begin{align}
	(\Psi_{v} \times \bar{\Psi}_{v-1})^{+} = \Disp + \dots\:, \quad (\Psi_{v} \times \bar{\Psi}_{v-1})^{-} = (\Psi_{v} \bar{\Psi}_{v-1})^{-} + \dots \:,\quad \D_{ (\Psi_{v} \bar{\Psi}_{v-1})^-}\geq 2~.
\end{align}
These assumptions are true in the $\varepsilon$-expansion, where one can see at leading order that the displacement operator appears in the OPE with the expected dimension $\D_\Disp=2$, see appendix~\ref{app:moremonodromy} for more details.

In figure~\ref{fig:MCBos_PhiibvsDispOPEvsv_L21P41} we plot the upper bound on $\D_{\pspsb^{\pm}}$ as a function of the $SO(2)_F$ charge $v$ and the OPE coefficient of the displacement operator $(\lambda_{\Psi_{v} \bar{\Psi}_{1-v} \Disp})^2$. 
Let us stress that $v$ enters the crossing equations through the dimension of the external operators, which we take to be the $\veps$-expansion prediction~\eqref{eq:dimPsiMon}.
Furthermore, since the crossing equations in eq.~\eqref{sys:monodr} are invariant under $v\leftrightarrow 1-v$, it suffices to consider the range $0<v\leq 1/2$, where the limiting case $v =1/2$ corresponds to the $\mathbb{Z}_2$ twist defect studied in~\cite{Gaiotto:2013nva}.
In the figure we observe a family of drops as we move along the ($\lambda_{\Psi_{v} \bar{\Psi}_{1-v} \Disp})^2$ direction, and as we increase $v$ towards $v=1/2$ these drops become sharper and move slightly to the right.  
It is promising that the $\veps$-expansion results are above and left of the free theory solutions, because this makes it possible for the theoretical prediction to saturate the bound. 
However, for the results with $\Lambda = 21$ derivatives shown in figure~\ref{fig:MCBos_PhiibvsDispOPEvsv_L21P41}(b), both the $\varepsilon$-expansion result and the free theory solution are still somewhat far from saturating the numerical bound. 

\begin{figure}
   \centering
	\hspace{2cm}\includegraphics[width = 0.8\textwidth]{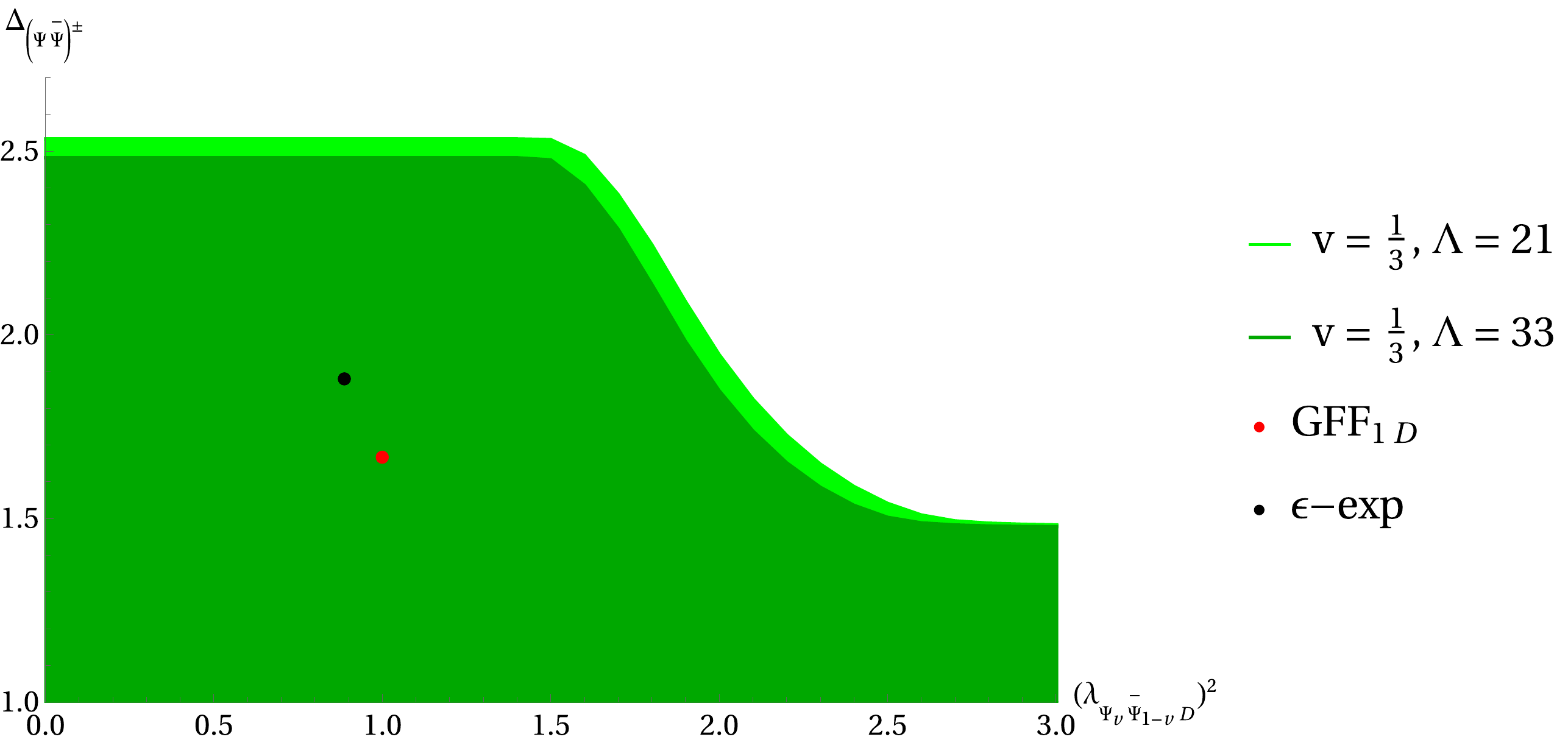}
	\caption{
		Bounds on the dimension of the first singlet in the $\Psi_s \times \bar{\Psi}_s$ OPE $\D_{\pspsb^{\pm}}$ versus the OPE coefficient of the displacement operator $(\lambda_{\Psi_v \bar{\Psi}_{1-v} D})^2$. $v = \frac{1}{3}$ and the gaps on the dimension of the lowest-lying operators in all other channels are set to the unitarity bound. Parity-even and odd operators are set to have the same gaps. $\Lambda = 21, P = 41$ and $\Lambda = 33, P = 53$.}
	\label{fig:MCBos_PhiibvsDispOPE_v033_L21P41}
\end{figure}

One possibility that we explore in figure~\ref{fig:MCBos_PhiibvsDispOPE_v033_L21P41} is whether increasing the number of derivatives can bring the bound closer to the analytical prediction.
For concreteness we focus on $v=1/3$,\footnote{The choice $v=1/3$ is particularly interesting for its connection to half-BPS defects in superconformal field theories. For example, we expect that the monodromy defect for the Wess-Zumino model~\cite{Gimenez-Grau:2021wiv} preserves supersymmetry whenever $v = 1/3$.} and we increase the number of derivatives to $\Lambda=33$. We observe that the kink moves slightly towards the left of the plot, but remains somewhat far from the analytical prediction.
Another possibility is that one needs to consider less agnostic gap assumptions in the channels besides $\Psi_{v} \times \bar{\Psi}_{1-v}$, or perhaps higher-order corrections from the $\veps$-expansion are needed to reconcile theory and numerics.
In either case, this provides a good motivation for a more detailed study of monodromy defects using the $\veps$-expansion, to which we hope to come back in future work.

\subsection{Bootstrapping the localized magnetic field line defect}
\label{sec:MCpinning}

\begin{figure}
	\centering
	\subfloat[]{
			\includegraphics[width=0.5\textwidth]{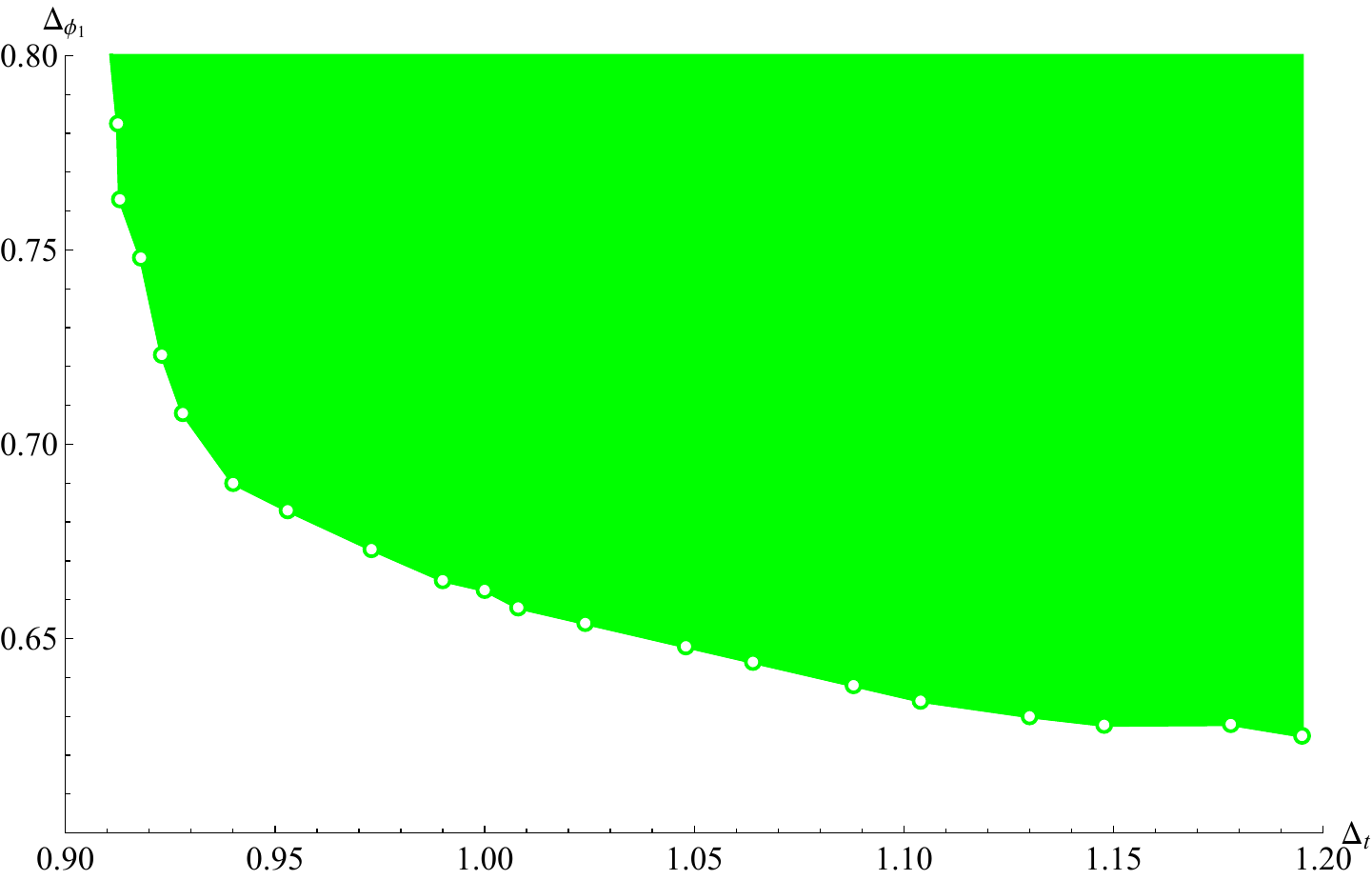}}
\subfloat[]{
			\includegraphics[width=0.5\textwidth]{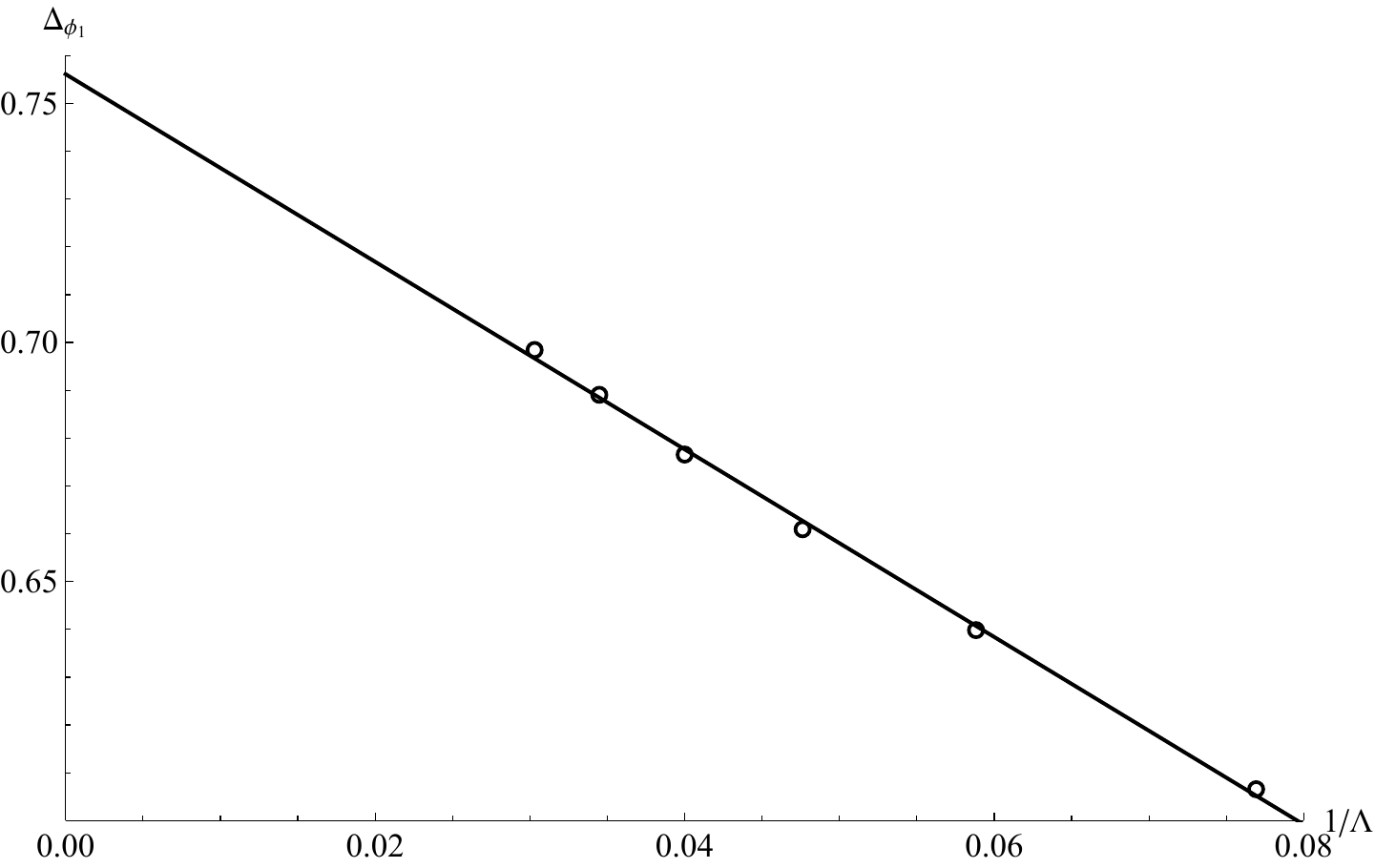}}
	\vspace{0.1cm}
	\caption{Bootstrapping the $O(3)$-breaking line defect. In (a), bounds on the scaling dimensions $\D_{\phi_1}$ and $\D_{t}$ for $\Lambda = 21, P =41$.  The gaps are given in eq.~\eqref{eq:gapsTTss}. Allowed values of $(\D_t,\D_{\phi_1})$ are given in green. In (b), the bound on $\D_{\phi_1}$ for $\D_t = 1$ as a function of the number of derivatives $\Lambda$, for which a fit is performed.}
	\label{fig:MCBos_dimBounds}
\end{figure}

This section studies the magnetic line defect of~\cite{Cuomo:2021kfm} in the bulk $O(3)$ CFT, combining the numerical bootstrap with the $\veps$-expansion results of section~\ref{sec:epsexp}. There are two features of the magnetic line defect which are important in our analysis: (i) the model is invariant under a $O(2)_F \subset O(3)$ flavor symmetry and, (ii) the defect spectrum features a tilt operator $t$ transforming in the vector representation of $O(2)_F$.
To exploit these features, we consider a bootstrap problem involving the tilt $t$ and the lowest-dimension neutral scalar $\phi_1$, so we consider the correlation functions
\begin{align}\label{sys:LMD}
	\langle \phi_1(\tau_1) \phi_1(\tau_2)  \phi_1(\tau_3)  \phi_1(\tau_4)  \rangle~\:, 
	&\quad	\langle  t(\tau_1)  \bar{t}(\tau_2)  t(\tau_3)  \bar{t}(\tau_4)  \rangle~\:,\nonumber\\
	\langle \phi_1(\tau_1) \phi_1(\tau_2)  &t(\tau_3)  \bar{t}(\tau_4)  \rangle~\:,
\end{align}
plus all other non-equivalent orderings. The corresponding crossing equations are given in eq.~\eqref{eq:cross-eq-real-cplx}.\footnote{Since we are not including the displacement operator, this system of correlation functions is agnostic about the co-dimension of the line defect. It will enter only implicitly via our gap assumptions which are determined by the $\veps$-expansion results of section~\ref{sec:epsexp}. The inclusion of the displacement operator is more involved.
Nevertheless, it is an interesting extension and we will leave it for future work.} 
There are five OPE channels that enter in our discussion
\begin{align}\label{eq:OPEmgl}
	\begin{split}
		&\phi_1 \times \phi_1\sim \id+\phi_1 + s_{-} + \cdots\:, \quad (t \times \bar{t})^{+} \sim \id+\phi_1 + s_{-} + \cdots\:, \quad 
		\\
		&(t \times \bar{t})^{-}  \sim A + \cdots, \quad
		t \times t  \sim T +\dots\:,	\quad
		t \times \phi_1  \sim t + V + \cdots\:,
	\end{split}
\end{align}
where further details on the operators exchanged can be found in section~\ref{sec:localized-overview}.
Note that the external operators are exchanged in some of the fusion channels, which allows us to impose the extra relations $\lambda_{\phi_1 t \bar{t} } = \lambda_{t \bar{t} \phi_1}$ in the crossing equations. Furthermore, we impose the following gaps
\begin{align}\label{eq:gapsTTss}
	\D_{s_{-}} = 2.36\:, \qquad \D_{A} = 3\:, \qquad \D_{T} = 2.18\:, \qquad \D_{V} = 3.18\:,
\end{align}
where the values are the $O(\veps)$ results for the scaling dimensions from eqs.~\eqref{eq:delta-spm}, \eqref{eq:dimV} and \eqref{eq:dimAT}.
With these assumptions we bound the scaling dimensions of $\phi_1$ and $t$. Although we could also fix the scaling dimensions of $t$ and $\phi_1$ using the perturbative calculation (in particular, $t$ has protected dimension $\D_t=1$), we keep them unfixed to see how they are constrained by the numerics. This logic is inspired by the search for the Ising model island, where (physically motivated) gap assumptions led to constraints on the external operators $\sigma$ and $\epsilon$~\cite{Kos:2014bka}. Our results are presented in figure~\ref{fig:MCBos_dimBounds}, where the allowed values in the  $(\D_t, \D_{\phi_1})$ plane are shown in green.

The lower bound for $t$ is strikingly close to 1, the numerics seems to be rediscovering the tilt.
For this value of $\D_t$, the bound goes up and cuts out a corner. It should also be noted that there are numerical instabilities in the region with large $\D_{\phi_1}$ outside the range of figure~\ref{fig:MCBos_dimBounds}(a). 
The lower bound on $\D_{\phi_1}$ is on the other hand weaker, its value being nowhere close to the Pad\'e extrapolation of $\D_{\phi_1}\simeq 1.55$ (and it is even further away from the crude $\veps$-expansion result $\D_{\phi_1}=2+O(\veps^2)$). It is nevertheless a non-trivial numerical result that such a lower bound exists at all. Note that the magnetic line defect has no relevant operators~\cite{Cuomo:2021kfm}, so $\D_{\phi_1}>1$ should be expected. Our $O(\veps)$ assumptions allow for a weaker numerical bound on $\D_{\phi_1}$. For the physically interesting value $\D_t=1$, figure~\ref{fig:MCBos_dimBounds}(b) displays an extrapolation to infinite number of derivatives of the lower bound, which converges to 
\begin{equation}
	\D_{\phi_1} \gtrsim 0.76\,.
\end{equation}
Again, we cannot completely rule out the region with $\Delta_{\phi_1} \leq 1$. It could be that our numerics is not strong enough to rule out the presence of relevant operators, so we should either change our assumptions or include more external operators to the system of eq.~\eqref{sys:LMD}. Another possibility which we cannot rule out is the existence of alternative models consistent with our assumptions but with one relevant scalar in the spectrum. Definitely, we think this result calls for a more systematic investigation, which nowadays can be efficiently performed with the help of the \emph{Navigator Function}~\cite{Reehorst:2021ykw,Reehorst:2021hmp} in order to search for bootstrap islands in a large parameter space. Finally, one may wonder if our gap assumptions allow for upper bounds on $\Delta_{\phi_1}$ as well. Clearly, $\D_{\phi_1}$ cannot be bigger than $\simeq 2.36$, which is when the next operator $\D_{s_{-}} $ appears, see eq.~\eqref{eq:gapsTTss}. Below this threshold there seems to be no upper bound, at least with the current number of derivatives.

\begin{figure}
	\centering
	\subfloat[]{\includegraphics[width=0.52\textwidth]{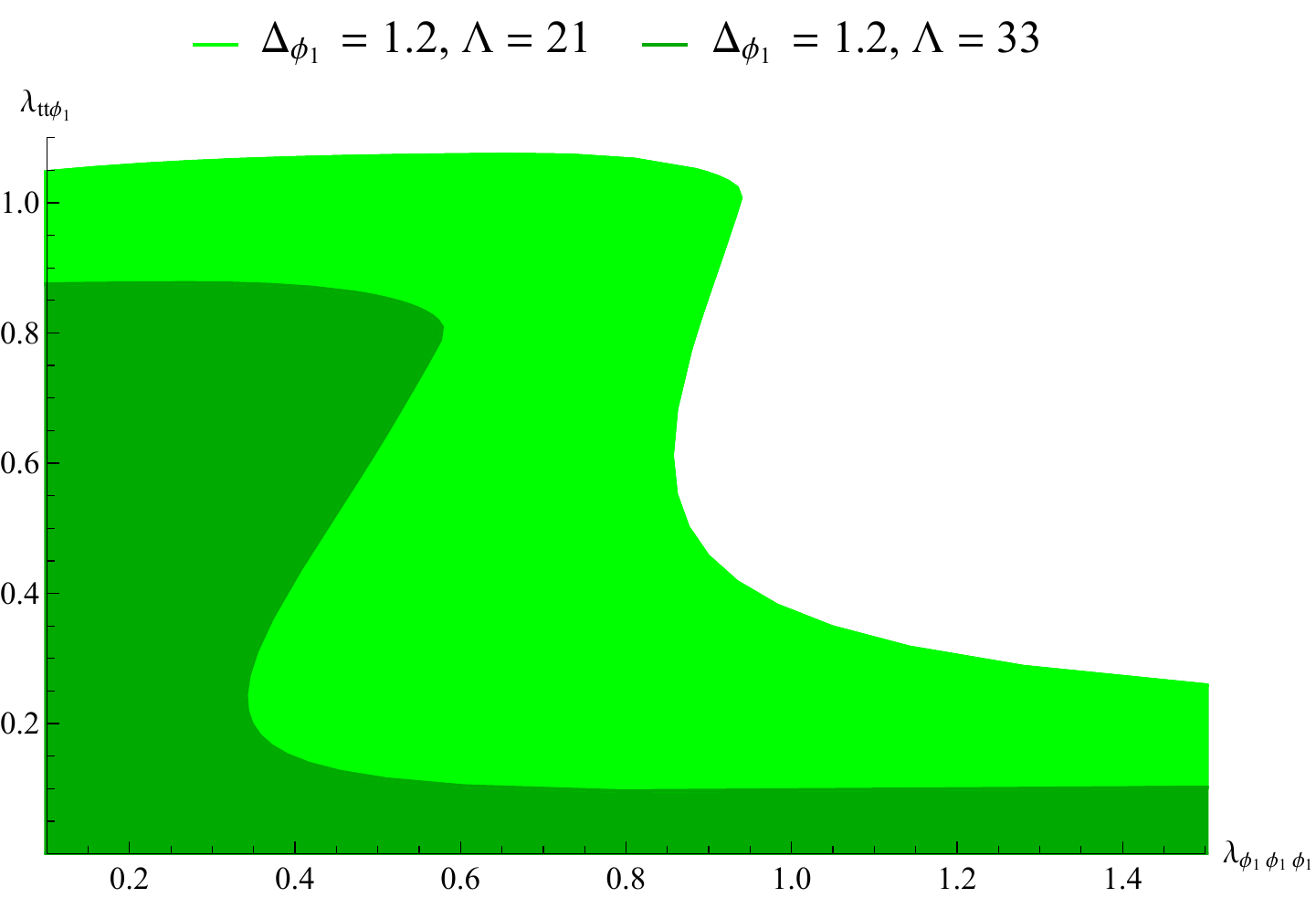}}
	\subfloat[]{\includegraphics[width=0.52\textwidth]{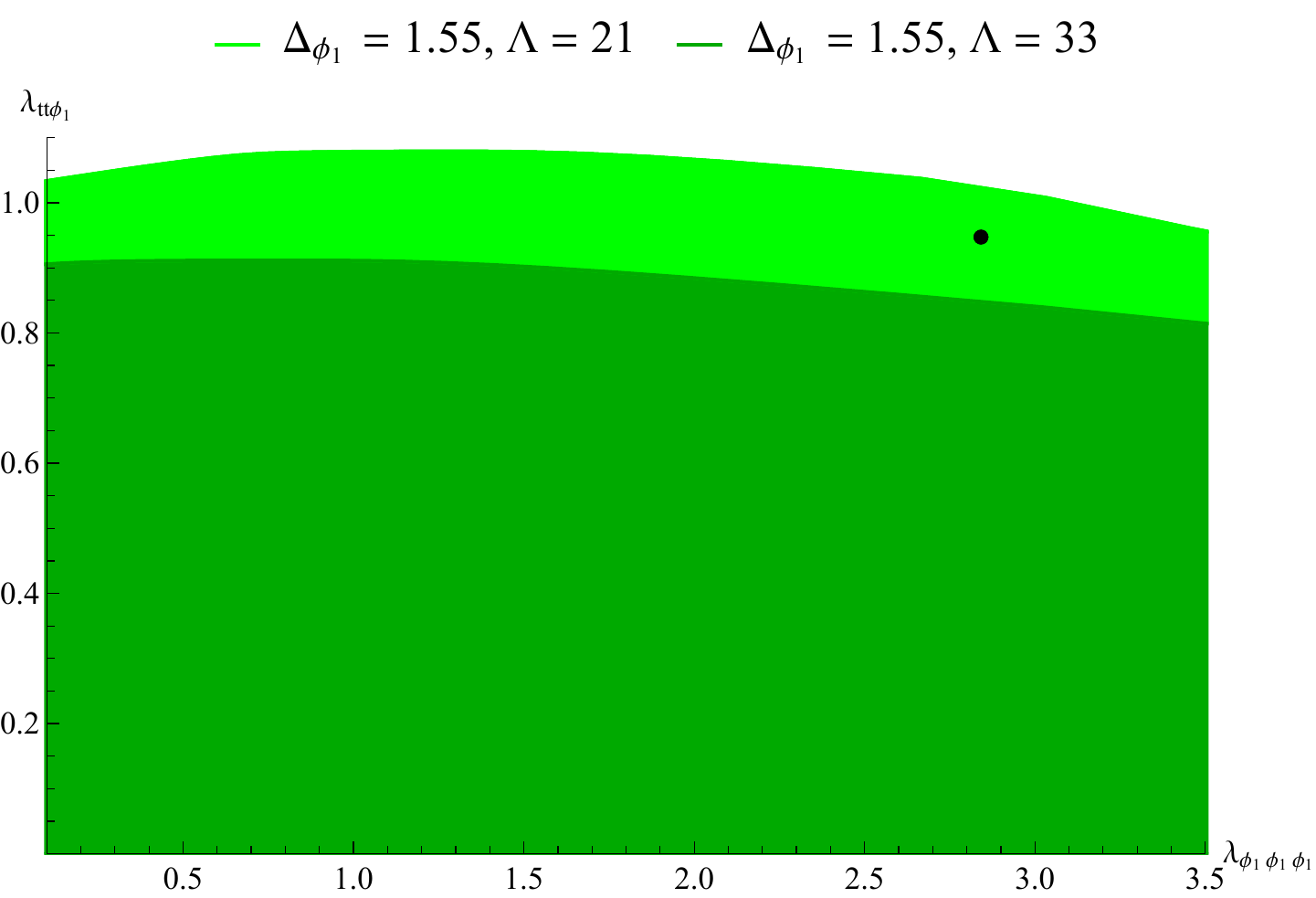}}\\
	\caption{Bootstrapping the $O(3)$-breaking magnetic line defect. Upper bounds on the OPE coefficient $\lambda_{\phi_1 \phi_1 \phi_1}$ as a function of $\lambda_{t \bar{t}\phi_1}$. $\Lambda=21, P = 41$ and $\Lambda = 33, P = 63$. In (a) the external scalar has dimension $\Delta_{\phi_1} = 1.2$. In (b) the external scalar has dimension $\Delta_{\phi_1} = 1.55 $. The $\veps$-expansion result is shown by the black dot.}
	\label{fig:MCBos_ttpp_slice}
\end{figure}

\paragraph{Bounds on OPE coefficients.} 
\begin{figure}[h]
	\centering
	\includegraphics[width=0.8\textwidth]{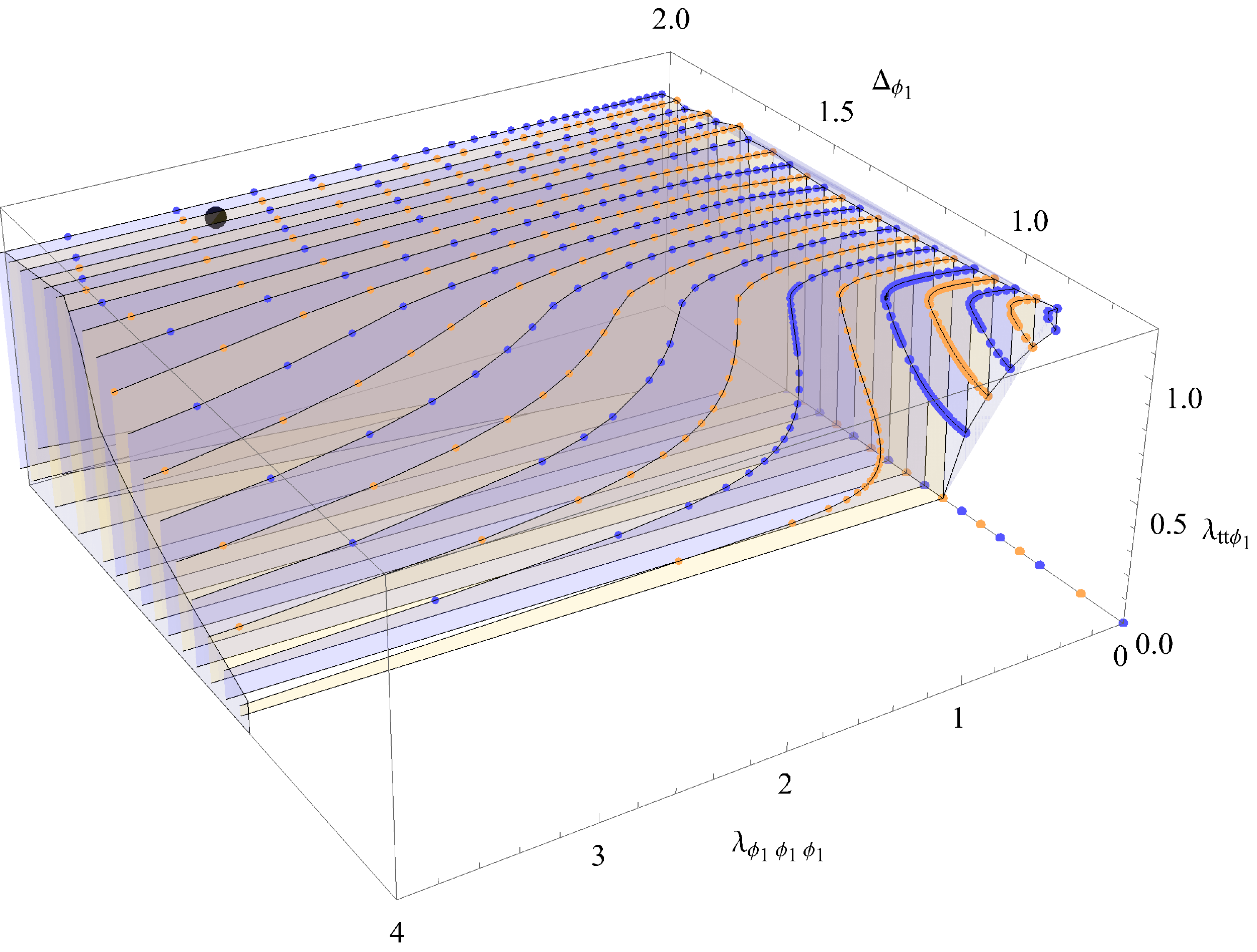} 
	\caption{Bounds on the OPE coefficients $\lambda_{t \bar{t}\phi_1 }$ and $\lambda_{\phi_1 \phi_1 \phi_1}$ as a function of the gap $\Delta_{\phi_1}$ for the $O(3)$-breaking magnetic line defect. The black dot is the prediction from the $\varepsilon$-expansion. The gaps are given in eq.~\eqref{eq:gapsTTss}. $\Lambda = 21, P = 41$. }
	\label{fig:MCBos_ttpp_dimscan}
\end{figure}
Having constrained the region $(\D_{\phi_1}, \D_t)$, we now set $\D_t$ to its physical value of 1, but continue to treat $\D_{\phi_1}$ as an external parameter. The goal is to bound the OPE coefficients $(\lambda_{t \bar{t} \phi_1 })^2$ and $(\lambda_{\phi_1 \phi_1  \phi_1 })^2$ as functions of $\Delta_{\phi_1 }$, which in turn were computed in the $\veps$-expansion in section~\ref{sec:epsexp}, see eqs.~\eqref{eq:dimPhi} and \eqref{eq:lambda-PPP}. To this end, we employ the strategy developed in~\cite{Kos:2014bka}, i.e. we introduce the OPE angle $\theta$ defined as
\begin{equation}\label{eq:theta}
	\tan \theta = \frac{\lambda_{\phi_1  \phi_1  \phi_1 }}{\lambda_{t \bar{t} \phi_1 }}\:,
\end{equation}
and we search for upper and lower bounds on the quantity $(\lambda_{\phi_1  \phi_1 \phi_1 })^2 + (\lambda_{t \bar{t} \phi_1 })^2$ as a function of $\theta \in [0, \pi)$ and of $\D_{\phi_1 }$. For concreteness we restrict to the case where $0 \leq \theta <\frac{\pi}{2}$ (i.e. $\lambda_{\phi_1  \phi_1 \phi_1 }\lambda_{t \bar{t} \phi_1 }\geq 0$), which is also compatible with the leading $\veps$-expansion prediction. The numerics is of course sensitive to the relative sign only.
The results in figure~\ref{fig:MCBos_ttpp_dimscan} show that for any value of the OPE coefficients $\lambda_{\phi_1 \phi_1 \phi_1}$ and $\lambda_{t \bar{t} \phi_1}$, there is an excluded region which is fully consistent with the lower bound from figure~\ref{fig:MCBos_dimBounds}(a). However, as it turns out, for $\lambda_{\phi_1 \phi_1 \phi_1}/\lambda_{t \bar{t} \phi_1} \geq 0$, the excluded region is larger and exists for $\D_{\phi_1} \lesssim 0.9$, while for $\lambda_{\phi_1 \phi_1 \phi_1}/\lambda_{t \bar{t} \phi_1} \leq 0$, only values up to $\D_{\phi_1} \lesssim 0.7$ are excluded.
For values between $0.7 \lesssim \Delta_{\phi_1} \lesssim 1.1$, the upper and lower bounds on $\lambda_{t \bar{t} \phi_1}$ approach each other until they meet, after which the upper bound goes to zero. This means that certain values of the OPE coefficent $\lambda_{\phi_1 \phi_1 \phi_1}$ are ruled out by the numerical bootstrap, even though the corresponding $\Delta_{\phi_1}$ values are allowed in figure~\ref{fig:MCBos_dimBounds}(a).
For $\Delta_{\phi_1} \gtrsim 1.1$, the bound on $ \lambda_{t \bar{t} \phi_1}$ is always positive. There is an intriguing feature visible starting around $\Delta_{\phi_1} \lesssim 0.9$ and continuing until $\Delta_{\phi_1} \simeq 1.4$, where the projection onto the $(\lambda_{\phi_1 \phi_1 \phi_1},\lambda_{t \bar{t} \phi_1})$ plane shows a pronunced cusp, see figure~\ref{fig:MCBos_ttpp_slice}(a). For higher values of $\Delta_{\phi_1}$ this feature disappears and the plot becomes smooth.
For the quadrant where the OPE coefficients have opposite sign this cusp is not present.
In order to better understand this feature, we have plotted two slices of the three-dimensional plot in figure~\ref{fig:MCBos_ttpp_slice}.
For figure~\ref{fig:MCBos_ttpp_slice}(a) we have chosen the value $\Delta_{\phi_1}=1.2$ somewhat arbitrarily, in order to highlight a region where the cusp is clearly visible.
It appears around $\lambda_{\phi_1 \phi_1 \phi_1} \simeq 0.9$, although its position shifts depending on the value of $\Delta_{\phi_1}$ and the number of derivatives.
In figure~\ref{fig:MCBos_ttpp_slice}(b) we have chosen $\Delta_{\phi_1} = 1.55$, which corresponds to the best estimate of the conformal dimension of $\phi_1$ using a Pad\'e approximation~\cite{Cuomo:2021kfm}.
For this value of $\Delta_{\phi_1}$ the features from figure~\ref{fig:MCBos_ttpp_slice}(a) are gone and the bound is smooth. 
We have added in figure~\ref{fig:MCBos_ttpp_slice}(b) the prediction coming from the $\veps$-expansion for reference. We should warn the reader that this prediction is valid up to $O(\veps)$, for which $\Delta_{\phi_1} = 2$. Our numerical plot was obtained with a different set of assumptions, in particular for $\D_{\phi_1} = 1.55$, which explains why the dot is outside the numerical exclusion region. 
One possibility is that the cusp is also present for $\Delta_{\phi_1} \simeq 1.55$, but was lost due to numerical precision. Indeed from figure~\ref{fig:MCBos_ttpp_slice}(a) the feature becomes more pronounced when we jump from $\Lambda=21$ to $\Lambda=33$ derivatives. It would be remarkable if the numerics could be pushed, such that figure~\ref{fig:MCBos_ttpp_slice}(b) starts looking more like figure~\ref{fig:MCBos_ttpp_slice}(a). The position of the cusp would then be a good candidate for the magnetic line defect.

%% file: sections/conclusions.tex
\section{Conclusions and future directions}\label{sec:conclusions}

In this paper we have studied $O(2)$ line defects in a 3d bulk CFT with $SO(2)$ flavor symmetry -- the monodromy defect -- or $O(3)$ flavor symmetry -- the magnetic line defect -- using the conformal bootstrap. After writing the corresponding crossing equations we applied the numerical machinery in order to obtain exclusion bounds valid for generic 1d defect CFTs. We concentrated mostly on two canonical operators: the displacement and the tilt. The displacement operator is truly universal in that it controls the breaking of translations due to the presence of the conformal defect. The tilt operator can exist when a conformal defect breaks a bulk global symmetry, which is still quite a generic phenomenon.

After obtaining general bounds that constrain the landscape of 1d theories, we changed gears and focused on two models of interest: the monodromy defect studied in~\cite{Soderberg:2017oaa,Giombi:2021uae,Bianchi:2021snj}, and the magnetic line defect described in~\cite{Cuomo:2021kfm}. In order to guide the numerics we complemented our bootstrap analysis with explicit analytic calculations in the $\veps$-expansion. For the monodromy defect, the results had already appeared in~\cite{Soderberg:2017oaa,Giombi:2021uae}, while for the magnetic line defect most of the calculations presented in section~\ref{sec:epsexp} are new. 

For the monodromy defect we have kept the gap assumptions to a minimum, using the dimension of the external operators from the $\veps$-expansion results and imposing the existence of the displacement operator. Hence the numerical results are quite general, and they are fully consistent with the analytical data. 
The results for the magnetic line defect are more constraining.
By imposing gaps coming from the $\veps$-expansion, the allowed region for certain OPE coefficients shows a series of intriguing cusps. The location of these cusps however, does not match the best numerical estimates of the conformal dimension of the scalar field $\D_{\phi_1} \simeq 1.55$. Nevertheless, we consider this result an encouraging sign. Ways in which our methods can be improved include higher order terms in the $\veps$-expansion, and higher precision in the numerics. The hope is that with these improvements, the region where the magnetic line defected is expected to sit also develops a cusp. 

There are several avenues in which our research can be extended. A natural question to ask is whether one could construct a monodromy defect that breaks a bulk $O(3)$ symmetry to an $O(2)$ symmetry on the line.
Such a defect would have a tilt operator, a displacement, and fundamental fields with transverse spin related to the charge of the monodromy. Mixed correlators with three external operators have already been studied in~\cite{Go:2019lke,Chester:2019ifh,Chester:2020iyt} in the context of $O(2)$ and $O(3)$ models. 

Another interesting possibilty is to reproduce the $\veps$-expansion results using purely bootstrap means, without explicit Feynman diagram calculations. This line of thinking has been particularly succesful, with numerous analytical results obtained using only symmetry arguments and basic structural constraints. A promising framework tailored for our setup is that of
analytic functionals, which were used in~\cite{Ghosh:2021ruh} to study the Ising twist line defect with $\mathbb{Z}_2$ global symmetry.

Furthermore, we could extend our results to line defects with $O(N)$ flavor symmetry. 
In addition to the analytical tools used in this work, we would gain access to the large $N$ expansion, giving us perturbative results in a different regime to compare with the numerics. 
At the same time, these line defects could break a bulk $O(N + m)$ symmetry
to a defect $O(N)$ symmetry, where $N, m \in \mathbb{Z}$. In this generalization, the tilt will no longer be a vector under $O(2)$ flavour symmetry, but will be part of another representation. 

Finally, our setup can generalized to study monodromy line defects in superconformal models, like the monodromy defect of the Wess-Zumino model studied in~\cite{Gimenez-Grau:2021wiv}. This model contains a displacement as well as a tilt operator due to the breaking of the $R$-symmetry to a $U(1)$ symmetry on the defect, which in turn sit inside protected multiplets of the superconformal algebra. It would be interesting to pursue a systematic study of this setup using the superconformal bootstrap and we leave it for the future~\cite{Gimenez-Grau:202X}.

%% file: sections/app_monodromy.tex
\section{More data on the \texorpdfstring{$O(N)$}{O(N)} monodromy defect in 3d}\label{app:moremonodromy}

In this appendix, we give a more detailed explanation of the calculation of the dimensions and OPE coefficients given in section~\ref{sec:monodromy} and computed in~\cite{Giombi:2021uae}.

\subsection{OPE coefficients with the displacement operator}

The displacement operator appears in the OPE of two defect modes $\Psi_{v} \times \bar{\Psi}_{v - 1} \sim \Disp + \cdots $.
Hence, we are interested in the correlators involving these defect modes, which are

\begin{align}
\begin{split}
	\langle \Psi_{v} (\vec{x}_1) \bar{\Psi}_{v} (\vec{x}_2) \Psi_{v} (\vec{x}_3) \bar{\Psi}_{v} (\vec{x}_4) \rangle\:, &\quad  \langle \Psi_{1-v} (\vec{x}_1) \bar{\Psi}_{1-v} (\vec{x}_2) \Psi_{1-v} (\vec{x}_3) \bar{\Psi}_{1-v} (\vec{x}_4) \rangle\:, \\
	\langle \Psi_{v} (\vec{x}_1) \bar{\Psi}_{1-v} &(\vec{x}_2) \Psi_{1-v} (\vec{x}_3) \bar{\Psi}_{v} (\vec{x}_4) \rangle \:. 
\end{split}
\end{align}

The results for the single-operator correlators were already computed in \cite{Gaiotto:2013nva} for a $\mathbb{Z}_2$ monodromy defect.
For correlators with two operators $\Psi_1$ and $\Psi_2$, the anomalous dimensions of the operators appearing in the OPEs are computed in \cite{Giombi:2021uae}, and from their results also the OPE coefficients can be readily derived.\footnote{Note that our conventions differ slightly from those used in \cite{Giombi:2021uae}; in particular, we normalize the two-point function as
\begin{equation}
 \langle \bar{\Psi}_{s_1} (\vec{x}_1) \Psi_{s_2} (\vec{x}_2) \rangle_{0} = \frac{\delta_{s_1,s_2}}{\vec{x}_{12}^{2\Delta_{s_1}}}\:,
\end{equation}
 without an extra factor of $\mathcal{C}_{\Delta_{s_1}}$ that was present in \cite{Giombi:2021uae}. This does not have any impact on the anomalous dimensions, but needs to be taken into account when computing the OPE coefficients.}
The anomalous dimensions and first-order corrections to the OPE coefficients appear in the four-point correlator as

\begin{align}
 \begin{split}
\label{eq:4ptMonDef}
 &\langle \Psi_{\alpha + v} (\vec{x}_1) \bar{\Psi}_{\alpha - s + v} (\vec{x}_2) \Psi_{\beta - s + v} (\vec{x}_3) \bar{\Psi}_{\beta + v} (\vec{x}_4) \rangle \in \sum_m \big(|\lambda^{(0)}_{\Psi \bar{\Psi} \bar{\Oop}_{\alpha}^{s,m}}|^2 + \varepsilon |\lambda^{(1)}_{\Psi \bar{\Psi} \bar{\Oop}_{\alpha}^{s,m}}|^2 \big)  \\
 &\times \left( \delta_{\alpha \beta}  + \frac{\varepsilon}{2} \Delta^{s,m}_{\alpha \beta}\partial_m \right) W_{d-2+s+2m,0}(\vec{x}_i)\:, 
 \end{split}
\end{align}
where 
\begin{equation}
 W_{\Delta, \ell} (\vec{x}_i) = \left( \frac{\vec{x}_{24}^{2}}{\vec{x}_{14}^2} \right)^{\frac{1}{2} \Delta_{12}} \left( \frac{\vec{x}_{14}^2}{\vec{x}_{13}^2} \right)^{\frac{1}{2} \Delta_{34}} \frac{G_{\Delta, \ell} (u,v)}{\vec{x}_{12}^{\frac{1}{2} (\Delta_1 + \Delta_2)} \vec{x}_{34}^{\frac{1}{2} (\Delta_{3} + \Delta_4)} }\:,
\end{equation}
and $G_{\Delta,\ell}$ are the conformal blocks as defined in \cite{Dolan:2011dv}.
The variables $\alpha,\beta$ in eq.~\eqref{eq:4ptMonDef} refer to degeneracies between operators that have the same tree-level conformal dimension and $SO(2)_T$ spin.
For the correlator of interest to us, we set $s = 1$ and since $\alpha, \beta = 0,1, \cdots, s-1 = 0$ \cite{Giombi:2021uae}, we will not have to worry about said degeneracies.

The anomalous dimensions will receive contributions from the logarithmic term of the four-point correlator, and from the $\mathcal{O}(\varepsilon)$ term of the dimension of the external operators.
The anomalous dimension of the external operators was given in eq.~\ref{eq:dimPsiMon}, and gives a contribution of 
\begin{equation}\label{eq:anomDimDisc}
 \Delta^{(1);\text{disc}} = \Delta_{s_1}^{(1)} + \Delta_{s_2}^{(1)} = \frac{v (v-1) (N + 2)}{2 (N + 8)} \left(\frac{1}{|s_1|}+ \frac{1}{|s_2|}\right) 
\end{equation}
to the anomalous dimension of the operators appearing in their OPEs.
At tree level, the four-point correlator is given by the free theory discussed in eq.~\eqref{eq:gff-MC}.
At $O(\varepsilon)$, one can compute the correction to the correlator through a contact Witten diagram with the four defect modes $\Psi_{s_i}$ as external operators.
These diagrams are described by the well-known D-functions \cite{DHoker:1999kzh,Liu:1998ty,Dolan:2000ut}. 
The first-order correction to the correlator in eq.~\eqref{eq:4ptMonDef} is given by \cite{Giombi:2021uae}
\begin{equation}
 \langle \Psi_{v} (\vec{x}_1) \bar{\Psi}_{v- 1} (\vec{x}_2) \Psi_{v - 1} (\vec{x}_3) \bar{\Psi}_{v} (\vec{x}_4) \rangle_1  = - \pi \lambda_{*} \times D_{\Delta_{\Psi_v} , \Delta_{\Psi_{v-1}}, \Delta_{\Psi_{v-1}}, \Delta_{\Psi_{v}}} (\vec{x}_i)\:, 
\end{equation}
where $\lambda_{*} \sim O(\varepsilon)$ is given in eq.~\eqref{eq:WFlambda}.
The D-function can be expanded in conformal blocks as \cite{Hijano:2015zsa,Jepsen:2019svc}:
\begin{equation}\label{eq:DfuncExp}
 D_{\Delta_{s_1} , \Delta_{s_2}, \Delta_{s_3}, \Delta_{s_4}} (\vec{x}_i) = \sum_{m} P_{1}^{(12)} (m,0) W_{\Delta_m,0} (\vec{x}_i) + \sum_n P_{1}^{(34)}(n,0) W_{\Delta_n,0} (\vec{x}_i)\:.
\end{equation}
Note that since $\lambda_{*} \sim O(\varepsilon)$, we can evaluate the conformal blocks at $p = d - 2 = 2$, and the dimensions $\Delta_m = \Delta_{s_1} + \Delta_{s_2} + 2m$ contribute at tree level.
The coefficients $P_1 (m,0)$ are given in \cite{Giombi:2021uae}.

When $\Delta_{s_1} + \Delta_{s_2} - \Delta_{s_3} - \Delta_{s_4} = 2k, k \in \mathbb{Z}$, where $\Delta_{s_i}$ are evaluated at tree level, the coefficients have a divergence and hence contribute to the anomalous dimension. 
For the correlator we study, $\Delta_{v} + \Delta_{v-1} - \Delta_{v-1} - \Delta_{v} = 0$.
In this case, the D-function can be written as \cite{Giombi:2021uae}:
\begin{align}
\begin{split}\label{eq:DfuncExp2}
 &D_{\Delta_{s_1}, \Delta_{s_2}, \Delta_{s_3}, \Delta_{s_4}} (\vec{x}_i) = \sum_{m} \frac{\pi \big(\Pi_{i = 1}^{4} (\Delta_{s_i})_m \big) (\Delta_{s_1} + \Delta_{s_2} + 2m - 1)^{2}_{-m}}{2 (m!)^2 (\Delta_{s_1} + \Delta_{s_2} + 2m - 1)} \Big(\frac{2}{\Delta_{s_1} + \Delta_{s_2} + 2m -1}\\
 &+ 2 H^m - 2 H^{m + \Delta_{s_1} + \Delta_{s_2} - 2} + 4 H^{2m + \Delta_{s_1} + \Delta_{s_2} -2} - \sum_{i = 1}^{4} H^{m + \Delta_{s_i} - 1} - \partial_m \Big) W_{\Delta_{s_1} + \Delta_{s_2} + 2m,0} (\vec{x}_i)\:,
 \end{split}
\end{align}
where $H^m$ are harmonic numbers.
Setting $s_1 = s_4 = v\:, \quad s_2 = s_3 = v-1$, we can extract the contribution to the anomalous dimension from the logarithmic part:
\begin{equation}\label{eq:anomDimCorr}
 \Delta^{(1);\text{con}} = \frac{4}{10 (2+ 2m)}\:,
\end{equation}
and the $O(\veps)$ correction to the OPE coefficients from the non-divergent part, for $m = 0$:
\begin{equation}
 |\lambda^{(1)}_{\Psi_v \bar{\Psi}_{v-1} \Oop_{0,0}^{1,0}}|^2 = \frac{(N+2) \left(2 H^{1-v}+2 H^{v}-3\right)}{2 N (N+8)}\:.
\end{equation}
Combining the contribution in eq.~\eqref{eq:anomDimCorr} and the contribution in eq.~\eqref{eq:anomDimDisc}, we get
\begin{equation}
 \Delta_{0,0}^{1,m} = \frac{1}{5 (1 + m)} - \frac{1}{5} \:,
\end{equation}
such that indeed, for $m = 0$, $\Delta_{0,0}^{1,0} = 0$ and the displacement, a protected operator, does not get any anomalous dimension.

\subsection{The leading singlet}
The anomalous dimension of the first singlet in the $\Psi_v \times \bar{\Psi}_v$, or the $\Psi_{v-1} \times \bar{\Psi}_{v-1}$ OPE, and its OPE coefficient, can be computed in the same way as for the $\Psi_{v} \times \bar{\Psi}_{v-1}$ OPE.
We still do not encounter degeneracies.
The correction coming from the anomalous dimensions of the external operators is now given by 
\begin{equation}\label{eq:anomDimSCDisc}
 2 \Delta_{s_i}^{(1);\text{disc}} = \frac{v (v-1) (N + 2)}{(N + 8)} \frac{1}{|s_i|}\:,
\end{equation}
and the tree-level dimensions and OPE coefficients of the operators in the $\Psi_{s_i} \times \bar{\Psi}_{s_i}$ OPE are given by the free theory results described in eq.~\eqref{eq:gff-PPbPPb}.
The first-order correction to the correlator is given by 
\begin{equation}
 \langle \Psi_{s_i} (\vec{x}_1) \bar{\Psi}_{s_i} (\vec{x}_2) \Psi_{s_i} (\vec{x}_3) \bar{\Psi}_{s_i} (\vec{x}_4) \rangle_1  = - 2 \pi \lambda_{*} \times D_{\Delta_{s_i} , \Delta_{s_i}, \Delta_{s_i}, \Delta_{s_i}} (\vec{x}_i)\:, 
\end{equation}
and we can use the same conformal block expansion of the D-functions in eq.~\eqref{eq:DfuncExp}.
Since all external dimensions are equal, the relation $\Delta_{s_1} + \Delta_{s_2} - \Delta_{s_3} - \Delta_{s_4} = 2k, k \in \mathbb{Z}$ holds. Hence, the coefficients $P_1 (m,0)$ have a divergence and the D-function is given by eq.~\eqref{eq:DfuncExp2}.
We extract the contributions to the anomalous dimensions from the logarithmic part of eq.~\eqref{eq:DfuncExp2} and obtain
\begin{equation}\label{eq:anomDimSCCon}
 \Delta^{(1);\text{con}}_{v}  = \frac{2(N+2)}{(N+8)(1 + 2v)}\:, \quad \Delta^{(1);\text{con}}_{v-1} =  \frac{2(N+2)}{(N+8)(3 - 2v)}\:.
\end{equation}
The $O(\veps)$ corrections to the OPE coefficients is now given by
\begin{align}
 |\lambda^{(1)}_{\Psi_v \bar{\Psi}_v \Oop_{0}}|^2 &= \frac{2 (N+2) \left(2 H^{v} -  H^{2 v}\right)}{(1 + 2v) N (N+8)}  - \frac{2 (N+2) }{(1 + 2 v)^2 N (N+8)}\:, \\
 |\lambda^{(1)}_{\Psi_{v-1} \bar{\Psi}_{v-1} \Oop_{0}}|^2 &= \frac{2 (N+2) \left(  2 H^{1 - v} -  H^{2 - 2 v}\right)}{(3 - 2v) N (N+8)} - \frac{2 (N+2)}{(3 - 2 v)^2 N (N+8)}\:.
\end{align}
Adding the results from eq.~\eqref{eq:anomDimSCCon} and eq.~\eqref{eq:anomDimSCDisc}, we obtain the following anomalous dimensions:
\begin{align}
 &\Delta^{0,m}_{0,0}|_{s_i = v} = \frac{2(N+2)}{(N+8)(1 + 2v)} + \frac{(N+2) (v-1)}{(N+8)}\:, \\
 &\Delta^{0,m}_{0,0}|_{s_i = v-1} = \frac{2(N+2)}{(N+8)(3 - 2v)} -  \frac{(N+2) v}{(N+8)}\:.
\end{align}
All results are summarized in table~\ref{tab:OPE-coeffs}.
\begin{table}
\centering
 \begin{tabular}{|c|c|c|c|}
OPE & $s$ &  $\Delta^{(1)}$ &  $\lambda^{(1)}$\\
\hline
\hline
$\Psi_v \times \bar{\Psi}_{v-1}$ & 1 & $\frac{N+2}{(N + 8)(2 + 2m)} - \frac{(N+2)}{2 (N+8)} $  & $ \frac{(N+2) \left(2 H^{1-v}+2 H^{v}-3\right)}{2 N (N+8)}$\\
$\Psi_{v-1} \times \bar{\Psi}_{v-1}$ & $0$  & $\frac{2(N+2)}{(N+8)(3 - 2v)} -  \frac{(N+2) v}{(N+8)}$  & $\frac{2 (N+2) \left(2 H^{1 - v} - H^{2 - 2 v}\right)}{(3 - 2 v) N (N+8)} - \frac{2 (N+2)}{(3 - 2 v)^2 N (N+8)}$\\
$\Psi_v \times \bar{\Psi}_v$ & $0$ & $ \frac{2(N+2)}{(N+8)(1 + 2v)} + \frac{(N+2) (v-1)}{(N+8)} $  & $\frac{2 (N+2) \left(2 H^{v} - H^{2 v}\right)}{(1 + 2 v) N (N+8)} - \frac{2 (N+2)}{(1 + 2 v)^2 N (N+8)}$
 \end{tabular}
 \caption{The $\varepsilon$-expansion results for various operators, $m = 0$}
\end{table}
We have only considered the $\Psi_{s_1} \times \bar{\Psi}_{s_2}$ channel.
One can also consider the $\Psi_{s_1} \times \Psi_{s_2}$ channel, which contains operators of the form $\Oop_{\alpha}^{s,m} = \Psi_{\alpha + v} (\vec{\partial}^2)^m \Psi_{s - \alpha - v}$ that have fractional spin $s = s_1 + s_2 = k + 2v \in \mathbb{Z} + 2v$. 
Since the anomalous dimensions have been given in \cite{Giombi:2021uae} and we have shown how to obtain the OPE coefficients from their results, we will not repeat the calculation for this channel in this work.

%% file: sections/app_crosseqs.tex
\section{Crossing vectors}\label{ap:crosseqs}

In this appendix, we give explicit formulas for the vectors that enter the crossing equations. 

\subsection{One complex scalar}

The crossing vectors for eq.~\eqref{eq:cross-one-cplx} read
\begin{equation}
  \vec{V}_{\Delta,S}^{\phi\bar\phi} = \begin{pmatrix}
  F_{-,\D}  \\
  (-1)^S F_{-,\D} \\
  (-1)^S F_{+,\D}  
  \end{pmatrix}\:,\qquad 
  \vec{V}_{\Delta}^{\phi\phi} = \begin{pmatrix}
  0 \\
  F_{-,\D}  \\
  -F_{+,\D}  
  \end{pmatrix} \: ,
 \label{eq:single-scalar-vecs}
\end{equation}
where the shorthand notation $F_{\pm,\Delta} = F_{\pm,\Delta}^{\phi\phi\phi\phi}(\xi)$ is understood.

\subsection{Tilt and displacement}
\label{sec:cv-tilt-disp}

The crossing equations for the tilt and displacement in eq.~\eqref{eq:cross-eq-tilt-disp} are written in terms of the following crossing vectors:
\begin{equation}
 \vec{V}^+_{\Delta} = \begin{pmatrix}
 \begin{pmatrix} F^{tt,tt}_{-,\D}  & 0 \\ 0 & 0 \end{pmatrix} \\
 \begin{pmatrix} F^{tt,tt}_{-,\D} & 0 \\ 0 & 0 \end{pmatrix} \\[1em]
 \begin{pmatrix} F^{tt,tt}_{+,\D} & 0 \\ 0 & 0 \end{pmatrix} \\
 \begin{pmatrix} 0 & 0 \\ 0 & F^{DD,DD}_{-,\D} \end{pmatrix} \\ 
 \begin{pmatrix} 0 & 0 \\ 0 & F^{DD,DD}_{-,\D} \end{pmatrix} \\[1em] 
 \begin{pmatrix} 0 & 0 \\ 0 & F^{DD,DD}_{+,\D} \end{pmatrix} \\
 0 \\
 \begin{pmatrix} 0 & \frac{1}{2} \\ \frac{1}{2} & 0 \end{pmatrix} F^{tt,DD}_{-,\D} \\[1em]
 \begin{pmatrix} 0 & \frac{1}{2} \\ \frac{1}{2} & 0 \end{pmatrix} F^{tt,DD}_{+,\D} \\
 \end{pmatrix} \, , \quad
\vec{V}^{t\bar t,-}_{\Delta} = \begin{pmatrix}
 F^{tt,tt}_{-,\D}  \\
 -F^{tt,tt}_{-,\D}  \\ 
 -F^{tt,tt}_{+,\D}  \\ 0 \\ 0 \\ 0 \\ 0 \\ 0 \\ 0 \end{pmatrix}\:, \quad
 \vec{V}^{D \bar D,-}_{\Delta} = \begin{pmatrix}
 0 \\ 0 \\ 0 \\ F^{DD,DD}_{-,\D} \\ 
 -F^{DD,DD}_{-,\D} \\ 
 -F^{DD,DD}_{+,\D}  \\ 0 \\ 0 \\ 0 \end{pmatrix}\:,
\end{equation}
\begin{equation}
\vec{V}^{tt}_{\Delta} = \begin{pmatrix}
 0 \\
  F^{tt,tt}_{-,\D}  \\ 
 -F^{tt,tt}_{+,\D}  \\ 0 \\ 0 \\ 0 \\ 0 \\ 0 \\ 0  \end{pmatrix}\:, \quad
 \vec{V}^{DD}_{\Delta} = \begin{pmatrix}
 0 \\ 0 \\ 0 \\ 0 \\ 
  F^{DD,DD}_{-,\D} \\ 
 -F^{DD,DD}_{+,\D} \\ 0 \\ 0 \\ 0  \end{pmatrix}\:, \quad
  \vec{V}^{tD}_{\Delta,S} = \begin{pmatrix}
  0 \\ 0 \\ 0 \\ 0 \\ 0 \\ 0 \\
  (-1)^S F^{tD,tD}_{-,\D}  \\ 
         F^{Dt,tD}_{-,\D}  \\ 
       - F^{Dt,tD}_{+,\D}  \end{pmatrix} \: .
\end{equation}

\subsection{One real and one complex scalar}
\label{sec:cv-real-cplx}

The crossing equations for one real scalar $\phi_1$ and one complex scalar $\phi_2$ appear in eq.~\eqref{eq:cross-eq-real-cplx}, with crossing vectors
\begin{align}
 & \vec{V}^+_{\Delta} = \begin{pmatrix}
 \begin{pmatrix} F^{11,11}_{-,\D} & 0 \\ 0 & 0  \end{pmatrix} \\[1em]
 \begin{pmatrix} 0 & 0 \\ 0 &  F^{22,22}_{-,\D} \end{pmatrix} \\[1em]
 \begin{pmatrix} 0 & 0 \\ 0 &  F^{22,22}_{-,\D} \end{pmatrix} \\[1em]
 \begin{pmatrix} 0 & 0 \\ 0 &  F^{22,22}_{+,\D} \end{pmatrix} \\
 0 \\
 \begin{pmatrix} 0 & \frac{1}{2} \\ \frac{1}{2} & 0 \end{pmatrix} F^{11,22}_{-,\D} \\[1em] 
 \begin{pmatrix} 0 & \frac{1}{2} \\ \frac{1}{2} & 0 \end{pmatrix} F^{11,22}_{+,\D}
 \end{pmatrix} \: , \qquad
 \vec{V}^-_{\Delta} 
 = \begin{pmatrix} 
 0 \\   
 F^{22,22}_{-,\D}  \\
 -F^{22,22}_{-,\D} \\
 -F^{22,22}_{+,\D} \\ 
 0 \\ 0 \\ 0  
 \end{pmatrix} \:,
 \end{align}
 \begin{align}
 & \vec{V}^{22}_{\Delta} = \begin{pmatrix}
 0 \\ 0 \\ F^{22,22}_{-,\D} \\ -F^{22,22}_{+,\D} \\ 0 \\ 0 \\ 0 \end{pmatrix} \:, \qquad
 \vec{V}^{12}_{\Delta,S} = \begin{pmatrix}
 0 \\ 0 \\ 0 \\ 0 \\ (-1)^S F^{12,12}_{-,\D} \\ F^{21,12}_{-,\D} \\ -F^{21,12}_{+,\D} \end{pmatrix} \: .
\end{align}
Here we are using $F_{\pm,\Delta}^{ij,kl} = F_{\pm,\Delta}^{\phi_i\phi_j,\phi_k\phi_l}(\xi)$, and we do not distinguish between $\phi_2$ and $\bar \phi_2$ because they have the same scaling dimension.

\subsection{Two complex scalars}
\label{sec:cv-cplx-cplx}

The crossing equations for two complex scalars given in eq.~\eqref{eq:cross-eq-cplx-cplx} are written in terms of the following crossing vectors:
\begin{equation}
 \vec{V}_{\Delta,S} = \begin{pmatrix}
 \begin{pmatrix} F^{11,11}_{-,\D}  & 0 \\ 0 & 0 \end{pmatrix} \\
 \begin{pmatrix} (-1)^S F^{11,11}_{-,\D} & 0 \\ 0 & 0 \end{pmatrix} \\[1em]
 \begin{pmatrix} (-1)^S F^{11,11}_{+,\D} & 0 \\ 0 & 0 \end{pmatrix} \\
 \begin{pmatrix} 0 & 0 \\ 0 & F^{22,22}_{-,\D} \end{pmatrix} \\ 
 \begin{pmatrix} 0 & 0 \\ 0 & (-1)^S F^{22,22}_{-,\D} \end{pmatrix} \\[1em] 
 \begin{pmatrix} 0 & 0 \\ 0 & (-1)^S F^{22,22}_{+,\D} \end{pmatrix} \\
 0 \\ 0 \\
 \begin{pmatrix} 0 & \frac{1}{2} \\ \frac{1}{2} & 0 \end{pmatrix} F^{11,22}_{-,\D} \\[1em]
 \begin{pmatrix} 0 & \frac{1}{2} \\ \frac{1}{2} & 0 \end{pmatrix} F^{11,22}_{+,\D} \\
 \begin{pmatrix} 0 & \frac{1}{2} \\ \frac{1}{2} & 0 \end{pmatrix} (-1)^S F^{11,22}_{-,\D} \\[1em]
 \begin{pmatrix} 0 & \frac{1}{2} \\ \frac{1}{2} & 0 \end{pmatrix} (-1)^S F^{11,22}_{+,\D} \\
 \end{pmatrix} \, , \quad
\vec{V}^{11}_{\Delta} = \begin{pmatrix}
 0 \\
  F^{11,11}_{-,\D}  \\ 
 -F^{11,11}_{+,\D}  \\ 0 \\ 0 \\ 0 \\ 0 \\ 0 \\ 0 \\ 0 \\ 0 \\ 0 \end{pmatrix}\:, \quad
 \vec{V}^{22}_{\Delta} = \begin{pmatrix}
 0 \\ 0 \\ 0 \\ 0 \\ 
  F^{22,22}_{-,\D} \\ 
 -F^{22,22}_{+,\D} \\ 0 \\ 0 \\ 0 \\ 0 \\ 0 \\ 0 \end{pmatrix}\:,
\end{equation}
\begin{equation}
  \vec{V}^{1\bar2}_{\Delta,S} = \begin{pmatrix}
  0 \\ 0 \\ 0 \\ 0 \\ 0 \\ 0 \\
  (-1)^S F^{12,12}_{-,\D} \\ 
 -(-1)^S F^{12,12}_{+,\D} \\ 
         F^{21,12}_{-,\D}  \\ 
       - F^{21,12}_{+,\D}  \\ 
 0 \\ 0 \end{pmatrix}\:, \quad
 \vec{V}^{12}_{\Delta,S} = \begin{pmatrix}
 0 \\ 0 \\ 0 \\ 0 \\ 0 \\ 0 \\
 (-1)^S F^{12,12}_{-,\D} \\ 
 (-1)^S F^{12,12}_{+,\D} \\ 
 0 \\ 0 \\ 
   F^{21,12}_{-,\D} \\ 
 - F^{21,12}_{+,\D} \end{pmatrix} \: .
\end{equation}